\documentstyle[12pt,epsfig]{article}
\begin{document}
\parskip=5pt plus 1pt minus 1pt

\begin{flushright}
\underline{\bf Int. J. Mod. Phys. A 19 (2004) 1} \\
\underline{\underline{\bf hep-ph/0307359}} 
%{\small BIHEP-TH-2003-25} 
\end{flushright}

\vspace{0.1cm}

\begin{center}
{\large\bf Flavor Mixing and CP Violation of Massive Neutrinos}
\footnote{Invited review article published in 
{\sf Int. J. Mod. Phys. A 19 (2004) 1.}}
\end{center}

\vspace{0.4cm}
\begin{center}
{\bf Zhi-zhong Xing}
\footnote{Electronic address: xingzz@mail.ihep.ac.cn} \\
{\it Institute of High Energy Physics, Chinese Academy of 
Sciences \\
P.O. Box 918 (4), Beijing 100039, China}
\end{center}

\vspace{2.5cm}
\begin{abstract}
We present an overview of recent progress in the phenomenological
study of neutrino masses, lepton flavor mixing and CP violation.
We concentrate on the {\it model-independent} properties of massive
neutrinos, both in vacuum and in matter. Current experimental
constraints on the neutrino mass spectrum and the lepton flavor
mixing parameters are summarized. The Dirac- and Majorana-like 
phases of CP violation, which are associated respectively with the 
long-baseline neutrino oscillations and the neutrinoless double 
beta decay, are discussed in detail. The seesaw mechanism, the 
leptogenesis scenario and the strategies to construct 
lepton mass matrices are briefly described. The features of flavor 
mixing between one sterile neutrino and three active neutrinos are 
also explored. 
\end{abstract}

\newpage

\tableofcontents
\setcounter{footnote}{0}

\newpage

\section{Introduction}
\setcounter{equation}{0}
\setcounter{figure}{0}

Neutrino physics is a peculiar part of flavor physics, because
neutrinos belong to {\it super-light} flavors in comparison with 
charged leptons and quarks. There are three central concepts in 
flavor physics: mass, flavor mixing and CP violation. 
\begin{itemize}
\item       {\it Mass} represents the inertial energy possessed 
by a particle when it exists at rest. A massless particle has
no way to exist at rest - instead, it must always move at
the speed of light. A massive fermion (lepton or quark) must exist 
in both left-handed and right-handed states, because the field 
operators responsible for the nonzero mass of a fermion have to be
bilinear products of the spinor fields which flip the fermion's 
handedness.
\item       {\it Flavor mixing} measures the mismatch between 
flavor eigenstates and mass eigenstates of leptons or quarks, 
caused by the Higgs interactions. Flavor 
eigenstates are the members of the weak isospin doublets that 
transform into each other via the interaction with $W^\pm$ 
bosons, while mass eigenstates are the states of definite masses 
created by the interaction with Higgs bosons. If neutrinos 
were massless, lepton flavor mixing would not exist.
\item       {\it CP violation} means that matter and 
antimatter or a reaction and its CP-conjugated process are
distinguishable. It may manifest itself in the weak interactions 
through nontrivial complex phases residing in the flavor 
mixing matrix of leptons or quarks. If CPT is invariant, 
CP violation will give rise to T violation, or vice versa.
\end{itemize}
The problems of fermion masses, flavor mixing and CP violation 
are correlated with one another and fundamentally important in
particle physics. The study of these problems will ultimately 
help us understand the nature of matter and the matter-antimatter
asymmetry of the universe. Since the early 1960's, a lot of 
experimental efforts have been made to measure the parameters 
of quark mixing and CP violation. The problem of lepton mixing 
and CP violation was not at the front of experimental particle
physics for a long time, however, partly because neutrinos were 
assumed to be massless in the successful theory of weak and 
electromagnetic interactions -- the standard electroweak 
model \cite{SM}. This situation has dramatically changed since 
the compelling evidence in favor of atmospheric and solar 
neutrino oscillations was achieved by Super-Kamiokande 
(SK) \cite{SK}, SNO \cite{SNO}, KamLAND \cite{KM} and 
K2K \cite{K2K} Collaborations in the past five years. 

Neutrino oscillation is a quantum phenomenon which can naturally
occur if neutrinos are massive and lepton flavors are mixed.
Thanks to the elegant SK, SNO, KamLAND and K2K experiments, we
are now convinced that the long-standing solar neutrino ($\nu_e$) 
deficit and the atmospheric neutrino ($\nu_\mu$) anomaly are both 
due to neutrino oscillations. The study of neutrino masses and 
lepton flavor mixing is therefore becoming one of the hottest 
fronts of today's particle physics and cosmology. 

Our present knowledge on the properties of neutrinos comes not 
only from a number of neutrino oscillation experiments, 
but also from the direct-mass-search experiments, the neutrinoless 
double beta decay experiments and the astrophysical or cosmological 
observations \cite{PDG02}. The following is a partial list of 
what we have known about neutrino masses and lepton flavor mixing.    
\begin{itemize}
\item      Neutrinos are massive but their masses are tiny. The 
mass scale of three active neutrinos ($\nu_e$, $\nu_\mu$ and 
$\nu_\tau$) is expected to be at or below ${\cal O}(1)$ eV.
\item      Two independent mass-squared differences of three 
neutrinos, which are associated separately with solar and 
atmospheric neutrino oscillations, are very small and have a 
strong hierarchy. Typically, 
$\Delta m^2_{\rm sun} \sim {\cal O}(10^{-5}) ~ {\rm eV}^2$ and 
$\Delta m^2_{\rm atm} \sim {\cal O}(10^{-3}) ~ {\rm eV}^2$ hold.
\item      Two lepton flavor mixing angles, which are related 
separately to solar and atmospheric neutrino oscillations, are 
much larger than the Cabibbo angle of quark mixing 
($\theta_{\rm C} \approx 13^\circ$). Typically, 
$\theta_{\rm sun} \sim 33^\circ$ and 
$\theta_{\rm atm} \sim 45^\circ$ hold.
\item      The other lepton flavor mixing angle, which is relevant 
to the CHOOZ reactor experiment for neutrino 
oscillations \cite{CHOOZ}, is very small and even vanishing. The
generous upper bound of this angle is 
$\theta_{\rm chz} \leq \theta_{\rm C}$ at present.
\end{itemize}
It is obvious that the mass spectrum of neutrinos and the mixing 
pattern of lepton flavors are very different from those of quarks.

However, there exist many open questions about massive neutrinos and
lepton flavor mixing. For example,
\begin{itemize}
\item      Are massive neutrinos Dirac or Majorana particles? If
massive neutrinos are Dirac particles, they can be distinguished 
from their antiparticles. By definition, a Majorana neutrino is
identical to its antiparticle. It is possible to identify the
Majorana nature of massive neutrinos through the observation of
the neutrinoless double beta decay of some even-even nuclei, in 
which the total lepton number is not conserved. 
\item      How many neutrino species are there? We have known that
there are three species of active neutrinos ($\nu_e$, $\nu_\mu$
and $\nu_\tau$), corresponding to three species of charged leptons.
It remains unclear whether the light {\it sterile} neutrinos, which 
have been assumed to interpret the controversial LSND data \cite{LSND} 
on $\overline{\nu}_\mu \rightarrow \overline{\nu}_e$ oscillations,
are really existent or not.
\item      Why are neutrino masses so tiny? The fact that the masses
of neutrinos are considerably smaller than the masses of charged 
leptons or quarks is a big puzzle to particle physicists. Although
a lot of theoretical models about neutrino masses have been proposed
at either low or high energy scales, none of them has proved to be 
very successful and conceivable.
\item      What is the absolute scale of neutrino masses? It is 
very important to know the absolute values of three neutrino masses,
because they are fundamental parameters of flavor physics. The 
mass scale of neutrinos is likely to indicate the energy scale of
new physics responsible for the generation of neutrino masses and 
lepton flavor mixing. Unfortunately, we are only aware of the upper 
bounds of neutrino masses. 
\item      Why are the mixing angles $\theta_{\rm sun}$ and
$\theta_{\rm atm}$ so big? The {\it bi-large} neutrino mixing 
pattern is also a mystery to many theorists, because it is 
``anomalously'' different from the familiar {\it tri-small} quark 
mixing pattern. Although the lepton flavor mixing angles are in
general expected to relate to the mass spectra of charged leptons 
and neutrinos, their specific relations have not convincingly
or model-independently been established.
\item      How small is the mixing angle $\theta_{\rm chz}$?
Current experimental data yield an upper bound 
$\theta_{\rm chz} \leq \theta_{\rm C}$, but the possibility 
$\theta_{\rm chz} =0^\circ$ cannot be excluded. The smallness of 
$\theta_{\rm chz}$ requires a good theoretical reason, so does
the vanishing of $\theta_{\rm chz}$. If $\theta_{\rm chz} =0^\circ$
held, there would be no chance to observe leptonic CP or T
violation in normal neutrino-neutrino and antineutrino-antineutrino 
oscillations.
\item      Is there leptonic CP violation? A necessary condition 
for the existence of CP and T violation in normal neutrino 
oscillations is $\theta_{\rm chz} \neq 0$. As CP violation has 
been discovered in the quark sector, we feel that there is no 
reason why CP should be conserved in the lepton sector. It is very 
difficult to detect the effects of CP or T violation in any 
realistic neutrino oscillation experiments, however. 
\item      Can the leptonic CP-violating phases be determined?
If lepton flavor mixing is correlated with CP violation, one has
to determine the relevant CP-violating phases through various 
possible experiments. The neutrinoless double beta decay and 
long-baseline appearance neutrino oscillations are expected to
be sensitive to the Majorana- and Dirac-like phases of CP violation,
respectively. To implement such measurements remains a big challenge 
to experimentalists.
\end{itemize}
Of course, much more experimental and theoretical efforts
are needed to make, in order to answer the important questions 
listed above. Much more phenomenological attempts are also 
needed to make, so as to bridge the gap between experimental
data and theoretical models.

The purpose of this article is to give an overview of recent 
progress in the phenomenological study of neutrino masses,
lepton flavor mixing and CP violation. We concentrate
on the {\it model-independent} properties of massive neutrinos,
although it is unavoidable to introduce two very attractive
ans$\rm\ddot{a}$tze in today's neutrino physics -- the seesaw 
mechanism and the thermal leptogenesis scenario.

The remaining parts of this article are organized as follows.
Section 2 is devoted to a review of the neutrino mass spectrum. 
First of all, we make a short introduction to the Dirac and 
Majorana neutrino masses, and to the well-known seesaw mechanism. 
Then current experimental constraints on neutrino masses, 
including those from the neutrinoless double beta decay, 
kinematic measurements, neutrino oscillations and cosmological 
observations, are briefly summarized. Finally some comments 
are given on the strategies to construct the phenomenological
textures of lepton mass matrices.

In section 3, the model-independent features of lepton flavor 
mixing are illustrated both in vacuum and in matter. We
present a classification of various parametrizations of the 
$3\times 3$ lepton flavor mixing matrix, and highlight one of 
them which is particularly useful for the study of neutrino 
oscillations. A few concise sum rules for neutrino masses
and lepton flavor mixing in matter are derived. Several 
constant mixing patterns of massive neutrinos are introduced.
Finally we point out the differences and similarities between
the phenomenon of lepton flavor mixing and that of quark 
flavor mixing. 

Section 4 is devoted to leptonic CP and T violation. The
rephasing invariants of CP violation and the unitarity
triangles are described both in vacuum and in matter. A few
salient features of CP and T violation in long-baseline 
neutrino oscillations are also discussed. We pay some special 
interest to the thermal leptogenesis scenario to interpret 
the observed baryon-antibaryon asymmetry of the universe. 
Finally we comment on the possible connection or disconnection 
between the leptonic CP-violating quantities at high and low 
energy scales.

The conclusion and outlook are presented in section 5.

It is worth remarking that the main body of this article 
deals with neutrino masses, lepton flavor mixing and CP 
violation in the scheme of three lepton families. To be
complete, the properties of flavor mixing between one sterile 
neutrino ($\nu_s$) and three active neutrinos ($\nu_e$, 
$\nu_\mu$ and $\nu_\tau$) are discussed in appendix A.

\section{Neutrino Mass Spectrum}
\setcounter{equation}{0}
\setcounter{figure}{0}

\subsection{Theoretical Preliminaries}

The standard model of electromagnetic and weak interactions is
based on the gauge group 
$\rm SU(2)_{\rm L} \times U(1)_{\rm Y}$ \cite{SM}. 
In this framework only the left-handed leptons (and quarks), 
which transform as $\rm SU(2)$ doublets, take part in the 
charged-current weak interactions:
\begin{equation}
-{\cal L}_{\rm cc} \; =\; \frac{g}{\sqrt{2}} ~ 
\overline{(\nu_e, \nu_\mu, \nu_\tau)^{~}_{\rm L}} ~ \gamma^\mu
\left (\matrix{
e \cr
\mu \cr
\tau \cr} \right )_{\rm L} W^+_\mu ~ + ~ {\rm h.c.} \; ,
%       (2.1)
\end{equation}
where $(e, \mu, \tau)$ and $(\nu_e, \nu_\mu, \nu_\tau)$ are
the flavor eigenstates of charged leptons and neutrinos, 
respectively. If $\rm SU(2)_{\rm L}$ were an exact symmetry,
all leptons would be massless and their flavor eigenstates
would be physically indistinguishable from their mass 
eigenstates. In reality, however, this gauge symmetry is
badly broken. A spontaneous breakdown of the $\rm SU(2)_{\rm L}$ 
symmetry is realized in the standard model by means of the Higgs
mechanism. After the symmetry breaking, the charged leptons (and
quarks) acquire their masses through the Yukawa interactions:
\begin{equation}
-{\cal L}_l \; =\; 
\overline{(e, ~\mu, ~\tau)^{~}_{\rm L}} ~ M_l \left (\matrix{
e \cr
\mu \cr
\tau \cr} \right )_{\rm R} + ~ {\rm h.c.} \; , 
%       (2.2)
\end{equation}
where $M_l$ denotes the charged lepton mass matrix and its scale 
is characterized by the electroweak symmetry breaking scale 
$v \approx 174$ GeV. The vanishing of
three neutrino masses follows as a straightforward consequence 
of the symmetry structure of the standard model, in which only a 
single Higgs doublet exists and the lepton number conservation is 
assumed
%%%%%%%%%%%%%%%%%%%%%%%%%%
\footnote{It is actually the $(B-L)$ symmetry that makes neutrinos 
exactly
massless in the standard model, where $B$ denotes the baryon number
and $L$ stands for the total lepton number. The reason is simply
that a neutrino $\nu$ and an antineutrino $\overline{\nu}$ have
different values of $(B-L)$ \cite{Witten}. Thus the naive argument
for massless neutrinos is valid to all orders in perturbation and
non-perturbation theories, if $(B-L)$ is an exact symmetry.}.
%%%%%%%%%%%%%%%%%%%%%%%%%

Note that the mass eigenstates of charged leptons can always be 
chosen to coincide with their flavor eigenstates through an 
appropriate but physically-irrelevant unitary transformation of 
the right-handed fields. In such a specific flavor basis the
coincidence between mass and flavor eigenstates of neutrinos can 
also be achieved, if neutrinos are assumed to be exactly massless 
Weyl particles. Hence there is no lepton flavor mixing within the 
framework of the standard electroweak model.

\subsubsection{Dirac and Majorana Masses}

However, the assumption of lepton number conservation or 
masslessness of neutrinos is not assured by any basic symmetry 
principle of particle physics. Most reasonable extensions of 
the standard model (such as the grand unified theories) 
do allow the absence of lepton
number conservation and the existence of nonvanishing neutrino 
masses. If neutrinos are really massive and their masses are 
non-degenerate, it will in general be impossible to find a basis 
of the flavor space in which the coincidence between flavor and
mass eigenstates holds both for charged leptons and for neutrinos. 
In other words, the flavor mixing phenomenon is naturally expected 
to appear between three charged leptons and three massive neutrinos, 
just like the flavor mixing between three up-type quarks and three 
down-type quarks \cite{CKM}. 

If neutrinos have nonzero masses, they may be either Dirac or 
Majorana particles. The field of a massive Dirac neutrino 
describes four independent states: left-handed and right-handed 
particle states ($\nu^{~}_{\rm L}$ and $\nu^{~}_{\rm R}$) as well 
as left-handed and right-handed antiparticle states 
($\bar{\nu}^{~}_{\rm L}$ and $\bar{\nu}^{~}_{\rm R}$). Among
them the $\nu^{~}_{\rm L}$ and $\bar{\nu}^{~}_{\rm R}$ states, 
which already
exist in the standard model, can take part in weak interactions.
The $\nu^{~}_{\rm R}$ and $\bar{\nu}^{~}_{\rm L}$ states 
need to be introduced into the standard model as necessary
ingredients to give the Dirac neutrino a mass, but they should
be ``sterile'' in the sense that they do not take part in
the normal weak interactions. A Dirac mass term, which conserves
the total lepton number but violates the law of individual
lepton flavor conservation, can be written as follows:
\begin{equation}
- {\cal L}_{\rm D} \; =\; 
\overline{(\nu_e, \nu_\mu, \nu_\tau)^{~}_{\rm L}} 
~ M_{\rm D} \left (\matrix{
\nu_e \cr
\nu_\mu \cr
\nu_\tau \cr} \right )_{\rm R} + ~ {\rm h.c.} \; , 
%        (2.3)
\end{equation}
where $M_{\rm D}$ denotes the $3\times 3$ Dirac neutrino mass 
matrix. The mass term in (2.3) is quite similar to the mass term 
of charged leptons in (2.2), hence the scale of $M_{\rm D}$ 
should also be characterized by the gauge symmetry breaking 
scale $v$. In this case the Yukawa coupling constants of three 
neutrinos must be extremely smaller than those of three charged 
leptons or six quarks, such that tiny neutrino masses can 
result. This dramatic difference between the Yukawa couplings of
neutrinos and charged leptons (or quarks) is commonly considered 
to be very unnatural in a sound theory of fermion mass generation, 
however. 

On the other hand, the neutrino $\nu$ may be a Majorana
particle, which has only two independent states 
of the same mass ($\nu^{~}_{\rm L}$ and $\bar{\nu}^{~}_{\rm R}$,
or $\nu^{~}_{\rm R}$ and $\bar{\nu}^{~}_{\rm L}$). By definition,
a Majorana neutrino is its own antiparticle:
$\nu^{\rm c} \equiv C \bar{\nu}^{\rm T} = 
e^{i\Theta}\nu$ \cite{Majorana},
where $C$ denotes the charge-conjugation operator and $\Theta$
is an arbitrary real phase. A Majorana mass term, which violates
both the law of total lepton number conservation and that of 
individual lepton flavor conservation, can be written either as
\begin{equation}
-{\cal L}_{\rm M(L)} \; =\; \frac{1}{2} ~
\overline{(\nu_e, \nu_\mu, \nu_\tau)^{~}_{\rm L}}
~ M_{\rm L} \left (\matrix{
\nu^{\rm c}_e \cr
\nu^{\rm c}_\mu \cr
\nu^{\rm c}_\tau \cr} \right )_{\rm R} + ~ {\rm h.c.} \; ,
%        (2.4)
\end{equation}
or as
\begin{equation}
-{\cal L}_{\rm M(R)} \; =\; \frac{1}{2} ~ 
\overline{(\nu^{\rm c}_e, \nu^{\rm c}_\mu, 
\nu^{\rm c}_\tau)^{~}_{\rm L}}
~ {M}_{\rm R} \left (\matrix{
\nu_e \cr
\nu_\mu \cr
\nu_\tau \cr} \right )_{\rm R} + ~ {\rm h.c.} \; ,
%        (2.5)
\end{equation}
where $M_{\rm L}$ and $M_{\rm R}$ denote the symmetric 
$3\times 3$ mass matrices of left-handed and right-handed
Majorana neutrinos, respectively. Note that the mass term
${\cal L}_{\rm M(L)}$ cannot naturally arise from a simple theory
of electroweak interactions which is invariant under the 
$\rm SU(2)_L \times U(1)_Y$ gauge transformation and has no 
$\rm SU(2)_L$ triplet field (like the case in the standard model). 
It is possible to incorporate the mass term 
${\cal L}_{\rm M(R)}$ into an electroweak theory with 
$\rm SU(2)_L \times U(1)_Y$ gauge symmetry, because right-handed 
neutrinos are $\rm SU(2)_L$ singlets. In general,
a neutrino mass Lagrangian may include all of the above-mentioned 
terms: ${\cal L}_{\rm D}$, ${\cal L}_{\rm M(L)}$ 
and ${\cal L}_{\rm M(R)}$, in which $M_{\rm D}$,
$M_{\rm L}$ and $M_{\rm R}$ are complex mass matrices.

A variety of specific theoretical and phenomenological models 
have been prescribed in the literature to interpret how lepton 
masses are generated and why neutrino masses are so tiny. 
Regardless of the energy scales at which those models are built, 
the mechanisms responsible for fermion mass generation and flavor 
mixing can roughly be classified into five different 
categories \cite{FXReview}:
(a) Radiative mechanisms \cite{RM}; (b) Texture zeros \cite{TZ};
(c) Family symmetries \cite{FS}; (d) Seesaw mechanisms \cite{SS}; 
and (e) Extra dimensions \cite{ED}. Among them, the seesaw 
mechanisms are particularly natural and interesting. A brief 
introduction to the seesaw idea will be presented in section 2.1.2.
For recent reviews of other interesting models and 
ans$\rm\ddot{a}$tze on neutrino masses, we refer the reader to 
Ref. \cite{FXReview} and Refs. \cite{Altarelli}--\cite{Review}. 

\subsubsection{The Seesaw Mechanism}

A simple extension of the standard model is to include one
right-handed neutrino in each of three lepton families, while the 
Lagrangian of electroweak interactions keeps invariant under the
$\rm SU(2)_L \times U(1)_Y$ gauge transformation. After spontaneous
symmetry breaking, the resultant lepton mass term 
${\cal L}_{\rm mass}$ consists of ${\cal L}_l$ (charged lepton term),
${\cal L}_{\rm D}$ (Dirac neutrino term) and ${\cal L}_{\rm M(R)}$
(righ-handed Majorana neutrino term), which have been given 
in (2.2), (2.3) and (2.5). To be explicit, we have
\begin{equation}
- {\cal L}_{\rm mass} \; = \; - {\cal L}_l ~ + ~ \frac{1}{2} 
\overline{(\nu, ~\nu^{\rm c} )^{~}_{\rm L}}
\left ( \matrix{
0     & M_{\rm D} \cr
M^T_{\rm D}    & M_{\rm R} \cr} \right ) 
\left ( \matrix{
\nu^{\rm c} \cr \nu \cr} \right )_{\rm R} \; ,
%       (2.6)
\end{equation}
where $\nu$ denotes the column vector of $\nu_e$, $\nu_\mu$ and
$\nu_\tau$ fields; i.e., 
$\nu^T \equiv (\nu_e, \nu_\mu, \nu_\tau )$ or $\nu^{\rm c} \equiv 
(\nu^{\rm c}_e, \nu^{\rm c}_\mu, \nu^{\rm c}_\tau )^T$.
In obtaining (2.6), we have used the relation
$\overline{\nu^{~}_{\rm L}} M_{\rm D} \nu^{~}_{\rm R} =
\overline{(\nu^{\rm c})_{\rm L}} M^T_{\rm D} (\nu^{\rm c})^{~}_{\rm R}$
as well as the properties of $\nu$ and $\nu^{\rm c}$.
As pointed out in section 2.1.1, the scale of $M_{\rm D}$ is 
characterized by the electroweak symmetry breaking scale $v$.
The scale of $M_{\rm R}$ can naturally be much higher than $v$,
because right-handed neutrinos are $\rm SU(2)_L$ singlets and their
corresponding mass term is not subject to the scale of gauge symmetry 
breaking. It has commonly been expected that the scale of $M_{\rm R}$ 
is not far away from the grand unified theory (GUT) scale 
$\Lambda_{\rm GUT} \sim 10^{16}$ GeV. In this case,
the smallness of left-handed neutrino masses ($m_i \ll v$ for $i=1,2,3$)
is essentially attributed to the largeness of right-handed neutrino 
masses ($M_i \gg v$ for $i=1,2,3$). Such an elegant idea, the 
so-called seesaw mechanism, was first proposed by Yanagida and by 
Gell-Mann, Ramond and Slansky in 1979 \cite{SS}. Fig. 2.1 illustrates
the seesaw mass term given in (2.6).
%%%%%%%%%%%%%%%%%%%% Fig. 2.1 %%%%%%%%%%%%%%%%
\begin{figure}[t]
\vspace{-2.9cm}
\epsfig{file=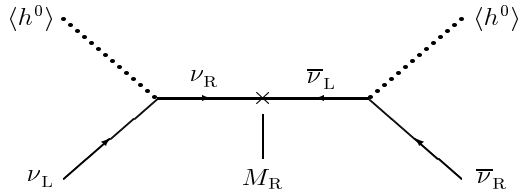,bbllx=1.5cm,bblly=2cm,bburx=19cm,bbury=31cm,%
width=15.5cm,height=22cm,angle=0,clip=}
\vspace{-16.2cm}
\caption{Illustrative plot for the seesaw mass term.}
\end{figure}
%%%%%%%%%%%%%%%%%%%%%%%%%%%%%%%%%%%%%%%%%%%%

To show how the seesaw mechanism works, we diagonalize the
symmetric $6\times 6$ neutrino mass matrix in (2.6) by use of the
following unitary transformation:
\begin{equation}
\left ( \matrix{
V  &  R \cr
S  &  U \cr} \right )^\dagger 
\left ( \matrix{
0     & M_{\rm D} \cr
M^T_{\rm D}    & M_{\rm R} \cr} \right ) 
\left ( \matrix{
V  &  R \cr
S  &  U \cr} \right )^*  = 
\left ( \matrix{
\overline{M}_\nu  &  0 \cr
0  &  \overline{M}_{\rm R} \cr} \right ) \; ,
%       (2.7)
\end{equation}  
where $R$, $S$, $U$ and $V$ are the $3\times 3$ 
matrices; $\overline{M}_\nu$ and $\overline{M}_{\rm R}$
denote the $3\times 3$ diagonal mass matrices with eigenvalues
$m_i$ and $M_i$ (for $i=1,2,3$), respectively. 
It is straightforward to obtain 
\begin{eqnarray}
V \overline{M}_\nu V^T & = & \Delta_V ~ - ~ 
M_{\rm D} M^{-1}_{\rm R} M^T_{\rm D} \; ,
\nonumber \\
U \overline{M}_{\rm R} U^T & = & \Delta_U ~ + ~ M_{\rm R} \; ,
%       (2.8)
\end{eqnarray}
where $\Delta_V \equiv V S^\dagger M^T_{\rm D} R^* U^T M^{-1}_{\rm R} 
U R^\dagger M_{\rm D} S^* V^T$ and
$\Delta_U \equiv M^T_{\rm D} R^* U^T + U R^\dagger M_{\rm D}$.
Note that both $R$ and $S$ in (2.7) are expected to be
of order $M_{\rm D}/M_{\rm R}$, as a result of the huge hierarchy
between the scales of $M_{\rm D}$ and $M_{\rm R}$. Hence
$\Delta_V \approx 0$ and $\Delta_U \approx 0$ are excellent
approximations for (2.8). We then arrive at
$M_{\rm R} \approx U \overline{M}_{\rm R} U^T$ for the heavy
(right-handed) neutrino mass matrix and
the well-known seesaw formula for the light (left-handed) neutrino 
mass matrix 
%%%%%%%%%%%%%%%%%%%%%%
\footnote{Note that the minus sign on the right-hand side of (2.9)
can always be absorbed via a redefinition of the approximately 
unitary matrix 
$V$. For instance, $\hat{V} \overline{M}_\nu \hat{V}^T \approx
M_{\rm D} M^{-1}_{\rm R} M^T_{\rm D}$ with $\hat{V} \equiv iV$ holds.},
%%%%%%%%%%%%%%%%%%%%%
\begin{equation}
M_\nu = V \overline{M}_\nu V^T \; \approx \; 
- M_{\rm D} M^{-1}_{\rm R} M^T_{\rm D} \; .
%       (2.9)
\end{equation}
The seesaw 
mechanism was actually motivated by the SO(10) GUTs \cite{SO10},
which include right-handed neutrinos automatically together with
other charged fermions in irreducible {\bf 16}-dimensional multiplets.
In such an attractive framework, $M_{\rm D}$ can simply be related to 
the up-type quark mass matrix $M_{\rm u}$, whose possible textures 
have well been restricted by current experimental data \cite{FXReview}. 
If both $M_{\rm D}$ ($= M_{\rm u}$) and $M_{\rm R}$ 
were assumed to be diagonal, 
$M_\nu$ would also be diagonal and its three mass eigenvalues would be
$m_1 = m^2_u/M_1$, $m_2 = m^2_c/M_2$ and $m_3 = m^2_t/M_3$. In this
illustrative case, $m_3 \sim 0.1$ eV may simply result from 
$M_3 \sim 10^{14}$ GeV, a scale not far away from the GUT scale 
$\Lambda_{\rm GUT}$. 

Without loss of generality, both the charged lepton mass matrix 
$M_l$ and the heavy Majorana neutrino mass matrix $M_{\rm R}$ can be  
arranged to be diagonal and have positive mass eigenvalues. In this 
specific flavor basis, the approximately 
unitary matrix $V$ in (2.9) links the
mass eigenstates of light neutrinos to their flavor eigenstates;
i.e., it is relevant to the phenomenon of lepton flavor mixing at 
low energies (see section 3.1 for more detailed discussions). A full
parametrization of $V$ requires three mixing angles and three
CP-violating phases. Thus $M_\nu$ consists of nine free
parameters. Note that the Dirac neutrino mass matrix $M_{\rm D}$ 
can be parametrized as $M_{\rm D} = i V \sqrt{\overline{M}_\nu} O
\sqrt{M_{\rm R}} ~$ \cite{Casas}, in
which $O$ is a complex orthogonal matrix containing three rotation
angles and three phase angles. It seems that $M_{\rm D}$ in this
parametrization consists of eighteen parameters: six from $V$,
three from $\overline{M}_\nu$, six from $O$ and three from 
$M_{\rm R}$. However, only fifteen of them (nine real parameters
and six CP-violating phases) are independent, due to the seesaw
relation. We conclude that the lepton mass matrices $M_l$, 
$M_{\rm D}$ and $M_{\rm R}$ totally have twenty-one free parameters:
three from $M_l$, fifteen from $M_{\rm D}$ and three from $M_{\rm R}$
in the chosen flavor basis
%%%%%%%%%%%%%%%%%%%%%%%%%
\footnote{The parameter counting is identical in models with and
without supersymmetry \cite{Raidal}.}.
%%%%%%%%%%%%%%%%%%%%%%%%%

It is worth mentioning that (2.9) has been referred to as the 
Type-I seesaw formula. A somehow similar relation, the so-called
Type-II seesaw formula \cite{Pal}, can be derived from the 
generalized lepton mass term
\begin{equation}
- {\cal L}_{\rm mass} \; = \; - {\cal L}_l ~ + ~ \frac{1}{2} 
\overline{(\nu, ~\nu^{\rm c} )^{~}_{\rm L}}
\left ( \matrix{
M_{\rm L}     & M_{\rm D} \cr
M^T_{\rm D}    & M_{\rm R} \cr} \right ) 
\left ( \matrix{
\nu^{\rm c} \cr \nu \cr} \right )_{\rm R} \; ,
%       (2.10)
\end{equation}
where $M_{\rm L}$ has been defined in (2.4). As $M_{\rm L}$ must
result from a new Yukawa-interaction term which violates the 
$\rm SU(2)_L \times U(1)_Y$ gauge symmetry in the standard   
model, its scale might be much lower than the electroweak symmetry
breaking scale $v$. The strong hierarchy between the scales of
$M_{\rm R}$ and $M_{\rm L}$ or $M_{\rm D}$ allow us to make some
safe approximations in diagonalizing the $6\times 6$ neutrino mass 
matrix in (2.10). We find that the heavy (right-handed) neutrino
mass matrix remains to take the form 
$M_{\rm R} \approx U \overline{M}_{\rm R} U^T$. In contrast, the
light (left-handed) neutrino mass matrix is given by
\begin{equation}
M_\nu \; = \; V \overline{M}_\nu V^T \; \approx \; 
M_{\rm L}  -  M_{\rm D} M^{-1}_{\rm R} M^T_{\rm D} \; .
%       (2.11)
\end{equation}
This result is just the Type-II seesaw formula. As emphasized 
in Ref. \cite{Pal}, the Type-II seesaw relation is a 
reflection of the left-right symmetry of the theory at high 
energies. For the phenomenological study of neutrino masses,
however, the Type-I seesaw mechanism is more popular and useful
because it involves fewer free parameters than the Type-II seesaw
mechanism.

Of course, there are some other versions of the seesaw
mechanism, which are either more complicated \cite{Pal} or 
simpler \cite{Frampton} than what we have discussed above. 
While the seesaw idea is qualitatively elegant to interpret
the smallness of left-handed neutrino masses, it cannot lead
to quantitative predictions unless specific textures of neutrino
mass matrices (e.g., $M_{\rm D}$ and $M_{\rm R}$) are assumed.
Many interesting ans$\rm\ddot{a}$tze of lepton mass matrices
have so far been proposed and incorporated with the seesaw
mechanism (for recent reviews with extensive references,
see Ref. \cite{FXReview} and Refs. \cite{Altarelli}--\cite{Review}), 
but they are strongly model-dependent and thus beyond the scope of 
the present review article. 

\subsection{Experimental Constraints}

\subsubsection{Neutrino Oscillations}

Neutrino oscillation is a quantum phenomenon which can 
naturally happen if neutrinos are massive and lepton flavors
are mixed. In a simple two-neutrino mixing scheme, the neutrino 
flavor eigenstates $\nu_\alpha$ and $\nu^{~}_\beta$ are 
linear combinations of the neutrino mass eigenstates $\nu_a$
and $\nu^{~}_b$ \cite{MNS}:
$\nu_\alpha = \nu_a \cos\theta + \nu^{~}_b \sin\theta$ 
and
$\nu^{~}_\beta = \nu^{~}_b \cos\theta - \nu_a \sin\theta$, where 
$\theta$ denotes the flavor mixing angle. Then the probabilities
of neutrino oscillations are governed by two characteristic 
parameters: one of them is the neutrino mass-squared difference
$\Delta m^2 \equiv m^2_b - m^2_a$ 
and the other is the flavor mixing factor 
$\sin^2 2\theta$. Corresponding to the ``disappearance'' and
``appearance'' neutrino experiments, the survival and conversion 
probabilities of a neutrino flavor eigenstate $\nu_\alpha$ can 
explicitly be expressed as 
\begin{eqnarray}
P(\nu_\alpha \rightarrow \nu_\alpha) & = & 1 - 
\sin^2 2\theta \sin^2 \left ( 1.27 \frac{\Delta m^2 L}{E} 
\right ) \; ,
\nonumber \\
P(\nu_\alpha \rightarrow \nu^{~}_\beta) & = & 
\sin^2 2\theta \sin^2 \left ( 1.27 \frac{\Delta m^2 L}{E} 
\right ) \; ,
%       (2.12)
\end{eqnarray}
where $\beta \neq \alpha$, $E$ is the neutrino beam energy (in
unit of GeV), and $L$ denotes the distance between the neutrino source 
and the neutrino detector (in unit of km). A lot of experimental
data, including those from solar, atmospheric and reactor neutrino
oscillation experiments, have been analyzed by use of (2.12).

(1) The first model-independent evidence for neutrino oscillations
was obtained from the Super-Kamiokande (SK) experiment \cite{SK} on 
atmospheric neutrinos, which are produced in the earth's atmosphere
by cosmic rays and are detected in an underground detector.
If there were no neutrino oscillation, the atmospheric neutrinos
entering and exiting the detector should have a spherical symmetry.
In other words, the downward-going and
upward-going neutrino fluxes should be equal to each other:
$\Phi_\alpha (\theta_z) = \Phi_\alpha (\pi - \theta_z)$ versus the 
zenith angle $\theta_z$ (for $\alpha = e$ or $\mu$). The SK Collaboration
has observed an approximate up-down flux symmetry for the atmospheric 
$\nu_e$ neutrinos and a significant up-down flux asymmetry for the 
atmospheric $\nu_\mu$ neutrinos. For instance,
\begin{equation}
\frac{\Phi_\mu (-1 \leq \cos\theta_z \leq -0.2)}
{\Phi_\mu (+0.2 \leq \cos\theta_z \leq +1)} \; = \; 
0.54 \pm 0.04 \; \neq \; 1 
%       (2.13)
\end{equation}
was measured for the multi-GeV $\nu_\mu$ neutrinos, and it is
in conflict with the up-down flux symmetry. This result
can well be interpreted in the assumption of 
$\nu_\mu \rightarrow \nu_\tau$ neutrino oscillations. Current
SK \cite{SK} and CHOOZ \cite{CHOOZ} data have ruled out the 
possibility that the atmospheric
neutrino anomaly is dominantly attributed to
$\nu_\mu \rightarrow \nu_e$ or $\nu_\mu \rightarrow \nu_s$
oscillations, where $\nu_s$ stands for a ``sterile'' neutrino
which does not take part in the normal electroweak interactions
(see appendix A for detailed discussions).

The hypothesis of atmospheric neutrino oscillations
has recently received very strong support from the K2K \cite{K2K}
long-baseline neutrino experiment, in which the $\nu_\mu$
beam is produced at the KEK accelerator and measured 250 km away
at the SK detector. The K2K Collaboration observed  
a reduction of the $\nu_\mu$ flux and a distortion of the 
$\nu_\mu$ energy spectrum, which must take place in the presence
of $\nu_\mu \rightarrow \nu_\tau$ oscillations. The possibility
that the K2K result is due to a statistical fluctuation instead
of neutrino oscillations is found to be less than $1\%$ \cite{K2K}.

It has been shown that the standard $\nu_\mu \rightarrow \nu_\tau$
oscillations provide the best description of the combined SK and K2K 
data \cite{Fogli03}, from which 
$\Delta m^2_{\rm atm} = (2.6 \pm 0.4) \times 10^{-3} ~ {\rm eV}^2$ 
and $\sin^2 2 \theta_{\rm atm} = 1.00^{+0.00}_{-0.05}$ are determined 
at the $1\sigma$ level. At the $90\%$ confidence level, we have
\begin{equation}
1.65 \times 10^{-3} ~ {\rm eV}^2 \leq \Delta m^2_{\rm atm} \leq
3.25 \times 10^{-3} ~ {\rm eV}^2 \; 
%	(2.14)
\end{equation}
and $0.88 \leq \sin^2 2\theta_{\rm atm} \leq 1.00$.

(2) The long-standing problem associated with solar neutrinos is that 
the flux of solar $\nu_e$ neutrinos measured in all experiments (e.g.,
the SK \cite{SK} and SNO \cite{SNO} experiments as well as the
earlier Homestake \cite{Home}, GALLEX-GNO \cite{Gallex} and
SAGE \cite{Sage} experiments) is
significantly smaller than that predicted by the standard solar
models \cite{BP}. The deficit of solar $\nu_e$'s is not the same 
in different experiments, implying that the relevant physical
effect is energy-dependent. It has been hypothesized that the solar
neutrino problem 
is due to the conversion of solar $\nu_e$'s into other active
or sterile neutrinos during their travel from the core 
of the sun to the detectors on the earth. The SNO experiment
has model-independently demonstrated that 
$\nu_e \rightarrow \nu_\mu$ and (or)
$\nu_e \rightarrow \nu_\tau$ transitions are dominantly responsible
for the solar neutrino deficit. 

What the SNO Collaboration has measured is the flux of solar
$^8{\rm B}$ neutrinos via the charged-current (CC),
neutral-current (NC) and elastic-scattering (ES) reactions:
$\nu_e + d \rightarrow e^- + p + p$,
$\nu_\alpha + d \rightarrow \nu_\alpha + p + n$ and
$\nu_\alpha + e^- \rightarrow \nu_\alpha + e^-$, where 
$\alpha = e$, $\mu$ or $\tau$. In the presence of flavor conversion,
the observed neutrino fluxes in different reactions satisfy
\begin{equation}
\Phi^{\rm CC} = \Phi_e \; , ~~~
\Phi^{\rm NC} = \Phi_e + \Phi_{\mu\tau} \; , ~~~
\Phi^{\rm ES} = \Phi_e + \frac{\sigma_\mu}{\sigma_e} \Phi_{\mu\tau} 
\approx \Phi_e + 0.154 \Phi_{\mu\tau} \; ,
%       (2.15)
\end{equation}
where $\sigma_\mu/\sigma_e \approx 0.154$ is the ratio of elastic
$\nu_e$-$e$ and $\nu_\mu$-$\mu$ scattering cross-sections, and
$\Phi_{\mu\tau}$ denotes the flux of active non-electron neutrinos.
Of course, $\Phi_{\mu\tau} = 0$ or equivalently
$\Phi^{\rm CC} = \Phi^{\rm NC} = \Phi^{\rm ES}$ would hold, 
if there were no flavor conversion. The SNO data \cite{SNO} yield
\begin{equation}
\Phi_e = (1.76 \pm 0.10) \times 10^6 
~ {\rm cm}^{-2}{\rm s}^{-1} \; ,
~~~~
\Phi_{\mu\tau} = (3.41^{+0.66}_{-0.64}) 
\times 10^6 ~ {\rm cm}^{-2}{\rm s}^{-1} \; ,
%       (2.16)
\end{equation}
a convincing evidence (at the $5.3\sigma$ level) for the existence
of $\nu_\mu$ and $\nu_\tau$ neutrinos in the flux of solar 
neutrinos onto the earth.

The flavor conversion of solar $\nu_e$ neutrinos is most likely 
due to neutrino oscillations. In the scheme of two-neutrino
oscillations, a global analysis of all available
experimental data on solar neutrinos (in particular, those from
the SK and SNO measurements) leads to several regions allowed for 
the parameters $\Delta m^2_{\rm sun}$ and $\tan^2 \theta_{\rm sun}$:
the SMA (small mixing angle), LMA (large mixing angle) and LOW
(low mass-squared difference) regions based on the 
Mikheev-Smirnov-Wolfenstein (MSW) mechanism \cite{MSW} as well as
the VO (vacuum oscillation) and other possible regions \cite{Solution},
among which the LMA region is most favored. Recently the LMA solution
has been singled out as the only acceptable solution to the solar
neutrino problem, thanks to the KamLAND \cite{KM} experiment.

The KamLAND Collaboration has measured the flux of 
$\overline{\nu}_e$'s from distant nuclear reactor via the inverse
$\beta$-decay reaction $\overline{\nu}_e + p \rightarrow e^+ + n$.
The typical baseline of this experiment is 180 km, allowing for
a terrestrial test of the LMA solution to the solar neutrino problem.
The ratio of the observed inverse $\beta$-decay events to the
expected number without $\overline{\nu}_e$ disappearance is 
$0.611 \pm 0.094$ for $E_\nu > 3.4$ MeV. Such a deficit can naturally
be interpreted in the hypothesis of neutrino oscillations, and the
parameter space of $(\Delta m^2, \sin^2 2\theta)$ is found to be
compatible very well with the LMA region. We are then led to the
conclusion that the LMA solution is the only correct solution to the
solar neutrino problem. A global analysis of the combined SK, SNO and 
KamLAND data yields \cite{SUN}
\begin{equation}
5.9 \times 10^{-5} ~ {\rm eV}^2 \leq \Delta m^2_{\rm sun} \leq
8.8 \times 10^{-5} ~ {\rm eV}^2
%       (2.17)
\end{equation}
and $0.25 \leq \sin^2 \theta_{\rm sun} \leq 0.40$ at the $90\%$
confidence level.        

(3) The purpose of the CHOOZ and Palo Verde reactor 
experiments \cite{CHOOZ} is
to search for $\overline{\nu}_e \rightarrow \overline{\nu}_e$ 
disappearance neutrino oscillations in the atmospheric range of
$\Delta m^2$. No indication in favor of neutrino oscillations was
found from both experiments, leading to a strong constraint on the
flavor mixing factor:
$\sin^2 2\theta_{\rm chz} < 0.10$ for 
$\Delta m^2_{\rm chz} > 3.5\times 10^{-3} ~ {\rm eV}^2$;
or $\sin^2 2\theta_{\rm chz} < 0.18$ for 
$\Delta m^2_{\rm chz} > 2.0\times 10^{-3} ~ {\rm eV}^2$.
The impact of this constraint on the lepton flavor mixing matrix
will be discussed in section 3.1.

Let us denote the mass eigenstates of $\nu_e$, $\nu_\mu$ and 
$\nu_\tau$ neutrinos as $(\nu_1, \nu_2, \nu_3)$, whose eigenvalues 
are $(m_1, m_2, m_3)$.
There are only two independent neutrino mass-squared differences:
$\Delta m^2_{21} \equiv m^2_2 - m^2_1$ and
$\Delta m^2_{32} \equiv m^2_3 - m^2_2$ or
$\Delta m^2_{31} \equiv m^2_3 - m^2_1$.
Without loss of generality, we make the identification
\begin{equation}
\Delta m^2_{\rm sun} \; = \; |\Delta m^2_{21}| 
\; \ll \; |\Delta m^2_{32}| \;
%\approx |\Delta m^2_{31}| 
= \; \Delta m^2_{\rm atm} \;\; ,
%       (2.18)
\end{equation}
where the big hierarchy between $\Delta m^2_{\rm sun}$ and 
$\Delta m^2_{\rm atm}$ has been taken into account. 
Indeed, $\Delta m^2_{\rm sun} \ll \Delta m^2_{\rm atm}$ together
with $\sin^2 2\theta_{\rm chz} \ll 1$ implies that solar and 
atmospheric neutrino oscillations are approximately decoupled. 
The deficits of solar and atmospheric neutrinos are
dominated respectively by $\nu_e\rightarrow \nu_\mu$ and
$\nu_\mu\rightarrow \nu_\tau$ transitions. With the help of (2.18),
$m_1$ and $m_2$ can be given in terms of $m_3$, 
$\Delta m^2_{\rm sun}$ and $\Delta m^2_{\rm atm}$ \cite{Xing02}:
\begin{eqnarray}
m_1 & = & \sqrt{m^2_3 + p \Delta m^2_{\rm atm} + 
q \Delta m^2_{\rm sun}} \;\; ,
\nonumber \\
m_2 & = & \sqrt{m^2_3 + p \Delta m^2_{\rm atm}} \;\; ,
%       (2.19)
\end{eqnarray}
where $p=\pm 1$ and $q=\pm 1$ stand for four possible 
neutrino mass spectra
%%%%%%%%%%%%%%%%%%%%%%%%%%
\footnote{The mass convention $m'_1 < m'_2 < m'_3$ has been used in 
some literature for the {\it redefined} neutrino mass eigenstates 
$\nu'_i$, which are related to the standard neutrino mass eigenstates
$\nu_i$ by an orthogonal transformation \cite{Aoki}. We shall not
adopt $\nu'_i$ in the present article.}:
%%%%%%%%%%%%%%%%%%%%%%%%%%
\begin{eqnarray}
(p,q) = (-1,-1): & & m_1 < m_2 < m_3 \; ,
\nonumber \\
(p,q) = (-1,+1): & & m_1 > m_2 < m_3 \; ,
\nonumber \\
(p,q) = (+1,-1): & & m_1 < m_2 > m_3 \; ,
\nonumber \\
(p,q) = (+1,+1): & & m_1 > m_2 > m_3 \; .
%       (2.20)
\end{eqnarray}
Current solar neutrino oscillation data favor $q =-1$; i.e.,
$m_1 < m_2$. The sign of $p$ may be determined from the future
long-baseline neutrino oscillation experiments.

Given the best-fit values 
$\Delta m^2_{\rm sun} = 7.3\times 10^{-5} ~{\rm eV}^2$ and 
$\Delta m^2_{\rm atm} = 2.5\times 10^{-3} ~{\rm eV}^2$, 
the numerical correlation of $m_1$, $m_2$ and $m_3$ is illustrated
in Fig. 2.2 by use of (2.19). We see that three masses become
nearly degenerate, if one of them is larger than 0.1 eV 
(i.e., $m_i > 0.1$ eV). For the $p=-1$ case, the lower bound of $m_3$ 
is $m_3 \geq \sqrt{\Delta m^2_{\rm atm} + \Delta m^2_{\rm sun}} ~$,
which leads to $m_2 \geq \sqrt{\Delta m^2_{\rm sun}} ~$.
A normal hierarchy $m_1 < m_2 < m_3$ may appear, if 
$m_1 \sim {\cal O}(10^{-2})$ eV or smaller. 
For the $p=+1$ case, however, $m_1 \approx m_2 > m_3$ 
always holds, no matter how small $m_3$ is taken (inverted
hierarchy). 
%%%%%%%%%%%%%%%%%%%% Fig. 2.2 %%%%%%%%%%%%%%%%
\begin{figure}[t]
\vspace{-2.75cm}
\epsfig{file=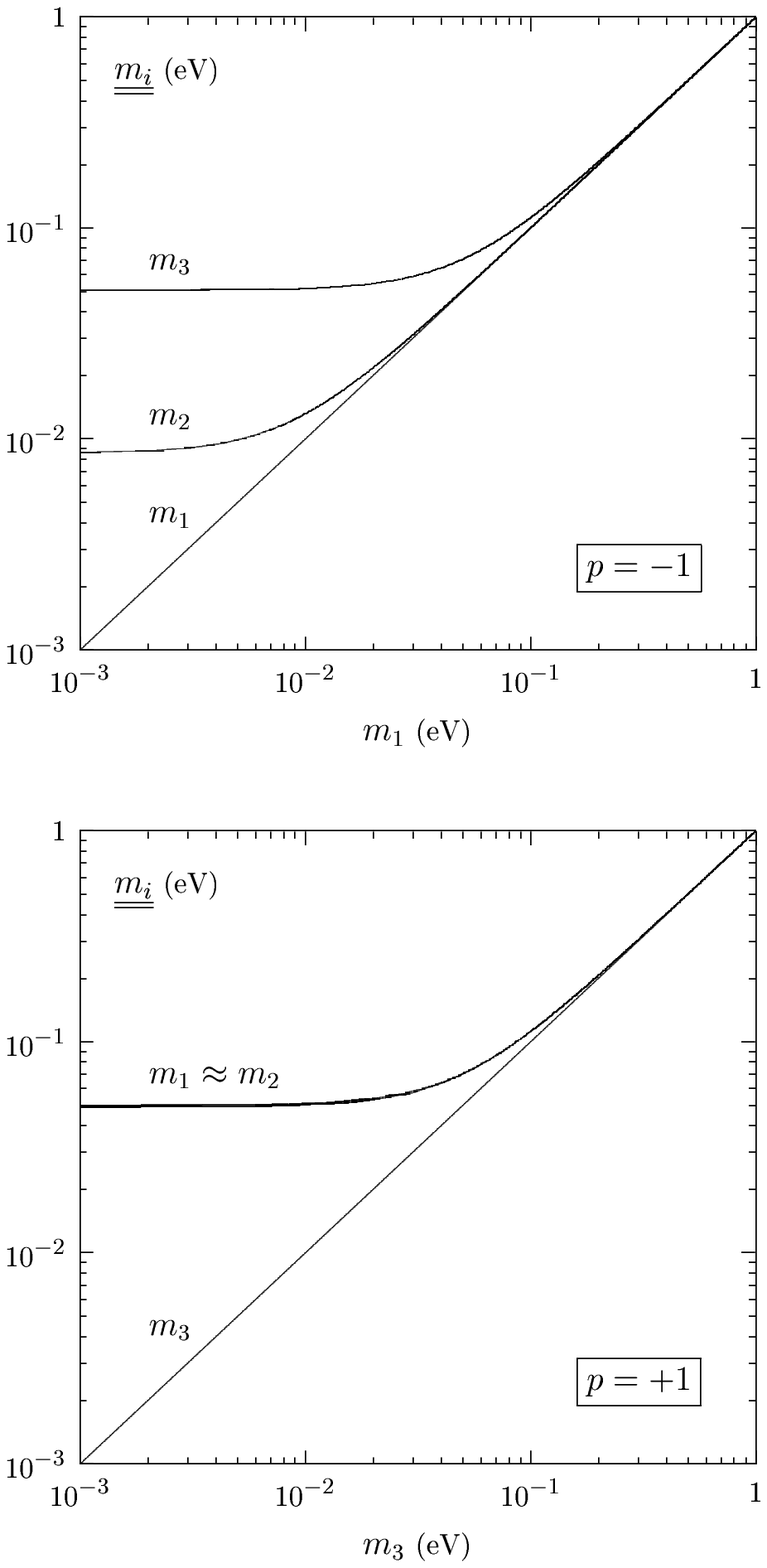,bbllx=-2cm,bblly=2cm,bburx=19cm,bbury=31cm,%
width=15.5cm,height=22cm,angle=0,clip=}
\vspace{-4.4cm}
\caption{Illustrative correlation of three neutrino masses
for $p=\pm 1$ cases.}
\end{figure}
%%%%%%%%%%%%%%%%%%%%%%%%%%%%%%%%%%%%%%%%%%%%

Finally it is worthwhile to mention the LSND evidence \cite{LSND} 
for $\bar{\nu}_\mu \rightarrow \bar{\nu}_e$ neutrino oscillations
with $\Delta m^2_{\rm LSND} \sim 1 ~ {\rm eV^2}$ and 
$\sin^2 2\theta_{\rm LSND} \sim 10^{-3} - 10^{-2}$. This observation 
was not confirmed by the KARMEN experiment \cite{KARMEN},
which is sensitive to most of the LSND parameter space. 
The disagreement between these two measurements will be resolved 
by the MiniBOONE experiment \cite{MB} at
Fermilab in the coming years. Before the LSND result is confirmed
on a solid ground, the conservative approach is to set it aside 
tentatively and to concentrate on solar and atmospheric neutrino 
oscillations in the mixing scheme of three lepton families.
A natural way to simultaneously accommodate solar, atmospheric
and LSND data is to assume the existence of a light sterile 
neutrino, which may oscillate with three active neutrinos.
Such a four-neutrino mixing scheme will be discussed in
appendix A. 

\subsubsection{Cosmological Bounds}

Neutrinos play an important role in cosmology and 
astrophysics \cite{Dolgov}. Useful information on neutrino 
masses can be obtained, for example, from studies of the
cosmological relic density and dark matter, the power
spectrum in large scale structure surveys, the Lyman $\alpha$
forest, the Big Bang nucleosynthesis, the supernovae, the
ultra high energy (UHE) cosmic rays, and the gamma ray 
bursts. For recent reviews of these interesting topics, we
refer the reader to Refs. \cite{Dolgov}--\cite{Bilenky}. Here we
focus our attention on the latest cosmological bound 
on the sum of light neutrino masses, obtained from 
the impressive data of the Wilkinson Microwave Anisotropy 
Probe (WMAP) \cite{WMAP}.

According to the Big Bang cosmology, neutrinos were in thermal
equilibrium with photons, electrons and positrons in the
early universe. When the universe cooled to temperatures of
${\cal O}(1)$ MeV, neutrinos decoupled from the primordial
$e^{\pm}\gamma$ plasma, leading to the fact that the present-day
number density of neutrinos is similar to that of photons. If 
neutrinos are massive, they contribute to the cosmological
matter density \cite{GZ},
\begin{equation}
\Omega_\nu h^2 \; = \; \sum_i \frac{m_i}
{93.5 ~{\rm eV}} \; ,
%       (2.21)
\end{equation}
where $\Omega_\nu$ denotes the neutrino mass density relative
to the critical energy density of the universe, and $h$ is the
Hubble constant in units of 100 km/s/Mpc. A tight limit 
$\Omega_\nu h^2 < 7.6\times 10^{-3}$ (at $95\%$ C.L.) has 
recently been extracted from the striking WMAP data \cite{WMAP},
combined with the additional cosmic microwave background (CMB)
data sets (CBI and ACBAR) and the observation of large scale
structure from 2dF Galaxy Redshift Survey (2dFGRS) \cite{CMB}.
This impressive bound leads straightforwardly to
%%%%%%%%%%%%%%%%%%%%%%%%%%%
\footnote{The exact value of this upper bound depends 
on other cosmological parameters, as emphasized by
Hannestad in Ref. \cite{Hannestad}, where a somewhat more
generous limit $m_1 + m_2 + m_3 < 1.0$ eV 
(at $95\%$ C.L.) has been obtained for $N_\nu =3$.}
%%%%%%%%%%%%%%%%%%%%%%%%%%%
\begin{equation}
\sum_i m_i \; < \; 0.71 ~ {\rm eV} \; , 
%       (2.22)
\end{equation}
which holds at the same confidence level. Some discussions 
are in order.

(a) A tight upper bound on $m_i$ can be achieved from (2.22)
together with current data on solar and atmospheric neutrino
oscillations. To see this point more clearly, we calculate
the dependence of $m_1+m_2+m_3$ on $m_3$ with the help of
(2.19) and (2.22). The numerical results are shown in Fig. 2.3,
where the best-fit values
$\Delta m^2_{\rm sun} = 7.3\times 10^{-5} ~ {\rm eV}^2$ and
$\Delta m^2_{\rm atm} = 2.5\times 10^{-3} ~ {\rm eV}^2$ have
typically been input. Note that only the $q=-1$ or 
$m_1 < m_2$ case, which is supported by current solar neutrino
oscillation data, is taken into account. For the $p=-1$ or
$m_2 < m_3$ case, $m_3$ has an interesting lower bound 
$m_3 \geq \sqrt{\Delta m^2_{\rm atm} + \Delta m^2_{\rm sun}} 
\approx 0.051$ eV; but for the $p=+1$ or $m_2 > m_3$ case, 
even $m_3 =0$ is allowed (inverted hierarchy). We see that 
these two cases become indistinguishable for $m_3 \geq 0.2$ eV, 
implying the near degeneracy of three neutrino masses. Once the 
WMAP limit in (2.22) is included, we immediately get 
$m_3 < 0.24$ eV. As a consequence, 
we have $m_i < 0.24$ eV for $i=1,2$ or $3$.
Similar results have been obtained in Refs. \cite{Hannestad,MAP}. 
%%%%%%%%%%%%%%%%%%%% Fig. 2.3 %%%%%%%%%%%%%%%%
\begin{figure}
\vspace{-2.75cm}
\epsfig{file=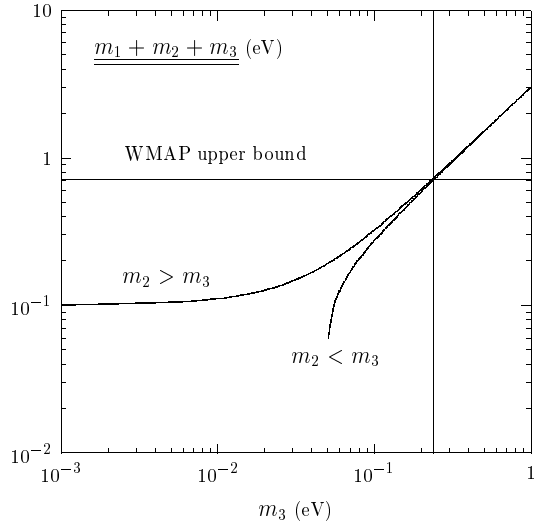,bbllx=-2cm,bblly=2cm,bburx=19cm,bbury=31cm,%
width=15.5cm,height=22cm,angle=0,clip=}
\vspace{-11.95cm}
\caption{Illustrative dependence of $m_1 + m_2 + m_3$ on $m_3$.
The WMAP result sets an upper limit on $m_3$; i.e.,
$m_3 < 0.24$ eV for both $m_3 > m_2$ and $m_3 < m_2$ cases.}
\end{figure}
%%%%%%%%%%%%%%%%%%%%%%%%%%%%%%%%%%%%%%%%%%%%

(b) The LSND \cite{LSND} evidence for neutrino oscillations is strongly 
disfavored, because $\Delta m^2_{\rm LSND} \sim 1 ~ {\rm eV}^2$
is essentially incompatible with (2.22). However, a marginal
agreement between the LSND data and the WMAP data is not 
impossible in the four-neutrino mixing scheme, if the latter's
limit on neutrino masses is loosened to some extent (e.g.,
$m_1 + m_2 + m_3 + m_4 < 1.4$ eV for $N_\nu =4$ \cite{Hannestad}).
In this sense, it might be premature to claim that the LSND 
result has been ruled out by the WMAP data \cite{MAP}. 
We hope that the MiniBOONE experiment \cite{MB}
can ultimately confirm or disprove the LSND result.
 
\subsubsection{Kinematic Measurements}

An essentially model-independent way to determine or constrain 
neutrino masses is to measure some typical weak decays whose
final-state particles include neutrinos. The kinematics of such 
a process in the case of non-zero neutrino masses is different 
from that in the case of zero neutrino masses. This provides an 
experimental opportunity to probe neutrino masses in a direct way.
In practice, direct neutrino mass measurements are based on the
analysis of the kinematics of charged particles produced together 
with the neutrino flavor eigenstates 
$|\nu_\alpha\rangle$ (for $\alpha = e, \mu, \tau$), which are
superpositions of the neutrino mass eigenstates 
$|\nu_i\rangle$ (for $i=1,2,3$):
$|\nu_\alpha\rangle = V_{\alpha 1} |\nu_1\rangle +
V_{\alpha 2} |\nu_2\rangle + V_{\alpha 3} |\nu_3\rangle$ with
$V$ being the $3\times 3$ lepton flavor mixing matrix. The 
effective masses of three neutrinos can then be defined as
\begin{equation}
\langle m\rangle_\alpha \; = \; \sqrt{\sum_i 
\displaystyle \left ( |V_{\alpha i}|^2 m^2_i \right )} \;\; ,
%       (2.23)
\end{equation}
for $\alpha = e, \mu$ and $\tau$. 
The kinematic limits on $\langle m\rangle_e$, $\langle m\rangle_\mu$
and $\langle m\rangle_\tau$ can respectively be obtained from the 
tritium beta decay 
$^3_1{\rm H} \rightarrow$$^3_2{\rm He} + e^- + \overline{\nu}_e$,
the $\pi^+ \rightarrow \mu^+ + \nu_\mu$ decay and the
$\tau \rightarrow 5\pi + \nu_\tau$ 
(or $\tau \rightarrow 3\pi + \nu_\tau$) decay \cite{PDG02}:
\begin{eqnarray}
\langle m\rangle_e & < & 2.2 ~ {\rm eV} \; ,
\nonumber \\
\langle m\rangle_\mu & < & 0.19 ~ {\rm MeV} \; ,
\nonumber \\
\langle m\rangle_\tau & < & 18.2 ~ {\rm MeV} \; .
%       (2.24)
\end{eqnarray}
We see that the experimental sensitivity for  
$\langle m\rangle_\mu$ is more than four orders of magnitude
smaller than that for $\langle m\rangle_e$, and the experimental 
sensitivity for $\langle m\rangle_\tau$ is two orders of
magnitude lower than that for $\langle m\rangle_\mu$.

Note that the unitarity of $V$ leads straightforwardly to a simple 
sum rule between $\langle m\rangle^2_\alpha$ and $m^2_i$ \cite{Xing03}:
\begin{equation}
\sum_\alpha \langle m\rangle^2_\alpha \; = \; 
\sum_i m^2_i \; < \; \left (\sum_i m_i \right )^2\; .
%       (2.25)
\end{equation}
As the sum of three relativistic neutrino masses has well been 
constrained by the recent WMAP data \cite{WMAP},
$m_1 + m_2 + m_3 < 0.71$ eV at the $95\%$ confidence level,
we are led to
\begin{equation}
\langle m\rangle^2_e + \langle m\rangle^2_\mu + 
\langle m\rangle^2_\tau \; < \; 0.50 ~ {\rm eV}^2 \; . ~
%       (2.26)
\end{equation}
This generous upper bound implies that 
$\langle m\rangle^2_\alpha < 0.50 ~ {\rm eV}^2$ or
$\langle m\rangle_\alpha < 0.71 ~ {\rm eV}$ holds for
$\alpha = e,\mu$ and $\tau$. One can see that the cosmological upper 
bound of $\langle m\rangle_e$ is about three times smaller than its 
kinematic upper bound given in (2.24). The former may be accessible 
in the future KATRIN experiment \cite{KATRIN}, whose sensitivity is 
expected to be about 0.3 eV. In contrast, the cosmological upper 
bound of $\langle m\rangle_\mu$ is more than five orders of magnitude 
smaller than its kinematic upper bound given in (2.24), and the upper 
limit of $\langle m\rangle_\tau$ set by the WMAP data is more than 
seven orders of magnitude smaller than its kinematic upper limit. It 
seems hopeless to improve the sensitivity of the kinematic 
$\langle m\rangle_\mu$ and $\langle m\rangle_\tau$ measurements 
to the level of 0.71 eV. 

If current neutrino oscillation data are taken into account,
however, more stringent upper limits can be obtained for the
effective neutrino masses $\langle m\rangle_e$, 
$\langle m\rangle_\mu$ and $\langle m\rangle_\tau$. Substituting
(2.19) into (2.23), we get
\begin{equation}
\langle m\rangle_\alpha \; =\; \sqrt{m^2_3 +
p \left (1 - |V_{\alpha 3}|^2 \right ) \Delta m^2_{\rm atm} +
q |V_{\alpha 1}|^2 \Delta m^2_{\rm sun}} \; \; .
%	(2.27)
\end{equation}
The present experimental data indicate 
$|V_{e3}|^2 \approx \sin^2\theta_{\rm chz} \ll 1$ \cite{CHOOZ}
and $\Delta m^2_{\rm sun} \ll \Delta m^2_{\rm atm}$.
Furthermore, $\sin^2 2\theta_{\rm atm} \approx 1$ is strongly
favored, leading to $|V_{\mu 3}| \approx \sin\theta_{\rm atm}
\approx 1/\sqrt{2} ~$ and $|V_{\tau 3}| \approx \cos\theta_{\rm atm}
\approx 1/\sqrt{2} ~$ for $|V_{e3}| \ll 1$. Therefore, (2.27) can be 
simplified as
\begin{eqnarray}
\langle m\rangle_e & \approx & \sqrt{m^2_3 + p
\Delta m^2_{\rm atm}} \;\; ,
\nonumber \\
\langle m\rangle_\mu & \approx & \sqrt{m^2_3 + \frac{p}{2}
\Delta m^2_{\rm atm}} \;\; ,
\nonumber \\
\langle m\rangle_\tau & \approx & \sqrt{m^2_3 + \frac{p}{2}
\Delta m^2_{\rm atm}} \;\; .
%	(2.28)
\end{eqnarray} 
We can see that $\langle m\rangle_\mu \approx \langle m\rangle_\tau$
is a consequence of current neutrino oscillation data.
In addition, (2.28) shows that $\langle m\rangle_e$ is 
slightly larger than $\langle m\rangle_\mu$ and $\langle m\rangle_\tau$ 
for $p=+1$ or $m_2 > m_3$; and it is slightly smaller than 
$\langle m\rangle_\mu$ and $\langle m\rangle_\tau$ for $p=-1$ or
$m_2 < m_3$. The numerical dependence of $\langle m\rangle_e$,
$\langle m\rangle_\mu$ and $\langle m\rangle_\tau$
on $m_3$ is illustrated in Fig. 2.4. We find that it is impossible
to distinguish between the $m_2 < m_3$ case and the $m_2 > m_3$ case 
for $m_3 \geq 0.2$ eV. Note that the dependence of 
$\langle m\rangle_\alpha$ on $m_3$ is similar to the dependence 
of $m_1 + m_2 + m_3$ on $m_3$ shown in Fig. 2.3.
Taking account of the upper limit $m_3 \leq 0.24$ eV obtained
in section 2.2.2, we arrive at $\langle m\rangle_\alpha \leq 0.24$ eV 
for $\alpha = e, \mu$ or $\tau$. This upper bound is suppressed by
a factor of three, compared to the upper bound obtained from
(2.26) which is independent of the neutrino oscillation data.
%%%%%%%%%%%%%%%%%%%% Fig. 2.4 %%%%%%%%%%%%%%%%
\begin{figure}[t]
\vspace{-2.75cm}
\epsfig{file=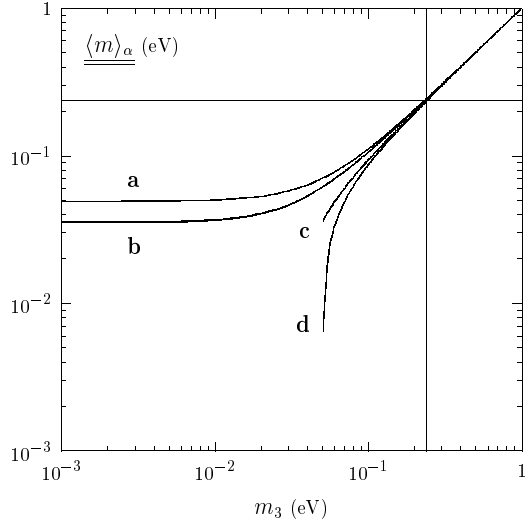,bbllx=-2cm,bblly=2cm,bburx=19cm,bbury=31cm,%
width=15.5cm,height=22cm,angle=0,clip=}
\vspace{-11.95cm}
\caption{Illustrative dependence of $\langle m\rangle_\alpha$ 
(for $\alpha = e, \mu, \tau$)
on $m_3$. Curve {\bf a}: $\langle m\rangle_e$ with $m_3 < m_2$;
Curve {\bf b}: $\langle m\rangle_\mu \approx \langle m\rangle_\tau$ 
with $m_3 < m_2$; 
Curve {\bf c}: $\langle m\rangle_\mu \approx \langle m\rangle_\tau$ 
with $m_3 > m_2$; and
Curve {\bf d}: $\langle m\rangle_e$ with $m_3 > m_2$. The WMAP
result leads to $\langle m\rangle_\alpha <0.24$ eV.}
\end{figure}
%%%%%%%%%%%%%%%%%%%%%%%%%%%%%%%%%%%%%%%%%%%%

The result 
$\langle m\rangle_\mu \approx \langle m\rangle_\tau < 0.24$ eV
implies that there is no hope to kinematically detect the 
effective masses of muon and tau neutrinos \cite{Xing03}. As the 
WMAP upper bound is in general valid for a sum of the masses of all 
possible relativistic neutrinos (no matter whether they are 
active or sterile), it seems unlikely to loosen the upper 
limit of $\langle m\rangle_\alpha$ obtained above in the
assumption of only active neutrinos. Therefore, the kinematic
measurements of $\langle m\rangle_\mu$ and $\langle m\rangle_\tau$
have little chance to reveal the existence of any exotic 
neutral particles with masses much larger than the light 
neutrino masses. 

\subsubsection{Neutrinoless Double-$\beta$ Decay}

The neutrinoless double beta decay of some even-even nuclei,
\begin{equation}
A(Z,N) \; \rightarrow \; A(Z+2, N-2) + 2e^- \; ,
%       (2.29)
\end{equation}
can occur through the exchange of a Majorana neutrino between 
two decaying neutrons inside a nucleus, as illustrated in Fig. 2.5.
It would be forbidden, however, if neutrinos were Dirac particles. 
Thus the neutrinoless double-$\beta$ decay provides us with a unique 
opportunity to identify the Majorana nature of massive neutrinos.

The rate of the neutrinoless double-$\beta$ decay depends both on 
an effective neutrino mass term $\langle m\rangle_{ee}$ and on
the associated nuclear matrix element. The latter can be 
calculated, but some uncertainties are involved in the
calculations \cite{Bilenky}. In the three-neutrino mixing scheme,
the effective Majorana mass $\langle m\rangle_{ee}$ is given by
\begin{equation}
\langle m \rangle_{ee} \; = \; \left |
m_1 V^2_{e1} + m_2 V^2_{e2} + m_3 V^2_{e3} \right | \; ,
%       (2.30)
\end{equation}
where $m_i$ (for $i=1,2,3$) denote the physical masses of three
neutrinos, and $V_{ei}$ stand for the elements in the first row
of the $3\times 3$ lepton flavor mixing matrix $V$. It is obvious
that $\langle m\rangle_{ee} = 0$ would hold, if $m_i = 0$ were
taken.
%%%%%%%%%%%%%%%%%%%% Fig. 2.5 %%%%%%%%%%%%%%%%
\begin{figure}[t]
\vspace{-3.2cm}
\epsfig{file=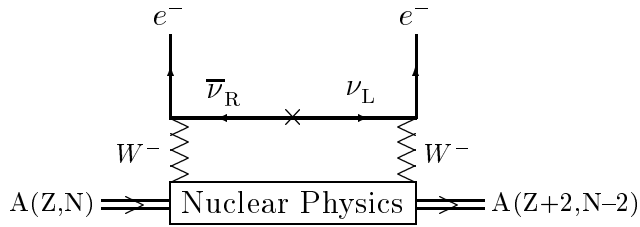,bbllx=3cm,bblly=8cm,bburx=18cm,bbury=30cm,%
width=15.5cm,height=22cm,angle=0,clip=}
\vspace{-15.3cm}
\caption{Illustrative plot for the neutrinoless double beta decay
of some even-even nuclei via the exchange of a virtual Majorana 
neutrino.}
\end{figure}
%%%%%%%%%%%%%%%%%%%%%%%%%%%%%%%%%%%%%%%%%%%%%%

Many experiments have been done to search for the neutrinoless 
double-$\beta$ decay \cite{Cremonesi}. Among them, the 
Heidelberg-Moscow \cite{HM} and IGEX \cite{IGEX} $^{76}{\rm Ge}$
experiments have the highest sensitivity and yield the most 
stringent upper bounds on $\langle m\rangle_{ee}$:
\begin{eqnarray}
\langle m\rangle_{ee} & \leq & (0.35 - 1.24) ~ {\rm eV} 
~~~~ ({\rm Heidelberg-Moscow}) \; , 
\nonumber \\
\langle m\rangle_{ee} & \leq & (0.33 - 1.35) ~ {\rm eV} 
~~~~ ({\rm IGEX}) \; , 
%       (2.31)
\end{eqnarray}
where the uncertainties associated with the nuclear matrix elements
have been taken into account
%%%%%%%%%%%%%%%%%%%%%%%%%%%
\footnote{After re-analyzing the data from the Heidelberg-Moscow 
experiment, Klapdor-Kleingrothaus {\it et al.} reported the first 
evidence for the neutrinoless double beta decay \cite{KK}. However, 
their result was strongly criticized by some authors \cite{KK2}.
Future experiments will have sufficiently high sensitivity to 
clarify the present debates \cite{KK3}, either to confirm or to 
disprove the alleged result in Ref. \cite{KK}.}.
%%%%%%%%%%%%%%%%%%%%%%%%%%%
A number of new experiments \cite{Cremonesi}, which may probe 
$\langle m\rangle_{ee}$ at the level of 10 meV to 50 meV, are in 
preparation. 

While $\langle m\rangle_{ee} \neq 0$ must imply that neutrinos
are Majorana particles, $\langle m\rangle_{ee} = 0$ does not 
{\it necessarily} imply that neutrinos are Dirac particles. The
reason is simply that the Majorana phases hidden in $V_{ei}$ may
lead to significant cancellations on the right-hand side of (2.30), 
making $\langle m\rangle_{ee}$ vanishing or too small to be 
detectable \cite{Bilenky2}. Hence much care has to be taken, if no 
convincing signal of the neutrinoless double beta decay can be 
experimentally established:
it may imply that (1) the experimental sensitivity is not high
enough; (2) the massive neutrinos are Dirac particles; or (3) the
vanishing or suppression of $\langle m\rangle_{ee}$ is due to
large cancellations induced by the Majorana CP-violating phases.
The third possibility has recently been examined \cite{Xing03a}
from a model-independent point of view and with the help of the
latest experimental data. 
It is found that current neutrino oscillation 
data {\it do} allow $\langle m\rangle_{ee} =0$ to hold, if the 
Majorana phases lie in two specific regions. To see this point
more clearly, we use $\rho$ and $\sigma$ to denote the Majorana 
CP-violating phases of $V$ and take the convention \cite{Xing02}
\begin{equation} 
\arg (V_{e1}) = \rho \; , ~~~
\arg (V_{e2}) = \sigma \; , ~~~
\arg (V_{e3}) = 0 \; .
%       (2.32)
\end{equation}
Of course, $\rho$ and $\sigma$ have nothing to do with CP and T 
violation in normal neutrino oscillations. In view of (2.30) and
(2.32), we find that $\langle m\rangle_{ee} = 0$ requires
\begin{eqnarray}
\frac{m_1}{m_2} & = & -\frac{|V_{e2}|^2}{|V_{e1}|^2}
\cdot \frac{\sin 2\sigma}{\sin 2\rho} \;\; ,
\nonumber \\
\frac{m_2}{m_3} & = & +\frac{|V_{e3}|^2}{|V_{e2}|^2}
\cdot \frac{\sin 2\rho}{\sin 2 (\sigma - \rho)} \;\; .
%       (2.33)
\end{eqnarray}
As $m_i >0$ holds, part of the $(\rho, \sigma)$ parameter space 
must be excluded. To pin down the whole ranges of $\rho$ and 
$\sigma$ allowed by current neutrino oscillation data under the 
$\langle m\rangle_{ee} = 0$ condition, we utilize (2.18) and 
(2.33) to calculate the ratio of solar and atmospheric neutrino 
mass-squared differences:
\begin{equation}
\frac{\Delta m^2_{\rm sun}}{\Delta m^2_{\rm atm}}
\; =\; \frac{|V_{e3}|^4}{|V_{e1}|^4} \cdot
\frac{\displaystyle \left | |V_{e1}|^4 \sin^2 2\rho -
|V_{e2}|^4 \sin^2 2\sigma \right |}
{\displaystyle \left | |V_{e2}|^4 \sin^2 2(\sigma -\rho) -
|V_{e3}|^4 \sin^2 2\rho \right |} \;\; .
%       (2.34)
\end{equation}
Then $\Delta m^2_{\rm sun}/\Delta m^2_{\rm atm} \sim 10^{-2}$ 
imposes quite strong constraints on the $(\rho, \sigma)$ parameter 
space, as numerically illustrated in Fig. 2 of Ref. \cite{Xing03a}.
One can in turn arrive at the allowed ranges of $m_1/m_2$ and
$m_2/m_3$ with the help of (2.33).
  
We certainly hope that $\langle m\rangle_{ee}$ is non-zero and
measurable. If the parameters of neutrino oscillations are
well determined, a measurement of $\langle m\rangle_{ee}$ will
allow us to extract valuable information about the Majorana 
phases of CP violation and the absolute scale of neutrino masses. 
Taking account of (2.19) and (2.32), one may rewrite (2.30) as
\begin{eqnarray}
\langle m\rangle_{ee} & = & \left | \sqrt{m^2_3 + 
p\Delta m^2_{\rm atm} + q\Delta m^2_{\rm sun}} ~ |V_{e1}|^2 
e^{2i\rho} \right . 
\nonumber \\
& & \left . + \sqrt{m^2_3 + p\Delta m^2_{\rm atm}} ~ |V_{e2}|^2
e^{2i\sigma} + m_3 |V_{e3}|^2 \right | \; ,
%       (2.35)
\end{eqnarray}
where $|V_{ei}|^2$ can be expressed in terms of the 
mixing factors of solar and CHOOZ reactor neutrino 
oscillations (see section 3.1.1 for details):
\begin{eqnarray}
|V_{e1}|^2 & = & \frac{1}{2} \left ( \cos^2\theta_{\rm chz}
+ \sqrt{\cos^4\theta_{\rm chz} - \sin^2 2\theta_{\rm sun}} \right ) \; ,
\nonumber \\
|V_{e2}|^2 & = & \frac{1}{2} \left ( \cos^2\theta_{\rm chz}
- \sqrt{\cos^4\theta_{\rm chz} - \sin^2 2\theta_{\rm sun}} \right ) \; ,
\nonumber \\
|V_{e3}|^2 & = & \sin^2 \theta_{\rm chz} \; .
%	(2.36)
\end{eqnarray}
We see that $\langle m\rangle_{ee}$ consists of three unknown parameters: 
$m_3$, $\rho$ and $\sigma$, which are unable to be determined
in any neutrino oscillation experiments. Once $\Delta m^2_{\rm sun}$,
$\Delta m^2_{\rm atm}$, $\theta_{\rm sun}$ and $\theta_{\rm chz}$ are 
measured to an acceptable degree of accuracy, one should be able to 
get a useful constraint on the absolute neutrino mass $m_3$ 
for arbitrary values of $\rho$ and $\sigma$ from the observation of
$\langle m\rangle_{ee}$. If the magnitude of $m_3$ could roughly be 
known from some cosmological constraints, it would be likely to obtain 
some loose but instructive information on the Majorana phases $\rho$ 
and $\sigma$ by confronting (2.35) with the experimental result of 
$\langle m\rangle_{ee}$. 

Let us illustrate the dependence of $\langle m\rangle_{ee}$ on 
$m_3$, $\rho$ and $\sigma$ in a numerical way. For simplicity,
we typically take $\Delta m^2_{\rm sun} = 5\times 10^{-5} ~ {\rm eV}^2$,
$\Delta m^2_{\rm atm} = 3\times 10^{-3} ~ {\rm eV}^2$, $q=-1$,
$\sin^2 2\theta_{\rm sun} = 0.8$ and $\sin^2 2\theta_{\rm chz} =0.05$
in our calculations.
We consider four instructive possibilities for the unknown Majorana
phases $\rho$ and $\sigma$: (1) $\rho = \sigma = 0$; (2) $\rho = \pi/4$ 
and $\sigma = 0$; (3) $\rho = 0$ and $\sigma = \pi/4$; and (4)
$\rho =\sigma = \pi/4$. The results for $\langle m\rangle_{ee}$ as a 
function of $m_3$ are shown in Fig. 2.6.
We find that it is numerically difficult to distinguish between  
possibilities (1) and (4), or between possibilities (2) and (3).
The reason is simply that the first two terms on the right-hand side of
(2.35) dominate over the third term, and they do not cancel each other
for the chosen values of $\rho$ and $\sigma$.
We also find that the changes of $\langle m\rangle_{ee}$ are rather
mild. If reasonable inputs of $\sin^2 2\theta_{\rm sun}$, 
$\sin^2 2\theta_{\rm atm}$ and $\sin^2 2\theta_{\rm chz}$ are taken, a 
careful numerical scan shows that the magnitude of $\langle m\rangle_{ee}$
does not undergo any dramatic changes for arbitrary $\rho$ and $\sigma$. 
Thus a rough but model-independent constraint on the absolute scale of 
neutrino masses can be obtained from the observation of 
$\langle m\rangle_{ee}$. Taking account of the alleged experimental 
region $0.05 ~ {\rm eV} \leq \langle m\rangle_{ee} \leq 0.84 ~ {\rm eV}$ 
in Ref. \cite{KK}, one may arrive at the conclusion that $m_3$ is most 
likely to lie in the range 0.1 eV $\leq m_3 \leq$ 1 eV (see Fig. 2.6). 
This result implies that $m_1 \approx m_2 \approx m_3$ and
$m_1 + m_2 +m_3 \approx 3 m_3 \approx (0.3 - 1.5)$ eV hold. 
Such a sum of three neutrino masses is only partly compatible with
the robust WMAP upper bound given in (2.22). 
%%%%%%%%%%%%%%%%%%%% Fig. 2.6 %%%%%%%%%%%%%%%%
\begin{figure}[t]
\vspace{-2.75cm}
\epsfig{file=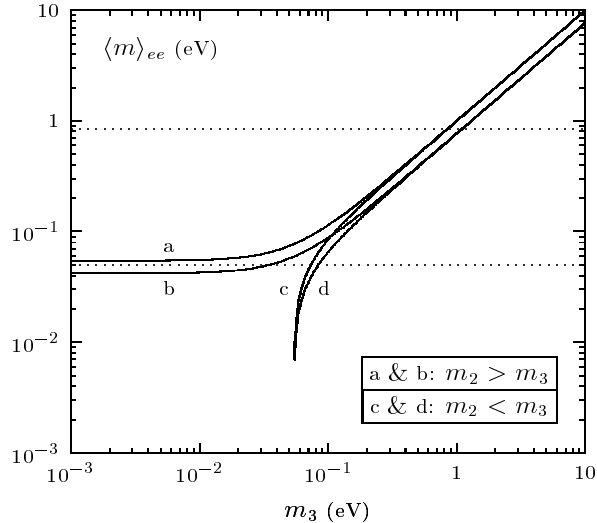,bbllx=1cm,bblly=2cm,bburx=19cm,bbury=31cm,%
width=15.5cm,height=22cm,angle=0,clip=}
\vspace{-11.95cm}
\caption{Illustrative dependence of $\langle m\rangle_{ee}$ on $m_3$ for 
the neutrino mass spectrum $m_2 > m_3$ (curves a and b) and the
neutrino mass spectrum 
$m_2 < m_3$ (curves c and d), where we have typically taken
$\{\rho, \sigma \} = \{0, 0 \}$ or $\{ \pi/4, \pi/4 \}$ (curves a and c) 
and $\{\rho, \sigma \} = \{ 0, \pi/4 \}$ or $\{ \pi/4, 0 \}$ 
(curves b and d). The region between two dashed lines corresponds to the
values of $\langle m\rangle_{ee}$ reported in Ref. \cite{KK}.}
\end{figure}
%%%%%%%%%%%%%%%%%%%%%%%%%%%%%%%%%%%%%%%%%%%%

Because of its importance, the neutrinoless double-$\beta$ decay has
attracted a lot of very delicate studies. For recent comprehensive 
analyses of the dependence of $\langle m\rangle_{ee}$ on neutrino masses 
and Majorana phases, we refer the reader to 
Refs. \cite{Bilenky,KK2,Petcov02} and references therein.

\subsection{Lepton Mass Matrices}

At low energy scales, the phenomenology of lepton masses and flavor
mixing can be formulated in terms of the charged lepton mass matrix
$M_l$ and the (effective) neutrino mass matrix $M_\nu$. They totally
involve twelve physical parameters:
\begin{itemize}
\item     three masses of charged leptons ($m_e$, $m_\mu$ and $m_\tau$),
which have precisely been measured \cite{PDG02};
\item     three masses of neutrinos ($m_1$, $m_2$ and $m_3$), whose 
relative sizes have roughly been known from solar and atmospheric 
neutrino oscillations;
\item     three angles of flavor mixing ($\theta_x$, $\theta_y$ and 
$\theta_z$), whose values have been determined or constrained to an
acceptable degree of accuracy from solar, atmospheric and reactor 
neutrino oscillations;
\item     three phases of CP violation ($\delta$, $\rho$ and $\sigma$),
which are completely unrestricted by current neutrino data.
\end{itemize}
The long-baseline neutrino oscillation experiments are expected to 
determine or constrain $\Delta m^2_{21}$, $\Delta m^2_{32}$, $\theta_x$, 
$\theta_y$, $\theta_z$ and $\delta$ (the Dirac-type CP-violating phase) 
to a high degree of accuracy in the future. The precision experiments 
for the tritium beta decay and the neutrinoless double-$\beta$ 
decay may help determine or constrain the absolute scale of neutrino 
masses. The possible regions of two Majorana-type CP-violating phases
($\rho$ and $\sigma$) could roughly be constrained by the neutrinoless
double-$\beta$ decay. It seems hopeless, at least for the presently
conceivable sets of feasible neutrino experiments, to fully determine
$(m_1, m_2, m_3)$ and $(\rho, \sigma)$ \cite{FGM}. In this case, to 
establish testable relations between the experimental quantities and the 
parameters of $M_l$ and $M_\nu$ has to rely on some phenomenological
hypotheses. 

The specific structures of $M_l$ and $M_\nu$ depend substantially on
the flavor basis to be chosen. There are two simple possibilities to
choose the flavor basis: (a) $M_l$ is diagonal and has positive 
eigenvalues; and (b) $M_\nu$ is diagonal and has positive eigenvalues. 
Let us focus on case (a) in the subsequent discussions, because it has 
popularly been taken in the phenomenological study of
lepton mass matrices
%%%%%%%%%%%%%%%%%%%%%%%%%%%%%%%
\footnote{Case (b) has been considered in Ref. \cite{FX96}, 
for example, to derive the {\it democratic} (nearly bi-maximal) neutrino 
mixing pattern.}.
%%%%%%%%%%%%%%%%%%%%%%%%%%%%%%
%%%%%%%%%%%%%%%%%%%%%% Table 2.1 %%%%%%%%%%%%%%%%%%%%%
\begin{table}[t]
\caption{Nine two-zero textures of the neutrino mass matrix, which
are compatible with current neutrino oscillation data. Note that 
the patterns in the same category (e.g., $\rm A_1$ and $\rm A_2$) 
have very similar phenomenological consequences \cite{Guo03}.}
\vspace{0.2cm}
\begin{center}
\begin{tabular}{lll|ll} \hline\hline
Pattern  & By current data && Pattern & By current data \\ \hline 
%-----------------------------------
$\rm A_1: {\small
\left ( \matrix{
{\bf 0} & {\bf 0} & \times \cr
{\bf 0} & \times & \times \cr
\times & \times 
& \times \cr} \right )}$
& favored &&
$\rm A_2: {\small
\left ( \matrix{
{\bf 0} & \times & {\bf 0} \cr
\times & \times & \times \cr
{\bf 0} & \times & \times \cr} \right )}$
& favored
\\ 
%-----------------------------------------
$\rm B_1: {\small
\left ( \matrix{
\times & \times & {\bf 0} \cr
\times & {\bf 0} & \times \cr
{\bf 0} & \times & \times \cr} \right )}$
& favored &&
$\rm B_2: {\small
\left ( \matrix{
\times & {\bf 0} & \times \cr
{\bf 0} & \times & \times \cr
\times & \times & {\bf 0} \cr} \right )}$
& favored
\\
%-----------------------------------------
$\rm B_3: {\small
\left ( \matrix{
\times & {\bf 0} & \times \cr
{\bf 0} & {\bf 0} & \times \cr
\times & \times & \times \cr} \right )}$
& favored &&
$\rm B_4: {\small
\left ( \matrix{
\times & \times & {\bf 0} \cr
\times & \times & \times \cr
{\bf 0} & \times & {\bf 0} \cr} \right )}$
& favored
\\
%-----------------------------------------
$\rm C: ~ {\small
\left ( \matrix{
\times & \times & \times \cr
\times & {\bf 0} & \times \cr
\times & \times & {\bf 0} \cr} \right )}$
& favored &&
&
\\
%-----------------------------------------------------------
$\rm D_1: {\small
\left ( \matrix{
\times & \times & \times \cr
\times & {\bf 0} & {\bf 0} \cr
\times & {\bf 0} & \times \cr} \right )}$
& marginally allowed &&
$\rm D_2: {\small
\left ( \matrix{
\times & \times & \times \cr
\times & \times & {\bf 0} \cr
\times & {\bf 0} & {\bf 0} \cr} \right )}$
& marginally allowed 
\\ \hline\hline
%-----------------------------------------------------------
\end{tabular}
\end{center}
\end{table}
%%%%%%%%%%%%%%%%%%%%%%%%%%%%%%%%%%%%%%%%%%%%%%%%%%%%%%%%%%%%

In the flavor basis where $M_l$ is diagonal (i.e., the flavor 
eigenstates of charged leptons are identified with their mass
eigenstates), the symmetric neutrino mass matrix $M_\nu$ can be
diagonalized by a single unitary transformation,
\begin{equation}
V^\dagger M_\nu V^* \; = \; \left (\matrix{
m_1 & 0 & 0 \cr
0 & m_2 & 0 \cr
0 & 0 & m_3 \cr} \right ) \; .
%        (2.37)
\end{equation}  
The $3\times 3$ unitary matrix $V$ is just the lepton flavor mixing 
matrix, consisting of three mixing angles and three CP-violating 
phases. It is straightforward to fix the form of $M_\nu$ from (2.37),
\begin{equation}
M_\nu \; = \; V \left (\matrix{
m_1 & 0 & 0 \cr
0 & m_2 & 0 \cr
0 & 0 & m_3 \cr} \right ) V^T \; ,
%        (2.38)
\end{equation}  
but extra assumptions have to be made for $M_\nu$ in order to 
relate neutrino masses to flavor mixing parameters. The general 
idea is to assume that some independent matrix elements of $M_\nu$ 
are actually dependent upon one another, caused by an underlying
(broken) flavor symmetry. In particular, this dependence becomes
very simple and transparent, if the relevant matrix elements 
are exactly equal to zero. Three general categories of 
$M_\nu$ with texture zeros have been classified and discussed
%%%%%%%%%%%%%%%%%%%%%%%%%%%
\footnote{Because $M_\nu$ is symmetric, a pair of off-diagonal texture 
zeros of $M_\nu$ have been counted as one zero.}:
%%%%%%%%%%%%%%%%%%%%%%%%%%%
\begin{itemize}
\item       The three-zero textures of $M_\nu$. There are totally
twenty possible patterns of $M_\nu$ with three independent vanishing
entries, but none of them is allowed by current neutrino oscillation
data \cite{Guo}.
\item       The two-zero textures of $M_\nu$. There are totally
fifteen possible patterns of $M_\nu$ with two independent vanishing
entries. Nine two-zero patterns of $M_\nu$, as illustrated in Table 2.1, 
are found to be compatible with current experimental data 
(but two of them are only marginally allowed \cite{Guo03}).
\item       The one-zero textures of $M_\nu$. There are totally
six possible patterns of $M_\nu$ with one vanishing entry, 
as shown in Table 2.2. It has been found in Ref. \cite{Xing03b}
that three one-zero textures of $M_\nu$ with $m_1 =0$ and four
one-zero textures of $M_\nu$ with $m_3 =0$ are compatible with the
present neutrino oscillation data.
\end{itemize}
Of course, the specific texture zeros of the neutrino mass matrix
may not be preserved to all orders or at any energy scales in the
unspecified interactions from which neutrino masses are generated.
Whether $m_1 =0$ or $m_3 =0$ is stable against radiative corrections
should also be taken into account in a concrete theoretical model.
%%%%%%%%%%%%%%%%%%%%%% Table 2.2 %%%%%%%%%%%%%%%%%%%%%
\begin{table}[t]
\caption{Six possible one-zero textures of the neutrino mass matrix.
They may agree or disagree with current neutrino oscillation data,
if one mass eigenvalue ($m_1$ or $m_3$) vanishes \cite{Xing03b}.}
\vspace{0.2cm}
\begin{center}
\begin{tabular}{lll|ll} \hline\hline
Pattern  & By current data && Pattern & By current data \\ \hline 
%-----------------------------------
$\rm A: {\small
\left ( \matrix{
{\bf 0} & \times & \times \cr
\times & \times & \times \cr
\times & \times 
& \times \cr} \right )}$
& $\matrix{m_1 =0 ~~ {\rm favored ~~~~} \cr 
m_3 =0 ~~ {\rm disfavored} \cr}$ &&
$\rm B: {\small
\left ( \matrix{
\times & {\bf 0} & \times \cr
{\bf 0} & \times & \times \cr
\times & \times & \times \cr} \right )}$
& $\matrix{m_1 =0 ~~ {\rm favored} \cr
m_3 =0 ~~ {\rm favored} \cr}$ 
\\ 
%-----------------------------------------
$\rm C: {\small
\left ( \matrix{
\times & \times & {\bf 0} \cr
\times & \times & \times \cr
{\bf 0} & \times & \times \cr} \right )}$
& $\matrix{m_1 =0 ~~ {\rm favored} \cr
m_3 =0 ~~ {\rm favored} \cr}$ &&
$\rm D: {\small
\left ( \matrix{
\times & \times & \times \cr
\times & {\bf 0} & \times \cr
\times & \times & \times \cr} \right )}$
& $\matrix{m_1 =0 ~~ {\rm disfavored ~~} \cr 
m_3 =0 ~~ {\rm favored ~~~~~} \cr}$ 
\\
%-----------------------------------------
$\rm E: {\small
\left ( \matrix{
\times & \times & \times \cr
\times & \times & {\bf 0} \cr
\times & {\bf 0} & \times \cr} \right )}$
& $\matrix{m_1 =0 ~~ {\rm disfavored ~} \cr 
m_3 =0 ~~ {\rm disfavored ~} \cr}$ &&
$\rm F: {\small
\left ( \matrix{
\times & \times & \times \cr
\times & \times & \times \cr
\times & \times & {\bf 0} \cr} \right )}$
& $\matrix{m_1 =0 ~~ {\rm disfavored ~~} \cr 
m_3 =0 ~~ {\rm favored ~~~~~} \cr}$ 
\\ \hline\hline
%-----------------------------------------------------------
\end{tabular}
\end{center}
\end{table}
%%%%%%%%%%%%%%%%%%%%%%%%%%%%%%%%%%%%%%%%%%%%%%%%%%%%%%%%%%%%

In general, both the charged lepton mass matrix $M_l$ and
the (effective) neutrino mass matrix $M_\nu$ are not diagonal.
A typical ansatz of this nature is the Fritzsch pattern \cite{F79}
of lepton mass matrices,
\begin{equation}
M_{l,\nu} \; = \; \left ( \matrix{
{\bf 0}	& ~ C_{l,\nu}	& ~ {\bf 0} \cr
C_{l,\nu}	& ~ {\bf 0}	& ~ B_{l,\nu} \cr
{\bf 0}	& ~ B_{l,\nu}	& ~ A_{l,\nu} \cr} \right ) \; ,
%	(2.39)
\end{equation}
in which six texture zeros are included.
It has been shown in Ref. \cite{XingF} that such a simple ansatz of 
lepton mass matrices can naturally predict a normal hierarchy 
of neutrino masses and a bi-large pattern of lepton flavor mixing. 
A generalized version of (2.39) with four texture zeros,
\begin{equation}
M_{l,\nu} \; = \; \left ( \matrix{
{\bf 0}	& ~ C_{l,\nu}	& ~ {\bf 0} \cr
C_{l,\nu}	& ~ \tilde{B}_{l,\nu}	& ~ B_{l,\nu} \cr
{\bf 0}	& ~ B_{l,\nu}	& ~ A_{l,\nu} \cr} \right ) \; ,
%	(2.40)
\end{equation}
has carefully been analyzed in Ref. \cite{Zhang}. The motivation
to consider the four-zero texture of lepton mass matrices is quite
clear. It is well known that the four-zero texture of
quark mass matrices is more successful than the six-zero texture of
quark mass matrices to interpret the strong hierarchy of quark 
masses and the smallness of flavor mixing angles \cite{40Q}. Thus 
the spirit of lepton-quark similarity motivates us to conjecture
that the lepton mass matrices might have the same texture zeros
as the quark mass matrices. Such a phenomenological
conjecture is indeed reasonable
in some specific models of grand unified theories \cite{GUTmodels}, 
in which the mass matrices of leptons and quarks are related to each 
other by a new kind of flavor symmetry. That is why the four-zero 
texture of lepton mass matrices has been considered as a typical
example in a number of interesting model-building works \cite{40L}. 

Taking account of the seesaw mechanism, one may investigate
how the texture zeros of $M_{\rm D}$ and $M_{\rm R}$ are related to
those of $M_\nu$ via (2.9). For example, it has been shown in 
Ref. \cite{Tanimoto02} that some two-zero textures of $M_\nu$ can 
easily be reproduced from $M_{\rm D}$ and $M_{\rm R}$ with several 
vanishing entries. Provided both $M_{\rm D}$ and 
$M_{\rm R}$ have texture zeros in their (1,1), (1,3) and (3,1) 
positions, $M_\nu$ must have the same texture zeros \cite{FXReview}. 
This seesaw invariance of lepton mass matrices implies that all 
lepton and quark mass matrices might have a universal texture like 
(2.40), resulting from a universal (broken) flavor symmetry at the 
energy scale where the seesaw mechanism works.

There are many other phenomenological ans$\rm\ddot{a}$tze of lepton
mass matrices, either at low energy scales or at high energy scales.
For recent reviews with extensive references, we refer the reader
to Ref. \cite{FXReview}, Refs. \cite{Altarelli}--\cite{Review},
and Ref. \cite{GUTmodels}. 

\section{Lepton Flavor Mixing}
\setcounter{equation}{0}
\setcounter{figure}{0}

\subsection{The Flavor Mixing Matrix in Vacuum}

The phenomenon of lepton flavor mixing arises from the mismatch between
the diagonalization of the charged lepton mass matrix $M_l$ and that
of the neutrino mass matrix $M_\nu$ in an arbitrary flavor basis. 
Without loss of any generality, one can choose to identify the flavor
eigenstates of charged leptons with their mass eigenstates. In this
specific basis with $M_l$ being diagonal, $M_\nu$ is in general not 
diagonal. If massive neutrinos are Dirac particles, it is always 
possible to diagonalize the arbitrary mass matrix $M_\nu$ by a 
bi-unitary transformation:
\begin{equation}
V^{\dagger} M_\nu V'
\; =\; \left (\matrix{
m^{~}_1 & 0 & 0 \cr
0 & m^{~}_2 & 0 \cr
0 & 0 & m^{~}_3 \cr} \right ) \; .
%        (3.1)
\end{equation}
If massive neutrinos are Majorana particles, one may diagonalize the
symmetric mass matrix $M_\nu$ by a single unitary transformation:
\begin{equation}
V^\dagger M_\nu V^*
\; = \; \left (\matrix{
m^{~}_1 & 0 & 0 \cr
0 & m^{~}_2 & 0 \cr
0 & 0 & m^{~}_3 \cr} \right ) \; .
%        (3.2)
\end{equation}
In (3.1) and (3.2), $m^{~}_i$ denote the (positive) mass eigenvalues of
three active neutrinos. In terms of the mass eigenstates of charged 
leptons and neutrinos, the Lagrangian of charged-current weak 
interactions in (2.1) can be rewritten as
\begin{equation}
-{\cal L}_{\rm cc} \; =\; \frac{g}{\sqrt{2}} ~ 
\overline{(\nu^{~}_1, \nu^{~}_2, \nu^{~}_3)^{~}_{\rm L}} ~ 
V^\dagger \gamma^\mu
\left (\matrix{
e \cr
\mu \cr
\tau \cr} \right )_{\rm L} W^+_\mu ~ + ~ {\rm h.c.} \; ,
%       (3.3)
\end{equation}
where the $3\times 3$ unitary matrix $V$ links the neutrino mass 
eigenstates $(\nu^{~}_1, \nu^{~}_2, \nu^{~}_3)$
to the neutrino flavor eigenstates $(\nu_e, \nu_\mu, \nu_\tau)$:
\begin{equation}
\left (\matrix{
\nu_e \cr
\nu_\mu \cr
\nu_\tau \cr} \right ) \; =\; \left (\matrix{
V_{e1} & V_{e2} & V_{e3} \cr
V_{\mu 1} & V_{\mu 2} & V_{\mu 3} \cr
V_{\tau 1} & V_{\tau 2} & V_{\tau 3} \cr} \right )
\left (\matrix{
\nu^{~}_1 \cr
\nu^{~}_2 \cr
\nu^{~}_3 \cr} \right ) \; .
%       (3.4)
\end{equation}
Obviously $V$ is just the flavor mixing matrix of charged leptons and
active neutrinos. The analogue of $V$ in the quark sector is well
known as the Cabibbo-Kobayashi-Maskawa (CKM) matrix \cite{CKM}. In some
literature, the lepton mixing matrix $V$ has been referred to
as the Maki-Nakagawa-Sakata (MNS) matrix \cite{MNS} or the
Pontecorvo-MNS (PMNS) matrix.

Note that massive neutrinos are commonly believed to be Majorana
particles, although there has not been any reliable 
experimental evidence for
that. For simplicity, we use $V$ to denote the lepton flavor mixing 
matrix in the Majorana case. Then $V$ can always be parametrized as a 
product of the Dirac-type flavor mixing matrix $U$ and a diagonal phase 
matrix $P$; i.e., $V = U P$ \cite{Valle}. The nontrivial CP-violating 
phases in $P$ characterize the Majorana nature of $V$. It is 
straightforward to show that the rephasing-invariant quantities
$V_{\alpha i} V_{\beta j} V^*_{\alpha j} V^*_{\beta i}$ and
$U_{\alpha i} U_{\beta j} U^*_{\alpha j} U^*_{\beta i}$ are identical 
for arbitrary indices ($\alpha,\beta$) over ($e, \mu, \tau$) and
($i, j$) over ($1, 2, 3$). This result implies that $V$ and $U$ have
the same physical effects in normal neutrino-neutrino and
antineutrino-antineutrino oscillations, whose probabilities depend 
only upon the rephasing-invariant quantities mentioned above. 
One then arrives at the conclusion that it is impossible to determine 
the nature of massive neutrinos by studying the phenomena of
neutrino oscillations. Only the experiments which probe transitions
between left-handed and right-handed neutrino states, like
the neutrinoless double beta decay, could tell whether massive neutrinos 
are of the Majorana type or not.

\subsubsection{Experimental Constraints}

Current experimental constraints on the lepton flavor mixing matrix
$V$ come mainly from solar, atmospheric and CHOOZ reactor neutrino
oscillations. These three types of neutrino oscillations all belong 
to the ``disappearance'' experiments, in which the survival 
probability of a flavor eigenstate $\nu_\alpha$ is given as 
\begin{equation}
P(\nu_\alpha \rightarrow \nu_\alpha) \; = \; 1 ~ - ~ 4 \sum_{i<j} 
\left (|V_{\alpha i}|^2 |V_{\alpha j}|^2 \sin^2 F_{ji} \right ) \; ,
%       (3.5)
\end{equation}
where $F_{ji} \equiv 1.27 \Delta m^2_{ji} L/E$ with
$\Delta m^2_{ji} \equiv m^2_j - m^2_i$, $L$ stands for the 
baseline length (in unit of km), and $E$ is the neutrino beam 
energy (in unit of GeV). Because of CPT invariance, the survival
probabilities of neutrinos and antineutrinos in vacuum are equal
to each other: 
$P(\overline{\nu}_\alpha \rightarrow \overline{\nu}_\alpha)
= P(\nu_\alpha \rightarrow \nu_\alpha)$. This equality may in
general be violated, if neutrinos and antineutrinos propagate in
matter \cite{XingCPT}.

As mentioned in section 2, the solar ($\nu_e \rightarrow \nu_e$)
and atmospheric ($\nu_\mu \rightarrow \nu_\mu$) neutrino 
oscillation data indicate 
\begin{equation}
|\Delta m^2_{21}| = \Delta m^2_{\rm sun} \ll \Delta m^2_{\rm atm} 
= |\Delta m^2_{32}| \approx |\Delta m^2_{31}| \; .
%        (3.6)
\end{equation} 
In addition, the CHOOZ 
($\overline{\nu}_e \rightarrow \overline{\nu}_e$) neutrino
oscillation data are obtained at the scale 
$\Delta m^2_{\rm chz} \approx |\Delta m^2_{32}|$. To a good degree 
of accuracy, (3.5) is simplified to
\begin{equation}
P(\nu_e \rightarrow \nu_e) \; \approx \; 1 - \sin^2 2\theta_{\rm sun}
\sin^2 \left ( 1.27 \frac{\Delta m^2_{\rm sun} L}{E} \right ) \;
%        (3.7)
\end{equation}
with 
\begin{equation}
\sin^2 2\theta_{\rm sun} \; = \; 4 |V_{e1}|^2 |V_{e2}|^2 \;
%        (3.8)
\end{equation}
for solar neutrino oscillations;
\begin{equation}
P(\nu_\mu \rightarrow \nu_\mu) \; \approx \; 1 - \sin^2 2\theta_{\rm atm}
\sin^2 \left ( 1.27 \frac{\Delta m^2_{\rm atm} L}{E} \right ) \;
%        (3.9)
\end{equation}
with 
\begin{equation}
\sin^2 2\theta_{\rm atm} \; = \; 4 |V_{\mu 3}|^2 \left ( 1 - 
|V_{\mu 3}|^2 \right ) \;
%        (3.10)
\end{equation}
for atmospheric neutrino oscillations; and
\begin{equation}
P(\overline{\nu}_e \rightarrow \overline{\nu}_e) \; \approx \; 
1 - \sin^2 2\theta_{\rm chz}
\sin^2 \left ( 1.27 \frac{\Delta m^2_{\rm chz} L}{E} \right ) \;
%        (3.11)
\end{equation}
with 
\begin{equation}
\sin^2 2\theta_{\rm chz} \; = \; 4 |V_{e 3}|^2 \left ( 1 - 
|V_{e 3}|^2 \right ) \;
%        (3.12)
\end{equation}
for the CHOOZ neutrino oscillation experiment. In view of the 
present experimental data from SK \cite{SK}, SNO \cite{SNO},
KamLAND \cite{KM}, K2K \cite{K2K} and CHOOZ \cite{CHOOZ}, we have 
\begin{eqnarray}
0.25 \; \leq & \sin^2\theta_{\rm sun} & \leq \; 0.4 \; 
~~ {\rm or} ~~~
30.0^\circ \; \leq \; \theta_{\rm sun} \; \leq \; 39.2^\circ \; ,
\nonumber \\
0.92 \; < & \sin^2 2\theta_{\rm atm} & \leq \; 1.0 \; 
~~ {\rm or} ~~~
36.8^\circ \; < \; \theta_{\rm atm} \; < \; 53.2^\circ \; ,
\nonumber \\
0 \; \leq & \sin^2 2\theta_{\rm chz} & < \; 0.1 \;
~~ {\rm or} ~~~~~~~
0^\circ \; \leq \; \theta_{\rm chz} \; < \; 9.2^\circ \; ,
%        (3.13)
\end{eqnarray}
at the $90\%$ confidence level. 
More precise data on $\theta_{\rm sun}$, $\theta_{\rm atm}$ and
$\theta_{\rm chz}$ will be available in the near future.

Taking account of the unitarity of $V$, one may reversely express
$|V_{e1}|$, $|V_{e2}|$, $|V_{e3}|$, $|V_{\mu 3}|$ and $|V_{\tau 3}|$
in terms of $\theta_{\rm sun}$, $\theta_{\rm atm}$ and
$\theta_{\rm chz}$ \cite{Xing02}:
\begin{eqnarray}
|V_{e1}| & = & \frac{1}{\sqrt 2} \sqrt{ \cos^2\theta_{\rm chz} +
\sqrt{\cos^4\theta_{\rm chz} - \sin^2 2\theta_{\rm sun}}} \;\; ,
\nonumber \\
|V_{e2}| & = & \frac{1}{\sqrt 2} \sqrt{ \cos^2\theta_{\rm chz} -
\sqrt{\cos^4\theta_{\rm chz} - \sin^2 2\theta_{\rm sun}}} \;\; ,
\nonumber \\
|V_{e3}| & = & \sin\theta_{\rm chz} \; ,
\nonumber \\
|V_{\mu 3}| & = & \sin\theta_{\rm atm} \; ,
\nonumber \\
|V_{\tau 3}| & = & \sqrt{\cos^2\theta_{\rm chz} -
\sin^2\theta_{\rm atm}} \; .
%        (3.14)
\end{eqnarray}
The other four matrix elements of $V$ (i.e., 
$|V_{\mu 1}|$, $|V_{\mu 2}|$, $|V_{\tau 1}|$ and $|V_{\tau 2}|$) 
are entirely unrestricted, however, unless one
of them or the rephasing invariant of CP violation of $V$ (defined
in section 4.1) is measured. A realistic way to get rough but useful
constraints on those four unknown elements is to allow the Dirac
phase of CP violation in $V$ to vary between 0 and
$\pi$ \cite{Fukugita}, such that one can find out the maximal and
minimal magnitudes of each matrix element. To see this point more
clearly, we adopt a simplified parametrization of $V$ in which the
Majorana phases of CP violation are omitted \cite{Maiani}:
\begin{equation}
V \; = \; \left ( \matrix{
c_x c_z & s_x c_z & s_z \cr
- c_x s_y s_z - s_x c_y e^{-i\delta} &
- s_x s_y s_z + c_x c_y e^{-i\delta} &
s_y c_z \cr
- c_x c_y s_z + s_x s_y e^{-i\delta} &
- s_x c_y s_z - c_x s_y e^{-i\delta} &
c_y c_z \cr } \right ) \; ,
%       (3.15)   
\end{equation}
where $s_x \equiv \sin\theta_x$, $c_x \equiv \cos\theta_x$, and
so on. The Dirac phase $\delta$ affects the magnitudes
of $V_{\mu 1}$, $V_{\mu 2}$, $V_{\tau 1}$ and $V_{\tau 2}$,
while the Majorana phases of CP violation do not have such effects
(see section 3.1.2 for more discussions).
Note that three mixing angles $(\theta_x, \theta_y, \theta_z)$, 
which are all arranged to lie in the first quadrant, can be written as
\begin{equation}
\tan\theta_x \; = \; \frac{|V_{e2}|}{|V_{e1}|} \; ,
~~~~~
\tan\theta_y \; = \; \frac{|V_{\mu 3}|}{|V_{\tau 3}|} \; ,
~~~~~
\sin\theta_z \; = \; |V_{e3}| \; .
%       (3.16)
\end{equation}
It is then straightforward to obtain
\begin{eqnarray}
|V_{\mu 1}| & = & \frac{\left | |V_{e2}| |V_{\tau 3}| +
|V_{e1}| |V_{e3}| |V_{\mu 3}| ~ e^{i\delta} \right |}
{1 - |V_{e3}|^2} \; ,
\nonumber \\
|V_{\mu 2}| & = & \frac{\left | |V_{e1}| |V_{\tau 3}| -
|V_{e2}| |V_{e3}| |V_{\mu 3}| ~ e^{i\delta} \right |}
{1 - |V_{e3}|^2} \; ,
\nonumber \\
|V_{\tau 1}| & = & \frac{\left | |V_{e2}| |V_{\mu 3}| -
|V_{e1}| |V_{e3}| |V_{\tau 3}| ~ e^{i\delta} \right |}
{1 - |V_{e3}|^2} \; ,
\nonumber \\
|V_{\tau 2}| & = & \frac{\left | |V_{e1}| |V_{\mu 3}| +
|V_{e2}| |V_{e3}| |V_{\tau 3}| ~ e^{i\delta} \right |}
{1 - |V_{e3}|^2} \; .
%       (3.17)
\end{eqnarray}
Varying the Dirac phase $\delta$ from 0 to $\pi$, we are led to
the {\it most generous} ranges of $|V_{\mu 1}|$, $|V_{\mu 2}|$,
$|V_{\tau 1}|$ and $|V_{\tau 2}|$ \cite{GX03}:
\begin{eqnarray}
\frac{|V_{e2}| |V_{\tau 3}| - |V_{e1}| |V_{e3}| |V_{\mu 3}|}
{1 - |V_{e3}|^2} & \leq & |V_{\mu 1}| \; \leq \;
\frac{|V_{e2}| |V_{\tau 3}| + |V_{e1}| |V_{e3}| |V_{\mu 3}|}
{1 - |V_{e3}|^2} \; ,
\nonumber \\
\frac{|V_{e1}| |V_{\tau 3}| - |V_{e2}| |V_{e3}| |V_{\mu 3}|}
{1 - |V_{e3}|^2} & \leq & |V_{\mu 2}| \; \leq \;
\frac{|V_{e1}| |V_{\tau 3}| + |V_{e2}| |V_{e3}| |V_{\mu 3}|}
{1 - |V_{e3}|^2} \; ,
\nonumber \\
\frac{|V_{e2}| |V_{\mu 3}| - |V_{e1}| |V_{e3}| |V_{\tau 3}|}
{1 - |V_{e3}|^2} & \leq & |V_{\tau 1}| \; \leq \;
\frac{|V_{e2}| |V_{\mu 3}| + |V_{e1}| |V_{e3}| |V_{\tau 3}|}
{1 - |V_{e3}|^2} \; ,
\nonumber \\
\frac{|V_{e1}| |V_{\mu 3}| - |V_{e2}| |V_{e3}| |V_{\tau 3}|}
{1 - |V_{e3}|^2} & \leq & |V_{\tau 2}| \; \leq \;
\frac{|V_{e1}| |V_{\mu 3}| + |V_{e2}| |V_{e3}| |V_{\tau 3}|}
{1 - |V_{e3}|^2} \; .
%       (3.18)
\end{eqnarray}
Note that the lower and upper bounds of each matrix element turn to
coincide with each other in the limit $|V_{e3}| \rightarrow 0$. 
Because of the smallness of $|V_{e3}|$, the ranges obtained in 
(3.18) should be quite restrictive. Hence it makes sense to recast 
the lepton flavor mixing matrix even in the absence of any 
experimental information on CP violation.

Using the values of $\theta_{\rm sun}$, $\theta_{\rm atm}$ and
$\theta_{\rm chz}$ given in (3.13), we calculate  
$|V_{e1}|$, $|V_{e2}|$, $|V_{e3}|$, $|V_{\mu 3}|$ and $|V_{\tau 3}|$
from (3.14). Then the allowed ranges of $|V_{\mu 1}|$, $|V_{\mu 2}|$,
$|V_{\tau 1}|$ and $|V_{\tau 2}|$ can be found with the help of
(3.18). Our numerical results are summarized as \cite{GX03}
\begin{equation}
|V| \; =\; \left ( \matrix{ 0.70-0.87 & 0.50-0.69 & <0.16 \cr
0.20-0.61 & 0.34-0.73 & 0.60-0.80 \cr 0.21-0.63 & 0.36-0.74 &
0.58-0.80 \cr} \right ) \; .
%       (3.19)
\end{equation}
This represents our present knowledge on the lepton flavor mixing
matrix.

\subsubsection{Standard Parametrization}

Flavor mixing among $n$ different lepton families is in general
described by a $n\times n$ unitary matrix $V$, whose number of 
independent parameters relies on the style of neutrinos. If neutrinos 
are Dirac particles, $V$ can be parametrized in terms of $n (n-1)/2$ 
rotation angles and $(n-1)(n-2)/2$ phase angles. If neutrinos are 
Majorana particles, however, a full parametrization of $V$ requires 
$n (n-1)/2$ rotation angles and the same number of phase angles
%%%%%%%%%%%%%%%%%%%%%%%%%%%
\footnote{No matter whether neutrinos are Dirac or Majorana particles,
the $n\times n$ flavor mixing matrix has $(n-1)^2(n-2)^2/4$ rephasing
invariants of CP violation \cite{Chau}.}.
%%%%%%%%%%%%%%%%%%%%%%%%%%
The flavor mixing of charged leptons and Dirac neutrinos is completely 
analogous to that of quarks, for which a number of different 
parametrizations have been proposed and classified in the 
literature \cite{FX98}. In this subsection we classify all possible 
parametrizations for the flavor mixing of charged leptons and
Majorana neutrinos with $n=3$. Regardless of the phase-assignment
freedom, we find that there are nine structurally
different parametrizations for the $3\times 3$ lepton flavor mixing
matrix $V$. Although all nine representations are mathematically
equivalent, one of them is likely to make the underlying physics of 
lepton mass generation (and CP violation) more transparent, or is 
more useful in the analyses of neutrino oscillation data. 

The $3\times 3$ lepton flavor mixing matrix $V$ defined in (3.4)
can be expressed as a product of three unitary matrices $O_1$, $O_2$ 
and $O_3$, which correspond to simple rotations in the complex (1,2), 
(2,3) and (3,1) planes:
\begin{eqnarray}
O_1(\theta_1, \alpha^{~}_1, \beta_1, \gamma^{~}_1) & = &
\left ( \matrix{
c_1 e^{i\alpha^{~}_1} &
s_1 e^{-i\beta_1} & 
0 \cr
-s_1 e^{i\beta_1} &
c_1 e^{-i\alpha^{~}_1} &
0 \cr
0 & 0 & e^{i\gamma^{~}_1} \cr} \right ) \; ,
\nonumber \\ \nonumber \\
O_2(\theta_2, \alpha^{~}_2, \beta_2, \gamma^{~}_2) & = &
\left ( \matrix{
e^{i\gamma^{~}_2} & 0 & 0 \cr
0 &
c_2 e^{i\alpha^{~}_2} &
s_2 e^{-i\beta_2} \cr
0 &
-s_2 e^{i\beta_2} &
c_2 e^{-i\alpha^{~}_2} \cr} \right ) \; ,
\nonumber \\ \nonumber \\
O_3(\theta_3, \alpha^{~}_3, \beta_3, \gamma^{~}_3) & = &
\left ( \matrix{
c_3 e^{i\alpha^{~}_3} &
0 &
s_3 e^{-i\beta_3} \cr
0 & e^{i\gamma^{~}_3} & 0 \cr
-s_3 e^{i\beta_3} &
0 &
c_3 e^{-i\alpha^{~}_3} \cr} \right ) \; ,
%       (3.20)
\end{eqnarray}
where $s_i \equiv \sin\theta_i$ and $c_i \equiv \cos\theta_i$
(for $i = 1, 2, 3$).
Obviously $O_i O^\dagger_i = O^\dagger_i O_i = {\bf 1}$ holds,
and any two rotation matrices do not commute with each other. We find 
twelve different ways to arrange the product of $O_1$, $O_2$ and
$O_3$, which can cover the whole $3\times 3$ space and provide a full
description of $V$. Explicitly, six of the twelve different 
combinations of $O_i$ belong to the type \cite{FX01}
\begin{equation}
V \; = \; O_i(\theta_i, \alpha^{~}_i, \beta_i, \gamma^{~}_i) \otimes
O_j(\theta_j, \alpha^{~}_j, \beta_j, \gamma^{~}_j) \otimes
O_i(\theta'_i, \alpha'_i, \beta'_i, \gamma'_i) \;
%       (3.21)
\end{equation}
with $i\neq j$, where the complex rotation matrix $O_i$ occurs twice; 
and the other six belong to the type
\begin{equation}
V  =  O_i(\theta_i, \alpha^{~}_i, \beta_i, \gamma^{~}_i) \otimes
O_j(\theta_j, \alpha^{~}_j, \beta_j, \gamma^{~}_j) \otimes
O_k(\theta_k, \alpha^{~}_k, \beta_k, \gamma^{~}_k) \; 
%       (3.22)
\end{equation}
with $i\neq j\neq k$, in which the rotations take place in three 
different complex planes. Note that the products
$O_i O_j O_i$ and $O_i O_k O_i$ (for $i\neq k$) in (3.21)
are correlated with each other, if the relevant phase parameters are
switched off. Hence only nine of the twelve parametrizations,
three from (3.21) and six from (3.22), are structurally different.

In each of the nine distinct parametrizations for $V$, there 
apparently exist nine phase parameters. Six of them or their 
combinations can be absorbed by redefining the arbitrary phases of 
charged lepton fields and a common phase of neutrino fields. If 
neutrinos are Dirac particles, one can also redefine the arbitrary 
phases of Dirac neutrino fields, so as to reduce the number of the 
remaining phase parameters from three to one. In this case $V$ 
contains a single CP-violating phase of the Dirac nature.
If neutrinos are Majorana particles, however, there is no freedom to 
rearrange the relative phases of Majorana neutrino fields. Hence 
three nontrivial phase parameters are present in the $3\times 3$ 
Majorana-like flavor mixing matrix $V$. Two of the three CP-violating 
phases in $V$ can always be factored out as the ``pure'' Majorana 
phases through a proper phase assignment of the charged lepton fields, 
and the remaining one is identified as the ``Dirac-like'' phase. 
Both CP- and T-violating effects in normal neutrino oscillations
depend only upon the Dirac-like phase.

Of course, different parametrizations of $V$
are mathematically equivalent, and adopting any of them does not point to
physical significance. It is quitely likely, however, that one particular
parametrization is more useful and transparent than the others in the
analyses of data from various neutrino experiments and (or) towards a 
deeper understanding of the underlying dynamics responsible for lepton 
mass generation and CP violation. We find that such an interesting  
parametrization does exist \cite{FX01}:
\begin{equation}
V \; = \; \left ( \matrix{
c_x c_z & s_x c_z & s_z \cr
- c_x s_y s_z - s_x c_y e^{-i\delta} &
- s_x s_y s_z + c_x c_y e^{-i\delta} &
s_y c_z \cr 
- c_x c_y s_z + s_x s_y e^{-i\delta} & 
- s_x c_y s_z - c_x s_y e^{-i\delta} & 
c_y c_z \cr } \right ) 
\left ( \matrix{
e^{i\rho}	& 0	& 0 \cr
0	& e^{i\sigma}	& 0 \cr
0	& 0	& 1 \cr} \right ) \; 
%     	(3.23)
\end{equation}
with $s_x \equiv \sin\theta_x$, $c_y \equiv \cos\theta_y$, and so on.
Without loss of generality, the three mixing angles 
($\theta_x, \theta_y, \theta_z$) can all be arranged to lie in
the first quadrant. Arbitrary values between $0$ and $2\pi$
are allowed for the Dirac CP-violating phase $\delta$ and the
Majorana CP-violating phases $\rho$ and $\sigma$.

The parametrization in (3.23) can be regarded as a straightforward
generalization of
the parametrization in (3.15) to include the Majorana phases of CP
violation. Thus the interesting relations obtained in (3.16) remain 
valid. A remarkable merit of this parametrization is that
its three mixing angles $(\theta_x, \theta_y, \theta_z)$
are directly related to the mixing angles of solar, atmospheric and
CHOOZ reactor neutrino oscillations:
\begin{equation}
\theta_x \; \approx \; \theta_{\rm sun} \; ,
~~~~~
\theta_y \; \approx \; \theta_{\rm atm} \; ,
~~~~~
\theta_z \; \approx \; \theta_{\rm chz} \; ,
%	(3.24)
\end{equation}
which can easily be derived from (3.8), (3.10), (3.12)
and (3.23). Other parametrizations of the lepton flavor mixing
matrix would not be able to result in such simple relations 
between the fundamental mixing quantities and the relevant
experimental observables.

The CP- and T-violating effects in normal neutrino oscillations are
measured by the well-known Jarlskog parameter $J$ \cite{Jarlskog}
(defined in section 4.1), which is proportional to $\sin\delta$ in 
the following way: 
\begin{equation}
J \; \approx \; \sin\theta_{\rm sun} \cos\theta_{\rm sun}
\sin\theta_{\rm atm} \cos\theta_{\rm atm}
\sin\theta_{\rm chz} \cos^2\theta_{\rm chz} \sin\delta \; .
%	(3.25)
\end{equation}
To see this point more clearly, we write out the transition 
probabilities
of different neutrino flavor eigenstates in vacuum \cite{FXReview}:
\begin{equation}
P(\nu_\alpha \rightarrow \nu^{~}_\beta) \; = \;
- 4 \sum_{i<j} \left [ {\rm Re} \left (V_{\alpha i} V_{\beta j} 
V^*_{\alpha j} V^*_{\beta i} \right ) \sin^2 F_{ji} \right ] ~ - ~ 
8 J \prod_{i<j} \left (\sin F_{ji} \right ) \; ,
%	(3.26)
\end{equation}
where $(\alpha,\beta)$ run over $(e,\mu)$, $(\mu,\tau)$ or $\tau,e)$;
$i$ and $j$ run over 1, 2, 3; and $F_{ji}$ and $\Delta m^2_{ji}$ 
have been defined below (3.5). In the assumption of 
CPT invariance, the transition probabilities 
$P(\nu^{~}_\beta \rightarrow \nu_\alpha)$ or
$P(\overline{\nu}_\alpha \rightarrow \overline{\nu}^{~}_\beta)$ 
can directly be read off from (3.26) through the replacement 
$J \Longrightarrow -J$ (i.e., $V \Longrightarrow V^*$). Therefore,
the CP-violating asymmetry between
$P(\nu_\alpha \rightarrow \nu^{~}_\beta)$ and
$P(\overline{\nu}_\alpha \rightarrow \overline{\nu}^{~}_\beta)$
amounts to the T-violating asymmetry between
$P(\nu_\alpha \rightarrow \nu^{~}_\beta)$ and
$P(\nu^{~}_\beta \rightarrow \nu_\alpha)$:
\begin{eqnarray}
\Delta P & \equiv & P(\nu_\alpha \rightarrow \nu^{~}_\beta) -
P(\overline{\nu}_\alpha \rightarrow \overline{\nu}^{~}_\beta) \;
\nonumber \\
& = & P(\nu_\alpha \rightarrow \nu^{~}_\beta) -
P(\nu^{~}_\beta \rightarrow \nu_\alpha) \; 
\nonumber \\
& = & -16 J \sin F_{21} \sin F_{31} \sin F_{32} \; .
%        (3.27)
\end{eqnarray}
The determination of $J$ from $\Delta P$ will allow us to extract the
CP-violating phase $\delta$, provided all three mixing angles
$(\theta_x, \theta_y, \theta_z)$ have been measured elsewhere.
In practical long-baseline neutrino oscillation experiments, however, 
all these measurable quantities may be
contaminated by the terrestrial matter effects.
Hence the fundamental parameters of lepton flavor mixing need be 
disentangled from the matter-corrected ones. Some discussions about 
the matter effects on lepton flavor mixing and CP violation will be
given in sections 3.2 and 4.1.

Regardless of the Majorana phases $\rho$ and $\sigma$, which 
have nothing to do with normal neutrino oscillations, we have located the
Dirac phase $\delta$ in such a way that the matrix elements in the
first row and the third column of $V$ are real. As a consequence, the
CP-violating phase $\delta$ does not appear in the effective mass term of
the neutrinoless double beta decay. Indeed, the latter reads:
\begin{eqnarray}
\langle m \rangle_{ee} & = & \left | m_1 V^2_{e1} + m_2 V^2_{e2} 
+ m_3 V^2_{e3} \right | 
\nonumber \\
& = & \sqrt{{\bf a} + {\bf b}\cos 2(\rho - \sigma) + {\bf c}\cos 2\rho 
+ {\bf d}\cos 2\sigma} \;\; ,
%	(3.28)
\end{eqnarray}
where 
\begin{eqnarray}
{\bf a} & = & m^2_1 c^4_x c^4_z + m^2_2 s^4_x c^4_z + m^2_3 s^4_z \; ,
\nonumber \\
{\bf b} & = & 2 m_1 m_2 s^2_x c^2_x c^4_z \; ,
\nonumber \\
{\bf c} & = & 2 m_1 m_3 c^2_x s^2_z c^2_z \; ,
\nonumber \\
{\bf d} & = & 2 m_2 m_3 s^2_x s^2_z c^2_z \; .
%	(3.29)
\end{eqnarray}
It becomes obvious that $\langle m \rangle_{ee}$ is independent of 
the Dirac phase $\delta$. On the other hand, the CP- or T-violating 
asymmetry $\Delta P$ in normal neutrino oscillations is independent 
of the Majorana phases $\rho$ and $\sigma$. Thus two different types 
of CP-violating phases can 
(in principle) be studied in two different types of experiments. 

A long-standing and important question is whether the two Majorana
phases $\rho$ and $\sigma$ can separately be determined by measuring
other possible lepton-number-nonconserving processes, in addition to 
the neutrinoless double beta decay. While the answer to this question 
is affirmative in principle, it seems to be negative in practice. The
key problem is that those lepton-number-violating processes, in which
the Majorana phases can show up, are suppressed in magnitude by an 
extremely small factor compared to normal weak 
interactions \cite{FX01,Kayser02}. 
Therefore it is extremely difficult, even impossible, to measure or 
constrain $\rho$ and $\sigma$ in any experiment other than the one 
associated with the neutrinoless double beta decay. 

On the theoretical side, how to predict or calculate those flavor 
mixing angles and CP-violating phases on a solid dynamical ground is 
also a big challenge. Phenomenologically, the parametrization
in (3.23) is expected to be very useful and convenient, and might even 
be able to provide some insight into the underlying physics of lepton 
mass generation. We therefore recommend it to experimentalists and 
theorists as the standard parametrization of the $3\times 3$ lepton 
flavor mixing matrix.

\subsubsection{Constant Mixing Patterns}

Combining (3.13) and (3.24), one can immediately observe the essential
feature of lepton flavor mixing: two mixing angles are large and
comparable in magnitude ($\theta_x \sim \theta_y \sim 1$), 
while the third one is very small ($\theta_z \ll 1$). Such a
``nearly bi-maximal'' flavor mixing pattern has attracted a lot of 
attention in model building. In particular, a number of constant mixing
patterns, which are more or less close to the experimentally-allowed
lepton flavor mixing matrix in (3.19), have been proposed. Possible
flavor symmetries and their spontaneous or explicit breaking
mechanisms hidden in those constant patterns might finally help us 
pin down the dynamics responsible for neutrino mass generation and
lepton flavor mixing. It is therefore worthwhile to have an overview of 
a few typical constant lepton mixing patterns existing in the literature
%%%%%%%%%%%%%%%%%%%%%%
\footnote{We only concentrate on those nearly bi-maximal flavor mixing
patterns, in which $\theta_x \sim \theta_y >> \theta_z$ holds. Hence
the interesting Cabibbo ansatz \cite{Cabibbo78}, which has
$|V_{\alpha i}| =1/\sqrt{3}$ (for $\alpha =e,\mu,\tau$ and $i=1,2,3$)
or $\theta_x = \theta_y = 45^\circ$ and $\theta_z \approx 35.3^\circ$,
will not be taken into account. Such a ``tri-maximal'' mixing pattern
has completely been ruled out by current experimental data.}.
%%%%%%%%%%%%%%%%%%%%%

The first example is the so-called ``democratic'' lepton flavor mixing
pattern, 
\begin{equation}
V \; = \; \left ( \matrix{
\frac{\sqrt{2}}{2}	& \frac{\sqrt{2}}{2}	& 0 \cr\cr
- \frac{\sqrt{6}}{6}	& \frac{\sqrt{6}}{6}	& \frac{\sqrt{6}}{3} \cr\cr
\frac{\sqrt{3}}{3}	& - \frac{\sqrt{3}}{3}	& \frac{\sqrt{3}}{3} \cr}
\right ) \; ,
%	(3.30)
\end{equation}
which was originally obtained by Fritzsch and Xing \cite{FX96}
from the breaking of flavor democracy of the charged lepton mass matrix
in the flavor basis where the neutrino mass matrix is diagonal.
The predictions of (3.30) for the mixing factors of solar and atmospheric 
neutrino oscillations, $\sin^2 2\theta_{\rm sun} = 1$ and 
$\sin^2 2\theta_{\rm atm} = 8/9$, are not favored by today's 
experimental data. Hence slight modifications of (3.30) have to be
made. In Ref. \cite{FX96} and Ref. \cite{Democracy}, 
some proper perturbative terms have
been introduced to break the $\rm S(3)_L \times S(3)_R$ symmetry
of the charged lepton mass matrix and the S(3) symmetry of the neutrino
mass matrix. The resultant flavor mixing matrix, whose leading term
remains to take the form of (3.30), is able to fit current solar,
atmospheric and CHOOZ reactor neutrino oscillation data very well.

The second example is the so-called ``bi-maximal'' 
neutrino mixing pattern, 
\begin{equation}
V \; = \; \left ( \matrix{
\frac{\sqrt{2}}{2}	& \frac{\sqrt{2}}{2}	& 0 \cr\cr
- \frac{1}{2}	& \frac{1}{2}	& \frac{\sqrt{2}}{2} \cr\cr
\frac{1}{2}	& - \frac{1}{2}	& \frac{\sqrt{2}}{2} \cr}
\right ) \; ,
%	(3.31)
\end{equation}
 proposed by Barger {\it et al} \cite{Barger} and by
Vissani \cite{Vissani}.
It predicts $\sin^2 2\theta_{\rm sun} = \sin^2 2\theta_{\rm atm} =1$ 
for solar and atmospheric neutrino oscillations. As the result
$\sin^2 2\theta_{\rm sun} = 1$ is disfavored by the LMA solution
to the solar neutrino problem, slight modifications of (3.31)  
are also required. A number of possibilities to modify the bi-maximal
mixing pattern into nearly bi-maximal mixing patterns have been
discussed in the literature \cite{BiMax}. 
The essential point is to introduce 
a perturbative term to (3.31), which may naturally be related to
the mass ratio(s) of charged leptons or to the Cabibbo angle of
quark flavor mixing, such that $\sin^2 2\theta_{\rm sun}$ deviates
from unity to a proper extent.
\setcounter{table}{0}
%%%%%%%%%%%%%%%%%%%%% Table 3.1 %%%%%%%%%%%%%%%%%%%%%%%
\begin{table}[t]
\caption{The mixing angles $(\theta_x, \theta_y, \theta_z)$
in four constant lepton flavor mixing patterns.}
\vspace{0.2cm}
\begin{center}
\begin{tabular}{c|c|c|c|c}\hline\hline 
%----------------------------------------------------------------------
Mixing Angle  & Pattern (3.30)	& Pattern (3.31)
& Pattern (3.32)  & Pattern (3.33) \\ \hline 
%---------------------------------------------------------------------
$\theta_x$ & $45^\circ$ & $45^\circ$ & $35.3^\circ$ & $30^\circ$ \\ 
%---------------------------------------------------
$\theta_y$ & $54.7^\circ$ & $45^\circ$ & $45^\circ$ & $45^\circ$ \\ 
%---------------------------------------------------
$\theta_z$ & $0^\circ$ & $0^\circ$ & $0^\circ$ & $0^\circ$ \\ \hline \hline
%----------------------------------------------------------------------
\end{tabular}
\end{center}
\vspace{-0.3cm}
\end{table}
%%%%%%%%%%%%%%%%%%%%%%%%%%%%%%%%%%%%%%%%%%%%%%%%%

Another interesting neutrino mixing pattern is
\begin{equation}
V \; = \; \left ( \matrix{
\frac{\sqrt 6}{3}	& \frac{\sqrt{3}}{3}	& 0 \cr\cr
- \frac{\sqrt{6}}{6}	& \frac{\sqrt{3}}{3}	& \frac{\sqrt{2}}{2} \cr\cr
\frac{\sqrt{6}}{6}	& - \frac{\sqrt{3}}{3}	& \frac{\sqrt{2}}{2} \cr}
\right ) \; .
%	(3.32)
\end{equation}
It was first conjectured by Wolfenstein long time ago \cite{Wolfenstein78}
%%%%%%%%%%%%%%%%%%%%%%%%
\footnote{The first and second columns of $V$ were interchanged 
in the original paper of Wolfenstein \cite{Wolfenstein78}. 
Such an interchange is allowed by 
current neutrino oscillation data, because $|V_{e1}|$ and $|V_{e2}|$
cannot separately be determined from the measurement of
$\sin^2 2\theta_{\rm sun} = 4 |V_{e1}|^2 |V_{e2}|^2$. In (3.14) and
(3.32), we have taken a common convention to set $|V_{e1}| \geq |V_{e2}|$.}.
%%%%%%%%%%%%%%%%%%%%%%%%
Possible (broken) flavor symmetries associated with (3.32) have recently 
been discussed by several authors \cite{TB}. 
Two straightforward consequences of (3.32) on
neutrino oscillations are $\sin^2 2\theta_{\rm sun} = 8/9$ and 
$\sin^2 2\theta_{\rm atm} = 1$, favored by the present solar and atmospheric 
neutrino oscillation data. Of course, nonvanishing $V_{e3}$ (or $\theta_z$) 
and CP-violating phases can be introduced into (3.32), if proper 
perturbations to the charged lepton and neutrino mass matrices are taken 
into account.

The last example for constant lepton flavor mixing is given as
\begin{equation}
V \; = \; \left ( \matrix{
\frac{\sqrt 3}{2}	& \frac{1}{2}	& 0 \cr\cr
- \frac{\sqrt{2}}{4}	& \frac{\sqrt 6}{4}	& \frac{\sqrt{2}}{2} \cr\cr
\frac{\sqrt{2}}{4}	& - \frac{\sqrt 6}{4}	& \frac{\sqrt{2}}{2} \cr}
\right ) \; ,
%	(3.33)
\end{equation}
which has recently been discussed by Giunti \cite{Giunti02},
Peccei \cite{Peccei} and Xing \cite{Xing02W}. It predicts
$\sin^2 2\theta_{\rm sun} = 3/4$ and $\sin^2 2\theta_{\rm atm} = 1$, 
which are in good agreement with the present solar and atmospheric 
neutrino oscillation data. Again slight modifications of (3.33) can be 
done \cite{Xing02W}, in order to accommodate nonvanishing $V_{e3}$ 
and leptonic CP violation.

We summarize the values of three flavor mixing angles for each of
the four constant patterns in Table 3.1. The similarity and difference 
between any two patterns can easily be seen. It is worth remarking
that a specific constant mixing pattern should be regarded 
as the leading-order approximation of the ``true'' lepton flavor mixing
matrix, whose mixing angles depend in general on both the ratios 
of charged lepton masses and those of neutrino masses. Naively, one
may make the following speculation:
\begin{itemize}
\item     Large values of $\theta_x$ and $\theta_y$ could arise from 
a weak hierarchy or a near degeneracy of the neutrino mass spectrum,
as the strong hierarchy of charged lepton masses implies that 
$m_e/m_\mu$ and $m_\mu/m_\tau$ are unlikely to contribute to 
$\theta_x$ and $\theta_y$ dominantly.
\item     Special values of $\theta_x$ and $\theta_y$ might stem from 
a discrete flavor symmetry of the charged lepton mass matrix or the
neutrino mass matrix. Then the contributions of lepton mass 
ratios to flavor mixing angles, due to flavor symmetry breaking, 
are only at the perturbative level.
\item     Vanishing or small $\theta_z$ could be a natural 
consequence of the explicit textures of lepton mass matrices. It might
also be related to the flavor symmetry which gives rise to sizable
$\theta_x$ and $\theta_y$.
\item	 Small perturbative effects on a constant flavor mixing 
pattern can also result from the renormalization-group equations of
leptons and quarks, e.g., from the high energy scales to the low energy 
scales or vice versa \cite{RGE}.
\end{itemize}
To be more specific, we refer the reader to 
Ref. \cite{FXReview} and Ref. \cite{Review}, 
from which a lot of theoretical models and phenomenological 
ans$\rm\ddot{a}$tze can be found for the interpretation of possible
spectra of lepton masses and possible patterns of lepton flavor mixing.

\subsection{The Flavor Mixing Matrix in Matter}

The effective Hamiltonian responsible for the propagation of neutrinos 
in vacuum can be written as ${\cal H}_{\rm eff} = \Omega_\nu/(2E)$ 
in the basis where the flavor eigenstates of charged leptons are 
identified with their mass eigensates. In ${\cal H}_{\rm eff}$, 
$E\gg m^{~}_i$ denotes the neutrino beam energy, and $\Omega_\nu$ is 
given by
\begin{equation}
\Omega_\nu \; =\; V \left (\matrix{
m^2_1 & 0 & 0 \cr
0 & m^2_2 & 0 \cr
0 & 0 & m^2_3 \cr} \right ) V^\dagger \; ,
%       (3.34)
\end{equation}
in which $V$ is the $3\times 3$ lepton flavor mixing matrix.
Taking account of (3.1) or (3.2), we obtain 
$\Omega_\nu = M_\nu M^\dagger_\nu$ \cite{Xing00a}, no matter whether 
massive neutrinos are Dirac or Majorana particles. A simple relation 
can therefore be established between the effective Hamiltonian 
${\cal H}_{\rm eff}$ and the neutrino mass matrix $M_\nu$.
The form of ${\cal H}_{\rm eff}$ has to be modified, however, 
in order to describe the propagation of neutrinos in matter.

If neutrinos travel through a normal material medium (e.g., the earth), 
which consists of electrons but of no muons or taus, they encounter
both charged- and neutral-current interactions with electrons. 
In this case, the effective Hamiltonian responsible for the 
propagation of neutrinos takes the form
$\tilde{\cal H}_{\rm eff} = \tilde{\Omega}_\nu/(2E)$, where \cite{MSW}
\begin{equation}
\tilde{\Omega}_\nu \; = \; \Omega_\nu ~ + ~ \left (\matrix{
A & 0 & 0 \cr
0 & 0 & 0 \cr
0 & 0 & 0 \cr} \right ) \; .
%       (3.35)
\end{equation}
The apparent deviation of $\tilde{\Omega}_\nu$ from $\Omega_\nu$ 
arises from the matter-induced effect; namely, 
$A = 2\sqrt{2} ~ G_{\rm F} N_e E$ describes the charged-current 
contribution to the $\nu_e e^-$ forward scattering, where
$N_e$ is the background density of electrons and $E$ stands for
the neutrino beam energy. The neutral-current contributions, which 
are universal for $\nu_e$, $\nu_\mu$ and $\nu_\tau$ neutrinos and
can only result in an overall unobservable phase, have been omitted.

Let us concentrate on the long-baseline neutrino oscillation 
experiments, which are designed to measure the parameters of lepton
flavor mixing and CP violation. We shall assume a constant earth 
density profile (i.e., $N_e$ = constant) in section 3.2.1 and 
section 3.2.3. Such an assumption is rather plausible and close to 
reality, only if the neutrino beam does not go through the earth's 
core \cite{Shrock}.

In the chosen flavor basis where $M_l$ is diagonal, 
$\tilde{\Omega}_\nu$ can be diagonalized by a unitary 
transformation:
\begin{equation}
\tilde{\Omega}_\nu = \tilde{V} \left (\matrix{
\tilde{m}^2_1 & 0 & 0 \cr
0 & \tilde{m}^2_2 & 0 \cr
0 & 0 & \tilde{m}^2_3 \cr} \right ) \tilde{V}^\dagger \; ,
%       (3.36)
\end{equation}
where $\tilde{m}_i$ are the {\it effective} neutrino masses
in matter. The unitary matrix $\tilde V$ is just the
matter-modified lepton flavor mixing matrix. Denoting the
effective neutrino mass matrix in matter as $\tilde{M}_\nu$,
one may get $\tilde{\Omega}_\nu = \tilde{M}_\nu \tilde{M}^\dagger_\nu$
similar to $\Omega_\nu = M_\nu M^\dagger_\nu$.
Phenomenologically, it is important to find the analytical relations 
between $\tilde{m}_i$ and $m_i$ as well as the exact relation between
$\tilde V$ and $V$. 

\subsubsection{Generic Analytical Formulas}

Combining (3.34), (3.35) and (3.36), we calculate $\tilde{m}_i$ in terms
of $m_i$ and the matrix elements of $V$. The results are
\begin{eqnarray}
\tilde{m}^2_1 & = & m^2_1 + \frac{1}{3} x - \frac{1}{3} \sqrt{x^2 - 3y}
\left [z + \sqrt{3 \left (1-z^2 \right )} \right ] \; , 
\nonumber \\ 
\tilde{m}^2_2 & = & m^2_1 + \frac{1}{3} x - \frac{1}{3} \sqrt{x^2 -3y}
\left [z - \sqrt{3 \left (1-z^2 \right )} \right ] \; , 
\nonumber \\ 
\tilde{m}^2_3 & = & m^2_1 + \frac{1}{3} x + \frac{2}{3} 
z \sqrt{x^2 - 3y} \;\; ,
%       (3.37)
\end{eqnarray}
where $x$, $y$ and $z$ are given by \cite{Barger80}
\begin{eqnarray}
x & = & \Delta m^2_{21} + \Delta m^2_{31} + A \; , 
\nonumber \\
y & = & \Delta m^2_{21} \Delta m^2_{31} + A \left [ 
\Delta m^2_{21} \left ( 1 - |V_{e2}|^2 \right ) 
+ \Delta m^2_{31} \left ( 1 - |V_{e3}|^2 \right ) \right ] \; , 
\nonumber \\
z & = & \cos \left [ \frac{1}{3} \arccos \frac{2x^3 -9xy + 27
A \Delta m^2_{21} \Delta m^2_{31} |V_{e1}|^2}
{2 \left (x^2 - 3y \right )^{3/2}} \right ] 
%       (3.38)
\end{eqnarray}
with $\Delta m^2_{21} \equiv m^2_2 - m^2_1$ and
$\Delta m^2_{31} \equiv m^2_3 - m^2_1$. 
Of course, $\tilde{m}^2_i = m^2_i$ can be reproduced from
(3.37) when $A =0$ is taken. Only the mass-squared differences 
$\Delta \tilde{m}^2_{21} \equiv \tilde{m}^2_2 - \tilde{m}^2_1$ 
and $\Delta \tilde{m}^2_{31} \equiv \tilde{m}^2_3 - \tilde{m}^2_1$,
which depend on $\Delta m^2_{21}$, $\Delta m^2_{31}$, $A$ and
$|V_{ei}|^2$ (for $i=1,2,3$), are
relevant to the practical neutrino oscillations in matter.

It is easy to prove that the sum of $\tilde{m}^2_i$ is related to
that of $m^2_i$ through
\begin{equation}
\sum^3_{i=1} \tilde{m}^2_i \; = \; \sum^3_{i=1} m^2_i ~ + ~ A \; .
%       (3.39)
\end{equation}
Also $m^2_i$, $|V_{ei}|^2$, $A$ and $\tilde{m}^2_i$ are correlated 
with one another through another instructive equation:
\begin{equation}
A \left (m^2_i - m^2_k \right ) \left (m^2_j - m^2_k \right )
|V_{ek}|^2 \; =\; \prod^3_{n=1} \left (\tilde{m}^2_n - m^2_k \right ) \; ,
%	(3.40)
\end{equation}
where $(i, j, k)$ run over $(1, 2, 3)$ with $i\neq j \neq k$. 
Assuming $|\Delta m^2_{31}| \gg |\Delta m^2_{21}|$ and taking
$k=3$, one can use (3.40) to derive the approximate analytical 
result for the correlation between $\tilde{m}^2_i$ and $m^2_i$ 
obtained in Ref. \cite{Pantaleone}.

The analytical relationship between the matrix elements of $\tilde V$ 
and those of $V$ can also be derived from (3.34), (3.35) and (3.36). 
After some lengthy but straightforward calculations \cite{Xing00b}, 
we arrive at 
\begin{eqnarray}
\tilde{V}_{\alpha i} \; = \; \frac{N_i}{D_i} V_{\alpha i} 
~ + ~ \frac{A}{D_i} V_{e i} \left [ \left (\tilde{m}^2_i - m^2_j \right )
V^*_{e k} V_{\alpha k} + \left (\tilde{m}^2_i - m^2_k \right )
V^*_{e j} V_{\alpha j} \right ] \; ,
%        (3.41)
\end{eqnarray}
where $\alpha$ runs over $(e, \mu, \tau)$ and $(i, j, k)$ run over 
$(1, 2, 3)$ with $i \neq j \neq k$, and
\begin{eqnarray}
N_i & = & \left (\tilde{m}^2_i - m^2_j \right ) 
\left (\tilde{m}^2_i - m^2_k \right )
- A \left [\left (\tilde{m}^2_i - m^2_j \right ) |V_{e k}|^2  
+ \left (\tilde{m}^2_i - m^2_k \right ) |V_{e j}|^2 \right ] \; ,
\nonumber \\
D^2_i & = & N^2_i + A^2 |V_{e i}|^2 \left [ 
\left (\tilde{m}^2_i - m^2_j \right )^2 |V_{e k}|^2  
+ \left (\tilde{m}^2_i - m^2_k \right )^2 |V_{e j}|^2 \right ] \; . 
%        (3.42)
\end{eqnarray}
Obviously, $A=0$ leads to $\tilde{V}_{\alpha i} = V_{\alpha i}$. 
This exact and compact formula shows clearly how the flavor mixing 
matrix in vacuum is corrected by the matter 
effects. Instructive analytical approximations can be made for (3.41),
once the spectrum of neutrino masses is experimentally known or
theoretically assumed.

The results obtained above are valid for neutrinos interacting with 
matter. As for antineutrinos, the corresponding formulas for the 
effective neutrino masses and flavor mixing matrix elements in matter 
can straightforwardly be obtained from (3.37)--(3.42) through the 
replacements $V \Longrightarrow V^*$ and $A \Longrightarrow -A$.

For illustration, we carry out a numerical analysis of the terrestrial
matter effects on lepton flavor mixing and CP violation for a
low-energy (100 MeV $\leq E \leq$ 1 GeV) and medium-baseline (100 km 
$\leq L \leq$ 400 km) neutrino oscillation experiment \cite{Xing00a}.
In view of current data, we typically take 
$\Delta m^2_{21} = 5\times 10^{-5} ~ {\rm eV}^2$ and
$\Delta m^2_{31} = 3\times 10^{-3} ~ {\rm eV}^2$ as well as 
$\theta_x \approx 35^\circ$, $\theta_y \approx 40^\circ$, 
$\theta_z \approx 5^\circ$ and $\delta \approx \pm 90^\circ$ in the 
standard parametrization of $V$. In addition, the dependence of the 
terrestrial matter effect on the neutrino beam energy is given as 
$A = 2.28\times 10^{-4} ~{\rm eV}^2 E$/[GeV] \cite{Shrock}.
With the help of (3.37) and (3.41), we are then able to compute the 
ratios $\Delta \tilde{m}^2_{i1}/\Delta m^2_{i1}$ (for $i=2,3$) and 
$|\tilde{V}_{\alpha i}|/|V_{\alpha i}|$ (for $\alpha = e, \mu, \tau$ 
and $i = 1, 2, 3$) as functions of $E$. The relevant numerical results
are shown respectively in Fig. 3.1 and Fig. 3.2. We see that
the earth-induced matter effect on $\Delta m^2_{21}$ is significant, 
but that on $\Delta m^2_{31}$ is negligibly small in the chosen range 
of $E$. The matter effects on $|V_{\mu 3}|$ and $|V_{\tau 3}|$ are 
negligible in the low-energy neutrino experiment. The smallest matrix 
element $|V_{e3}|$ is weakly sensitive to the matter effect.
In contrast, the other six matrix elements of $V$ are significantly 
modified by the terrestrial matter effects. 
%%%%%%%%%%%%%%%%%%%% Fig. 3.1 %%%%%%%%%%%%%%%%
\begin{figure}[t]
\vspace{-0.4cm}
\epsfig{file=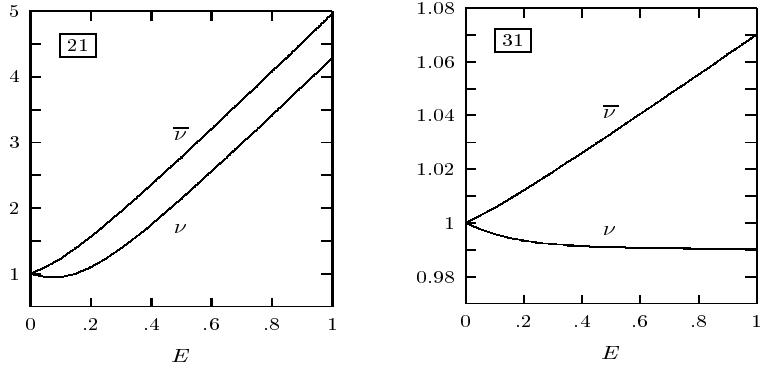,bbllx=1cm,bblly=0cm,bburx=19cm,bbury=28.1cm,%
width=15cm,height=19cm,angle=0,clip=}
\vspace{-13.5cm}
\caption{Ratios $\Delta \tilde{m}^2_{21}/\Delta m^2_{21}$
and $\Delta \tilde{m}^2_{31}/\Delta m^2_{31}$ changing with the 
beam energy $E$ (in unit of GeV) for neutrinos ($\nu$) and 
antineutrinos ($\overline{\nu}$),
in which $\Delta m^2_{21} = 5\times 10^{-5} ~ {\rm eV}^2$,
$\Delta m^2_{31} = 3\times 10^{-3} ~ {\rm eV}^2$,
$|V_{e1}| = 0.816$ and $|V_{e2}| = 0.571$ 
have typically been input.}
\end{figure}
%%%%%%%%%%%%%%%%%%%%%%%%%%%%%%%%%%%%%%%%%%%%
%%%%%%%%%%%%%%%%%%%% Fig. 3.2 %%%%%%%%%%%%%%%%
\begin{figure}[t]
\vspace{-3.3cm}
\epsfig{file=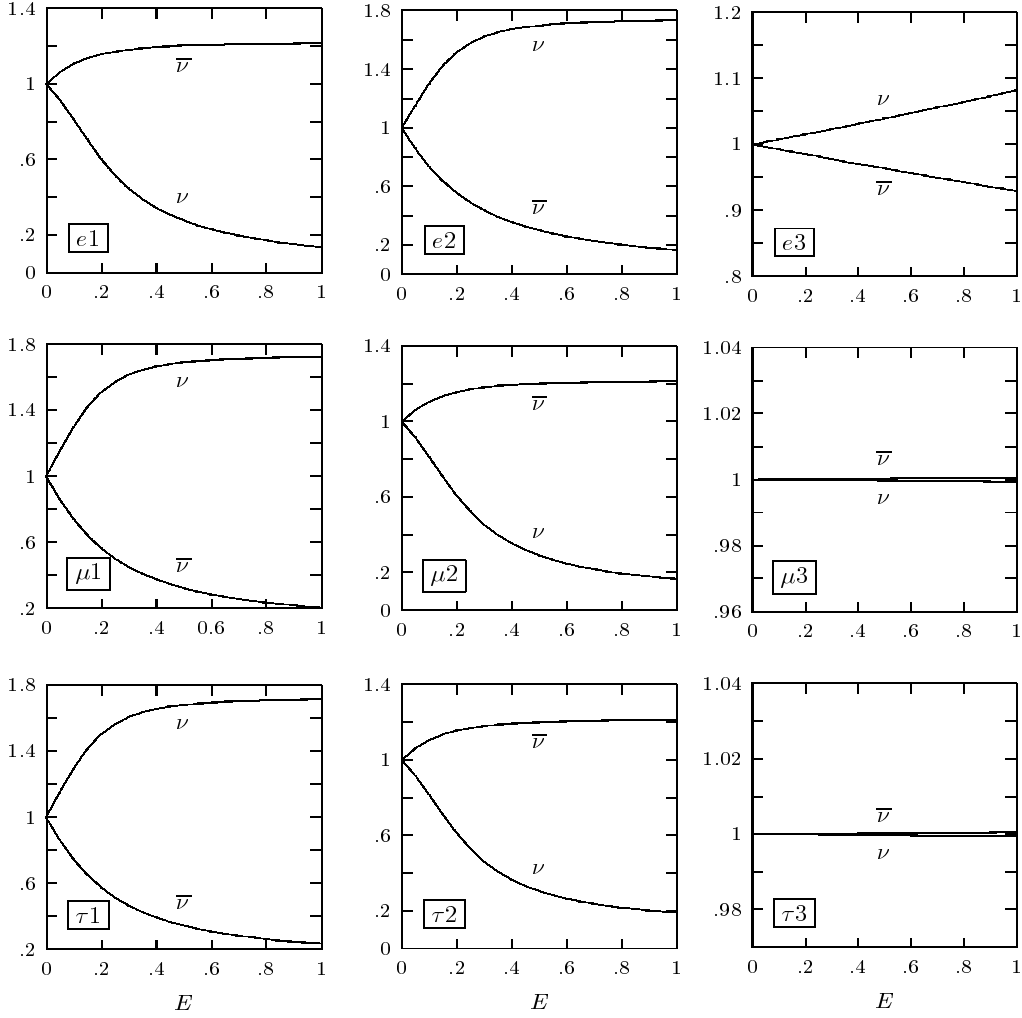,bbllx=1cm,bblly=6cm,bburx=19cm,bbury=31.5cm,%
width=15cm,height=19cm,angle=0,clip=}
\vspace{-2cm}
\caption{Ratios $|\tilde{V}_{\alpha i}|/|V_{\alpha i}|$ 
(for $\alpha = e, \mu, \tau$ and $i=1,2,3$)
changing with the beam energy $E$ (in unit of GeV) for neutrinos ($\nu$) 
and antineutrinos ($\overline{\nu}$),
in which $\Delta m^2_{21} = 5\times 10^{-5} ~ {\rm eV}^2$,
$\Delta m^2_{31} = 3\times 10^{-3} ~ {\rm eV}^2$,
$\theta_x \approx 35^\circ$, $\theta_y \approx 40^\circ$, 
$\theta_z \approx 5^\circ$ and $\delta \approx \pm 90^\circ$ 
have typically been input.}
\end{figure}
%%%%%%%%%%%%%%%%%%%%%%%%%%%%%%%%%%%%%%%%%%%%

\subsubsection{Commutators and Sum Rules}

If the charged lepton mass matrix $M_l$ and the neutrino mass
matrix $M_\nu$ could simultaneously be diagonalized in a given
flavor basis, there would be no lepton flavor mixing. A very instructive 
measure of the mismatch between the diagonalization
of $M_l$ and that of $M_\nu$ in vacuum should be the commutators of 
lepton mass matrices \cite{Xing00a}, defined as 
\begin{eqnarray}
\left [ M_l M^\dagger_l ~ , ~ M_\nu M^\dagger_\nu \right ] 
& \equiv & i X \; ,
\nonumber \\
\left [ M_l M^\dagger_l ~ , ~ M^\dagger_\nu M_\nu \right ] 
& \equiv & i X' \; ,
\nonumber \\
\left [ M^\dagger_l M_l ~ , ~ M_\nu M^\dagger_\nu \right ] 
& \equiv & i Y \; ,
\nonumber \\
\left [ M^\dagger_l M_l ~ , ~ M^\dagger_\nu M_\nu \right ] 
& \equiv & i Y' \; .
%        (3.43)
\end{eqnarray}
As for neutrinos propagating in matter, the corresponding commutators
of lepton mass matrices are defined as
\begin{eqnarray}
\left [ M_l M^\dagger_l ~ , ~ \tilde{M}_\nu \tilde{M}^\dagger_\nu \right ] 
& \equiv & i \tilde{X} \; ,
\nonumber \\
\left [ M_l M^\dagger_l ~ , ~ \tilde{M}^\dagger_\nu \tilde{M}_\nu \right ] 
& \equiv & i \tilde{X}' \; ,
\nonumber \\
\left [ M^\dagger_l M_l ~ , ~ \tilde{M}_\nu \tilde{M}^\dagger_\nu \right ] 
& \equiv & i \tilde{Y} \; ,
\nonumber \\
\left [ M^\dagger_l M_l ~ , ~ \tilde{M}^\dagger_\nu \tilde{M}_\nu \right ] 
& \equiv & i \tilde{Y}' \; .
%        (3.44)
\end{eqnarray} 
It is obvious that these commutators must
be traceless Hermitian matrices. Without loss of generality, one may
again choose to identify the flavor eigenstates of charged leptons 
with their mass eigenstates. In this specific basis, where $M_l$ takes
the diagonal form $D_l \equiv {\rm Diag} \{ m_e, m_\mu, m_\tau \}$,
we obtain $X = Y$, $X' = Y'$ and
$\tilde{X} = \tilde{Y}$, $\tilde{X}' = \tilde{Y}'$.

Taking account of (3.1), (3.2), (3.34) and (3.36), we obtain 
\begin{eqnarray}
M_\nu M^\dagger_\nu & = & V D^2_\nu V^\dagger \; = \; \Omega_\nu \; ,
\nonumber \\
\tilde{M}_\nu \tilde{M}^\dagger_\nu & = & \tilde{V} \tilde{D}^2_\nu 
\tilde{V}^\dagger \; = \; \tilde{\Omega}_\nu  \; 
%	(3.45)
\end{eqnarray}
in the chosen flavor basis, 
where $D_\nu \equiv {\rm Diag} \{ m_1, m_2, m_3 \}$ and
$\tilde{D}_\nu\equiv {\rm Diag} \{ \tilde{m}_1,\tilde{m}_2,\tilde{m}_3 \}$.
In contrast, there is no direct relationship between $M^\dagger_\nu M_\nu$ 
(or $\tilde{M}^\dagger_\nu \tilde{M}_\nu$) and $\Omega_\nu$ (or
$\tilde{\Omega}_\nu$). Hence we are only interested in the commutators 
$X$ and $\tilde{X}$, whose explicit expressions read as follows:
\begin{eqnarray}
X & = & i \left ( \matrix{
0 & \Delta_{\mu e} Z_{e \mu} & \Delta_{\tau e} Z_{e \tau} \cr
\Delta_{e \mu} Z_{\mu e} & 0 & \Delta_{\tau \mu} Z_{\mu \tau} \cr
\Delta_{e \tau} Z_{\tau e} & 
\Delta_{\mu \tau} Z_{\tau \mu} & 0 \cr} \right ) \; ,
\nonumber \\ \nonumber \\
\tilde{X} & = & i \left ( \matrix{
0 & \Delta_{\mu e} \tilde{Z}_{e \mu} & \Delta_{\tau e} \tilde{Z}_{e \tau} \cr
\Delta_{e \mu} \tilde{Z}_{\mu e} & 0 & 
\Delta_{\tau \mu} \tilde{Z}_{\mu \tau} \cr
\Delta_{e \tau} \tilde{Z}_{\tau e} 
& \Delta_{\mu \tau} \tilde{Z}_{\tau \mu} & 0 \cr} \right ) \; ,
%	(3.46)
\end{eqnarray}
where $\Delta_{\alpha \beta} \equiv m^2_\alpha - m^2_\beta$ for
$\alpha \neq \beta$ running over $(e, \mu, \tau)$, and
\begin{eqnarray}
Z_{\alpha \beta} & \equiv & \sum^3_{i=1} \left ( m^2_i V_{\alpha i} 
V^*_{\beta i} \right ) \; ,
\nonumber \\
\tilde{Z}_{\alpha \beta} & \equiv & \sum^3_{i=1} \left ( \tilde{m}^2_i
\tilde{V}_{\alpha i} \tilde{V}^*_{\beta i} \right ) \; .
%	(3.47)
\end{eqnarray}
One can see that $\Delta_{\beta \alpha} = - \Delta_{\alpha \beta}$,
$Z_{\beta \alpha} = Z^*_{\alpha \beta}$ and
$\tilde{Z}_{\beta \alpha} = \tilde{Z}^*_{\alpha \beta}$ hold. To find out 
how $\tilde{Z}_{\alpha\beta}$ is related to $Z_{\alpha\beta}$,
we need to establish the relation between $X$ and $\tilde{X}$. 
With the help of (3.35), it is easy to prove that $X$ and $\tilde{X}$ 
are identical to each other:
\begin{equation}
\tilde{X} \; = \; i \left [\tilde{\Omega}_\nu ~ , ~ D^2_l \right ] 
\; = \; i \left [\Omega_\nu ~ , ~ D^2_l \right ] \; = \; X \; .
%	(3.48)
\end{equation}
This interesting result, whose validity is independent of the specific 
flavor basis chosen above, implies that the commutator of lepton mass
matrices defined in vacuum is invariant under matter effects. 
As a straightforward consequence of $\tilde{X} = X$, we arrive at
$\tilde{Z}_{\alpha \beta} = Z_{\alpha \beta}$ from (3.46). Then a set 
of concise sum rules of neutrino masses emerge \cite{Xing01}:
\begin{equation}
\sum^3_{i=1} \left ( \tilde{m}^2_i \tilde{V}_{\alpha i} 
\tilde{V}^*_{\beta i} \right ) \; = \; 
\sum^3_{i=1} \left ( m^2_i V_{\alpha i} V^*_{\beta i} \right ) \; .
%	(3.49)
\end{equation}
Although we have derived these sum rules in the three-neutrino mixing 
scheme, they may simply be generalized to hold for an arbitrary 
number of neutrino families. In addition, we emphasize that the useful
results obtained above are independent of the profile of matter density
(i.e., they are valid everywhere in the sun or in the earth).

It should be noted that $Z_{\alpha \beta}$ and $\tilde{Z}_{\alpha \beta}$
are sensitive to a redefinition of the phases of charged lepton fields. 
The simplest rephasing-invariant equality is 
$|\tilde{Z}_{\alpha \beta}| = |Z_{\alpha \beta}|$, where
$(\alpha, \beta)$ may run over $(e, \mu)$, $(\mu, \tau)$ or $(\tau, e)$. 
Another useful rephasing-invariant relationship is
\begin{equation}
Z_{e \mu} Z_{\mu \tau} Z_{\tau e} \; =\; 
\tilde{Z}_{e \mu} \tilde{Z}_{\mu \tau} \tilde{Z}_{\tau e} \; .
%	(3.50)
\end{equation}
As one can see in section 4.1, 
the imaginary parts of $Z_{e \mu} Z_{\mu \tau} Z_{\tau e}$ and
$\tilde{Z}_{e \mu} \tilde{Z}_{\mu \tau} \tilde{Z}_{\tau e}$ are related 
respectively to leptonic CP violation in vacuum and that in matter.

It should also be noted that three additional sum rules can similarly 
be derived from (3.35). We find that $\tilde{\Omega}^{\alpha\beta}_\nu
= \Omega^{\alpha\beta}_\nu$ holds, if and only if 
$(\alpha, \beta) \neq (e,e)$. Thus the following equalities can be
obtained:
\begin{eqnarray}
\tilde{\Omega}^{e\tau}_\nu \tilde{\Omega}^{\tau\mu}_\nu -
\tilde{\Omega}^{e\mu}_\nu \tilde{\Omega}^{\tau\tau}_\nu & = &
\Omega^{e\tau}_\nu \Omega^{\tau\mu}_\nu -
\Omega^{e\mu}_\nu \Omega^{\tau\tau}_\nu \; ,
\nonumber \\
\tilde{\Omega}^{e\mu}_\nu \tilde{\Omega}^{\mu\tau}_\nu -
\tilde{\Omega}^{e\tau}_\nu \tilde{\Omega}^{\mu\mu}_\nu & = &
\Omega^{e\mu}_\nu \Omega^{\mu\tau}_\nu -
\Omega^{e\tau}_\nu \Omega^{\mu\mu}_\nu \; ,
\nonumber \\
\tilde{\Omega}^{\mu e}_\nu \tilde{\Omega}^{e\tau}_\nu -
\tilde{\Omega}^{\mu\tau}_\nu \tilde{\Omega}^{ee}_\nu & = &
\Omega^{\mu e}_\nu \Omega^{e\tau}_\nu -
\Omega^{\mu\tau}_\nu \left (\Omega^{ee}_\nu + A \right ) \; .
%       (3.51)
\end{eqnarray}
By use of (3.51) and the unitarity of $V$, one may arrive at
\begin{eqnarray}
\sum_{i\neq j \neq k}\left (\tilde{m}^2_i \tilde{m}^2_j
\tilde{V}_{ek} \tilde{V}^*_{\mu k} \right ) & = & 
\sum_{i\neq j \neq k}\left (m^2_i m^2_j
V_{ek} V^*_{\mu k} \right ) \; ,
\nonumber \\
\sum_{i\neq j \neq k}\left (\tilde{m}^2_i \tilde{m}^2_j
\tilde{V}_{ek} \tilde{V}^*_{\tau k} \right ) & = & 
\sum_{i\neq j \neq k}\left (m^2_i m^2_j
V_{ek} V^*_{\tau k} \right ) \; ,
\nonumber \\
\sum_{i\neq j \neq k}\left (\tilde{m}^2_i \tilde{m}^2_j
\tilde{V}_{\mu k} \tilde{V}^*_{\tau k} \right ) & = & 
\sum_{i\neq j \neq k}\left [\left (m^2_i m^2_j - Am^2_k \right )
V_{\mu k} V^*_{\tau k} \right ] \; .
%       (3.52)
\end{eqnarray}
These relations, similar to those obtained previously in Ref. \cite{Kimura}, 
imply that $\tilde{V}_{\alpha i} \tilde{V}^*_{\beta i}$ (for
$\alpha \neq \beta$) can be expressed in terms of 
$V_{\alpha i} V^*_{\beta i}$, $\Delta m^2_{ij}$ and $A$ \cite{Ohlsson}. 
In particular, we are able to get the analytical formula for the sides 
of three leptonic unitarity triangles in matter (see section 4.2 for
the detail).

Let us remark that the results obtained above are only valid for
neutrinos propagating in vacuum and in matter. As for antineutrinos, 
the corresponding formulas can straightforwardly be written out from 
(3.46)--(3.52) through the replacements $V\Longrightarrow V^*$ and
$A \Longrightarrow -A$.

\subsubsection{Effective Mixing Parameters}

Although the present non-accelerator neutrino experiments have yielded
some impressive constraints on three lepton flavor mixing angles 
($\theta_x$, $\theta_y$, and $\theta_z$), a precise determination
of them and a measurement of the Dirac-type CP-violating phase $\delta$ 
have to rely on a new generation of accelerator experiments with 
very long baselines, including the possible neutrino factories. In
such long- or medium-baseline neutrino experiments the terrestrial 
matter effects, which may deform the oscillating behaviors of neutrinos 
in vacuum and even fake the genuine CP-violating signals, must be taken 
into account. It is therefore desirable to figure out how the standard 
parametrization in (3.23) will be modified by the matter effects, in 
particular in the case of a constant earth density profile.

The analytical relationship between the elements of the effective
flavor mixing matrix $\tilde V$ in matter and those of the fundamental
flavor mixing matrix $V$ in vacuum has been given in (3.41).
Adopting the parametrization of $V$ in (3.23), we explicitly obtain
\begin{equation}
\tilde{V}_{\alpha i} \; =\; \frac{N_i}{D_i} V_{\alpha i} ~ + ~
\frac{A}{D_i} \sum_k \left (T_{\alpha k} P_{ki} \right ) \; ,
%	(3.53)
\end{equation}
where the expressions of $N_i$ and $D_i$ have been given in (3.42), 
$A$ denotes the matter parameter, 
$P \equiv {\rm Diag} \{e^{i\rho}, e^{i\sigma}, 1 \}$ is
the diagonal matrix of Majorana phases, and the matter-associated
matrix elements $T_{\alpha i}$ read as follows \cite{X01}
%%%%%%%%%%%%%%%%%%%%%%%%%%%%%%
\footnote{Note that there is a typing error in Eq. (18) of 
Ref. \cite{X01}, associated with the sign of 
$\Delta \tilde{m}^2_{21} \tilde{s}_1 \tilde{c}_1 \tilde{s}_2
e^{-i\tilde{\delta}}$ in the expression of $\tilde{T}_{\tau 3}$.}:
%%%%%%%%%%%%%%%%%%%%%%%%%%%%%%
\begin{eqnarray}
T_{e1} & = & +c_x c_z \left [ \left ( \tilde{m}^2_1 - m^2_2 \right ) 
s^2_z + \left ( \tilde{m}^2_1 - m^2_3 \right ) s^2_x c^2_z \right ] \; ,
\nonumber \\
T_{e2} & = & +s_x c_z \left [ \left ( \tilde{m}^2_2 - m^2_1 \right ) 
s^2_z + \left ( \tilde{m}^2_2 - m^2_3 \right ) c^2_x c^2_z \right ] \; ,
\nonumber \\
T_{e3} & = & +s_z c^2_z \left [ \left ( \tilde{m}^2_3 - m^2_1 \right ) 
s^2_x + \left ( \tilde{m}^2_3 - m^2_2 \right ) c^2_x \right ] \; ,
\nonumber \\
T_{\mu 1} & = & +c_x c^2_z 
\left [ \left ( \tilde{m}^2_1 - m^2_2 \right ) s_y s_z
- \left ( \tilde{m}^2_1 - m^2_3 \right ) \left ( s^2_x s_y s_z
- s_x c_x c_y e^{-i\delta} \right ) \right ] \; ,
\nonumber \\
T_{\mu 2} & = & +s_x c^2_z 
\left [ \left ( \tilde{m}^2_2 - m^2_1 \right ) s_y s_z
- \left ( \tilde{m}^2_2 - m^2_3 \right ) \left ( c^2_x s_y s_z +
s_x c_x c_y e^{-i\delta} \right ) \right ] \; ,
\nonumber \\
T_{\mu 3} & = & -s_z c_z \left [ \left ( \tilde{m}^2_3 - m^2_1 \right )
s^2_x s_y s_z + \left ( \tilde{m}^2_3 - m^2_2 \right ) c^2_x s_y s_z 
- \Delta m^2_{21} s_x c_x c_y e^{-i\delta} \right ] \; ,
\nonumber \\
T_{\tau 1} & = & +c_x c^2_z 
\left [ \left ( \tilde{m}^2_1 - m^2_2 \right ) c_y s_z
- \left ( \tilde{m}^2_1 - m^2_3 \right ) \left ( s^2_x c_y s_z
+ s_x c_x s_y e^{-i\delta} \right ) \right ] \; ,
\nonumber \\
T_{\tau 2} & = & +s_x c^2_z 
\left [ \left ( \tilde{m}^2_2 - m^2_1 \right ) c_y s_z
- \left ( \tilde{m}^2_2 - m^2_3 \right ) \left ( c^2_x c_y s_z
- s_x c_x s_y e^{-i\delta} \right ) \right ] \; ,
\nonumber \\
T_{\tau 3} & = & - s_z c_z \left [ \left ( \tilde{m}^2_3 - m^2_1 \right )
s^2_x c_y s_z + \left ( \tilde{m}^2_3 - m^2_2 \right ) c^2_x c_y s_z
+ \Delta m^2_{21} s_x c_x s_y e^{-i\delta} \right ] \; . ~~~~~~~~~~
%     	(3.54)
\end{eqnarray}
Clearly $\tilde{V}_{\alpha i} = V_{\alpha i}$ holds in the case of $A =0$. 

The results given in (3.53) and (3.54) indicate that the diagonal 
Majorana phase matrix $P$ on the right-hand side of $V$ is not affected 
by the matter effect. As a natural consequence, the effective flavor 
mixing matrix $\tilde{V}$ in matter can be parametrized in the same form 
as $V$:
\begin{equation}
\tilde{V} \; = \; \left ( \matrix{
\tilde{c}_x \tilde{c}_z & 
\tilde{s}_x \tilde{c}_z &
\tilde{s}_z \cr
- \tilde{c}_x \tilde{s}_y \tilde{s}_z - 
\tilde{s}_x \tilde{c}_y e^{-i\tilde{\delta}} &
- \tilde{s}_x \tilde{s}_y \tilde{s}_z + 
\tilde{c}_x \tilde{c}_y e^{-i\tilde{\delta}} &
\tilde{s}_y \tilde{c}_z \cr 
- \tilde{c}_x \tilde{c}_y \tilde{s}_z + 
\tilde{s}_x \tilde{s}_y e^{-i\tilde{\delta}} & 
- \tilde{s}_x \tilde{c}_y \tilde{s}_z - 
\tilde{c}_x \tilde{s}_y e^{-i\tilde{\delta}} & 
\tilde{c}_y \tilde{c}_z \cr } \right ) 
\left ( \matrix{
e^{i\rho} & 0	& 0 \cr
0	& e^{i\sigma} & 0 \cr
0       & 0       & 1} \right ) \; 
%     	(3.55)
\end{equation}
with $\tilde{s}_x \equiv \sin\tilde{\theta}_x$ and
$\tilde{c}_x \equiv \cos\tilde{\theta}_x$, and so on.
It should be noted that the matrix elements 
$\tilde{V}_{\mu 3}$ and $\tilde{V}_{\tau 3}$ in (3.53) are complex and 
dependent on the Dirac-type phase $\delta$, as one can see from (3.54).
Hence a proper redefinition of the phases for muon and tau fields is
needed, in order to make $\tilde{V}_{\mu 3}$ and $\tilde{V}_{\tau 3}$
real in the course from (3.53) to (3.55). The instructive relations 
between the effective mixing angles
in matter ($\tilde{\theta}_x, \tilde{\theta}_y, \tilde{\theta}_z$) 
and the fundamental mixing angles in vacuum 
($\theta_x, \theta_y, \theta_z$) are found to be
%%%%%%%%%%%%%%%%%%
\footnote{Here we have only presented the next-to-leading order 
expression for $\tan\tilde{\theta}_y/\tan\theta_y$, because the exact 
result is rather complicated and less instructive. The former works to 
a high degree of accuracy, and it is identical to the exact result 
provided that $\theta_y = 45^\circ$ and $\delta = \pm 90^\circ$ hold 
(i.e., one obtains $\tilde{\theta}_y = \theta_y = 45^\circ$ in this
special but interesting case).}
%%%%%%%%%%%%%%%%%
\begin{eqnarray}
\frac{\tan\tilde{\theta}_x}{\tan\theta_x} & = &
\frac{\displaystyle
N_2 + A \left [ \left ( \tilde{m}^2_2 - m^2_1 \right ) s^2_z + 
\left ( \tilde{m}^2_2 - m^2_3 \right ) c^2_x c^2_z \right ] }
{\displaystyle
N_1 + A \left [ \left ( \tilde{m}^2_1 - m^2_2 \right ) s^2_z + 
\left ( \tilde{m}^2_1 - m^2_3 \right ) s^2_x c^2_z \right ] }
\cdot \frac{D_1}{D_2} \; ,
\nonumber \\
\frac{\tan\tilde{\theta}_y}{\tan\theta_y} & = &
1 + \frac{ A \Delta m^2_{21} s_x c_x s_z \cos\delta
/\left ( s_y c_y \right )}
{\displaystyle
N_3 - A \left [ \left ( \tilde{m}^2_3 - m^2_1 \right ) s^2_x +
\left ( \tilde{m}^2_3 - m^2_2 \right ) c^2_x \right ] s^2_z } \; ,
\nonumber \\
\frac{\sin\tilde{\theta}_z}{\sin\theta_z} & = &
\frac{N_3}{D_3} ~ + ~ \frac{A}{D_3} 
\left [ \left ( \tilde{m}^2_3 - m^2_1 \right )
s^2_x + \left ( \tilde{m}^2_3 - m^2_2 \right ) c^2_x \right ] c^2_z \; .
%	(3.56)
\end{eqnarray}
Once the relation between $\tilde{\theta}_y$ and $\theta_y$ is 
established, one can obtain the relation between the effective 
CP-violating phase $\tilde{\delta}$ in matter and the genuine 
CP-violating phase $\delta$ in vacuum by use of the Toshev 
identity \cite{Toshev},
\begin{equation}
\frac{\sin\tilde{\delta}}{\sin\delta} \; =\; 
\frac{\sin 2\theta_y}{\sin 2\tilde{\theta}_y} \; = \;
\frac{s_y}{\tilde{s}_y} \cdot \frac{c_y}{\tilde{c}_y} \;\; .
%	(3.57)
\end{equation}
The detailed proof of this elegant identity has been done in 
Ref. \cite{Freund}. We see that $\tilde{\delta} \approx \delta$ holds as 
a consequence of $\tilde{\theta}_y \approx \theta_y$ in the leading-order 
approximation. Hence both $\theta_y$ and $\delta$ are insensitive to the 
matter effects. Note that the results obtained above are only valid for 
neutrinos propagating in matter. As for antineutrinos, the corresponding 
expressions can straightforwardly be obtained from (3.53)--(3.57) 
through the replacements 
$\delta\Longrightarrow -\delta$ and $A \Longrightarrow -A$. Such 
formulas should be very useful for the purpose of recasting the 
fundamental parameters of lepton flavor mixing from the matter-corrected 
ones \cite{X01}, which can be extracted from a variety of long- and 
medium-baseline neutrino oscillation experiments in the near future.
%%%%%%%%%%%%%%%%%%%%% Fig. 3.3 %%%%%%%%%%%%%%%%
\begin{figure}[t]
\vspace{-2.1cm}
\epsfig{file=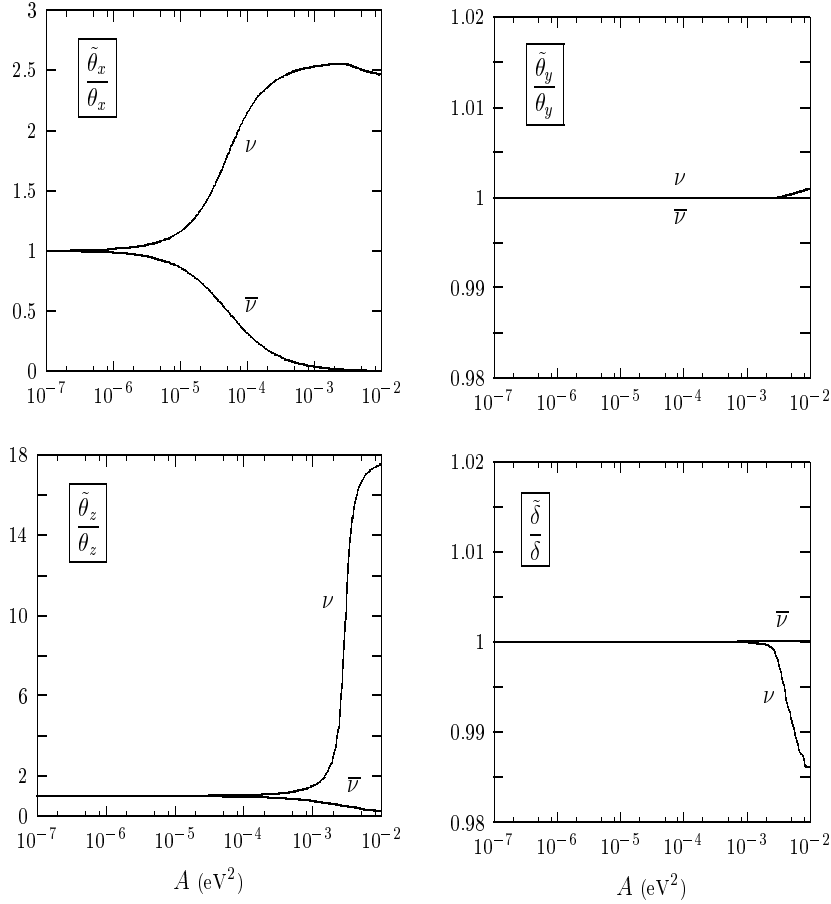,bbllx=-1.5cm,bblly=5.8cm,bburx=19cm,bbury=30.cm,%
width=15cm,height=20cm,angle=0,clip=}
\vspace{-5.8cm}
\caption{\small Ratios $\tilde{\theta}_x/\theta_x$, 
$\tilde{\theta}_y/\theta_y$, $\tilde{\theta}_z/\theta_z$, 
and $\tilde{\delta}/\delta$ changing with the matter parameter $A$
for neutrinos ($\nu$) and antineutrinos ($\overline{\nu}$),
in which $\Delta m^2_{21} = 5\times 10^{-5} ~ {\rm eV}^2$, 
$\Delta m^2_{31} = 3\times 10^{-3} ~ {\rm eV}^2$, 
$\theta_x \approx 35^\circ$, $\theta_y \approx 40^\circ$, 
$\theta_z \approx 5^\circ$ and $\delta \approx \pm 90^\circ$
have typically been input.}
\end{figure}
%%%%%%%%%%%%%%%%%%%%%%%%%%%%%%%%%%%%%%%%%%%%

To illustrate the dependence of $\tilde{\theta}_x$, $\tilde{\theta}_y$,
$\tilde{\theta}_z$, and $\tilde{\delta}$ on the matter effects, we 
typically take 
$\Delta m^2_{21} \approx \Delta m^2_{\rm sun} 
\approx 5\times 10^{-5} ~ {\rm eV}^2$ and 
$\Delta m^2_{31} \approx \Delta m^2_{\rm atm} 
\approx 3\times 10^{-3} ~ {\rm eV}^2$. In addition, 
we adopt $\theta_x \approx 35^\circ$, $\theta_y \approx 40^\circ$,
$\theta_z \approx 5^\circ$, and $\delta \approx \pm 90^\circ$. Such a
choice for the values of the input parameters is indeed favored by 
current experimental data on solar and atmospheric
neutrino oscillations. With the help of (3.56) and (3.57) as
well as the relevant formulas of $\tilde{m}^2_i$, $N_i$ and $D_i$ 
given in section 3.2.1, we are then able to calculate
the ratios $\tilde{\theta}_x/\theta_x$, $\tilde{\theta}_y/\theta_y$,
$\tilde{\theta}_z/\theta_z$, and $\tilde{\delta}/\delta$ as functions 
of the matter parameter $A$ in the typical region 
$10^{-7} ~ {\rm eV}^2 \leq A \leq 10^{-2} ~ {\rm eV}^2$, which 
corresponds to the neutrino beam energy $E$ in the region 
$5 ~{\rm MeV} \leq E \leq 50 ~{\rm GeV}$ for 
$N_e \approx 1.5 ~ {\rm g/cm^3}$ \cite{Earth}. The numerical results 
are shown in Fig. 3.3. We observe that the matter effects on $\theta_y$ 
and $\delta$ are negligibly small for 
$A\leq 5\times 10^{-2} ~ {\rm eV}^2$, just as indicated by our 
analytical results. In contrast, $\theta_z$ is 
significantly modified when $A > 10^{-3} ~ {\rm eV}^2$; and 
$\theta_x$ is sensitive to the matter effect even for quite small 
values of $A$. 

\subsection{Comparison with Quark Flavor Mixing}

We have seen that the $3\times 3$ lepton flavor mixing matrix $V$ is of
a {\it bi-large} mixing pattern, and most of its matrix elements do not
have a strong hierarchy in magnitude. In contrast, the $3\times 3$ quark 
flavor mixing matrix $V_{\rm CKM}$ \cite{CKM} is of a {\it tri-small} 
mixing pattern and has a strongly hierarchical structure. Hence 
$V_{\rm CKM}$ can be expanded in powers of a small parameter 
$\lambda \approx 0.22$, the so-called Wolfenstein 
parameter \cite{Wolfenstein83}:
\begin{equation}
V_{\rm CKM} \; \approx \; \left ( \matrix{
1 - \frac{1}{2} \lambda^2  & \lambda & A\lambda^3 (\rho - i\eta) \cr
- \lambda & 1 - \frac{1}{2} \lambda^2  & A\lambda^2 \cr
A\lambda^3 (1 - \rho - i\eta) & -A\lambda^2  & 1 \cr} \right ) \; ,
%        (3.58)
\end{equation}
where the terms of or below ${\cal O}(\lambda^4)$ have been 
neglected \cite{Higher}. Current experimental data yield 
$A \approx 0.83$, $\rho \approx 0.17$ and $\eta \approx 0.36$ \cite{PDG02}. 
The spirit of lepton-quark similarity motivates us to consider whether
the lepton flavor mixing matrix can also be expanded in powers of a
Wolfenstein-like parameter, whose magnitude may be somehow larger than 
$\lambda$. We find that it is indeed possible to expand $V$ in powers of 
$\Lambda \equiv |V_{\mu 3}| \sim 1/\sqrt{2}$ \cite{Xing02W}, in view of 
current neutrino oscillation data.

The parameter $\Lambda$ measures the strength of flavor mixing in 
atmospheric neutrino oscillations. The magnitude of flavor mixing in 
solar neutrino oscillations is then characterized by $A\Lambda^2$, where
$A$ is a positive coefficient of ${\cal O}(1)$. Because $|V_{e3}| < 0.16$ 
is required by the CHOOZ data \cite{CHOOZ}, we may take 
$|V_{e3}| \sim {\cal O}(\Lambda^8) \sim 0.06$ as a typical possibility for 
$\Lambda \sim 1/\sqrt{2} ~$. Smaller values of $|V_{e3}|$ are certainly
allowed. In some interesting models of lepton flavor 
mixing \cite{FXReview}, $|V_{e3}| \sim \sqrt{m_e/m_\mu} \sim 0.07$ 
is naturally predicted. Hence ${\cal O}(\Lambda^8)$ could be the plausible 
order of $|V_{e3}|$. Let us fix the matrix elements $V_{e2}$, $V_{e3}$ and
$V_{\mu 3}$ by use of four independent parameters: 
$V_{\mu 3} = \Lambda$, $V_{e2} = A\Lambda^2$ and 
$V_{e3} = B \Lambda^8 e^{-i\delta}$, where the positive coefficient $B$ 
is of ${\cal O}(1)$ or smaller and $\delta$ denotes the Dirac phase of 
leptonic CP violation. Then one may make use of the unitarity of $V$ to 
work out the exact analytical expressions for the other six matrix 
elements. We find that it is more instructive to approximate $V$ as
\footnotesize
\begin{equation}
V \; =\; \left ( \matrix{
\sqrt{1-A^2 \Lambda^4}  & A \Lambda^2  & B\Lambda^8 e^{-i\delta} \cr\cr
-A \Lambda^2 \sqrt{1-\Lambda^2}  & \sqrt{\left (1-\Lambda^2 \right )
\left (1-A^2\Lambda^4 \right )}  & \Lambda \cr\cr
\Lambda^3 \left [A - B\Lambda^5 \sqrt{\left (1-\Lambda^2 \right )
\left (1-A^2\Lambda^4 \right )} ~ e^{i\delta} \right ]
& -\Lambda \sqrt{1-A^2\Lambda^4}  & \sqrt{1-\Lambda^2} \cr}
\right ) \; .
%       (3.59)
\end{equation}
\normalsize
In this approximation, the unitary normalization relations of $V$ keep
valid to ${\cal O}(\Lambda^{11}) \sim 2\%$. Therefore (3.59) is 
sufficiently accurate to describe lepton flavor mixing, not only in solar 
and atmospheric neutrino oscillations, but also in some of the proposed 
long-baseline neutrino oscillation experiments. As the unitary 
orthogonality relations of $V$ are valid to 
${\cal O}(\Lambda^8) \sim 6\%$, the leptonic unitarity triangles and 
CP violation can also be described by (3.59) to an acceptable degree of 
accuracy. The off-diagonal asymmetries 
of the lepton flavor mixing matrix $V$ \cite{Xing02a} read as
\begin{eqnarray}
{\cal A}_{\rm L} & \equiv & |V_{e2}|^2 - |V_{\mu 1}|^2 = 
|V_{\mu 3}|^2 - |V_{\tau 2}|^2 = |V_{\tau 1}|^2 - |V_{e3}|^2
\nonumber \\
& = & A^2 \Lambda^6 \; ,
\nonumber \\
{\cal A}_{\rm R} & \equiv & |V_{e2}|^2 - |V_{\mu 3}|^2 = 
|V_{\mu 1}|^2 - |V_{\tau 2}|^2 = |V_{\tau 3}|^2 - |V_{e1}|^2
\nonumber \\
& = & \Lambda^2 \left (A^2 \Lambda^2 -1 \right ) \; .
%       (3.60)
\end{eqnarray}
We see that ${\cal A}_{\rm L} >0$ holds definitely. In comparison, the sign
of ${\cal A}_{\rm R}$ cannot be fixed from the present
experimental data. It is actually possible to obtain ${\cal A}_{\rm R} =0$,
when $A^2\Lambda^2 =1$ is satisfied. In this interesting case, the lepton 
flavor mixing matrix $V$ is exactly symmetric about its 
$V_{e3}$-$V_{\mu 2}$-$V_{\tau 1}$ axis \cite{Xing02a}. Note that the
CKM matrix of quark flavor mixing is approximately symmetric about
its $V_{ud}$-$V_{cs}$-$V_{tb}$ axis \cite{Xing95}. This is another
difference between the flavor mixing matrices of leptons and quarks.

Note that the Majorana CP-violating phases have been omitted in (3.59).
To incorporate $V$ with two Majorana phases $\rho$ and $\sigma$, we simply
multiply $V$ on its right-hand side with a pure phase matrix; i.e., 
$V \Longrightarrow V Q$ with 
$Q = {\rm Diag}\{ e^{i\rho}, e^{i\sigma}, e^{i\delta} \}$.
The chosen phase convention of $Q$ is to make the Dirac CP-violating phase 
$\delta$ not to manifest itself in the effective mass 
term of the neutrinoless double beta decay:
\begin{equation}
\langle m\rangle_{ee} \; = \; \left | m_1 \left (1 - A^2\Lambda^4
\right ) e^{2i\rho} + m_2 A^2 \Lambda^4 e^{2i\sigma} +
m_3 B^2 \Lambda^{16} \right | \; ,
%       (3.61)
\end{equation}
where $m_i$ (for $i=1,2,3$) are physical neutrino masses. This result
can somehow get simplified, if a specific pattern of the neutrino mass
spectrum is assumed. The present experimental upper bound is 
$\langle m\rangle_{ee} < 0.35 ~ {\rm eV}$ 
(at the $90\%$ confidence level \cite{HM}), from which no constraint 
on $\rho$ and $\sigma$ can be got.  

Now let us establish the direct relations between 
($\Lambda$, $A$, $B$) and ($\theta_{\rm atm}$, $\theta_{\rm sun}$,
$\theta_{\rm chz}$). With the help of (3.14) and (3.59), we find
\begin{eqnarray}
\Lambda & = & \sin\theta_{\rm atm} \; ,
\nonumber \\
A & = & \frac{\sqrt{\cos^2\theta_{\rm chz} - \sqrt{\cos^4\theta_{\rm chz}
- \sin^2 2\theta_{\rm sun}}}}{\sqrt{2} \sin^2\theta_{\rm atm}} \; ,
\nonumber \\
B & = & \frac{\sin\theta_{\rm chz}}{\sin^8\theta_{\rm atm}} \; .
%       (3.62)
\end{eqnarray}
Once the mixing angles $\theta_{\rm atm}$, $\theta_{\rm sun}$ and 
$\theta_{\rm chz}$ are precisely measured, we may use (3.62) to
determine the magnitudes of $\Lambda$, $A$ and $B$. For the purpose of
illustration, we typically take 
$0.25 \leq \sin^2\theta_{\rm sun} \leq 0.40$ \cite{SUN},
$\sin^2 2\theta_{\rm atm} > 0.92$ \cite{SK} and 
$\sin^2 2\theta_{\rm chz} < 0.1$ \cite{CHOOZ} to calculate the 
allowed regions of $\Lambda$, $A$ and $B$. Then we obtain
$0.6 \leq \Lambda \leq 0.8$, $0.8 \leq A \leq 1.94$ and
$0 \leq B \leq 9.5$. Although the allowed region
of $B$ is rather large, we expect that $B \sim {\cal O}(1)$ or a bit
smaller is most likely and most suitable for our parametrization.
It is also worthwhile to connect ($\Lambda$, $A$, $B$) to 
($\theta_x$, $\theta_y$, $\theta_z$), which are three mixing 
angles in the standard parametrization of $V$ in (3.23). We find 
$\sin\theta_x \approx A\Lambda^2$, $\sin\theta_y \approx \Lambda$
and $\sin\theta_z = B\Lambda^8$. In addition, the Dirac phase of CP 
violation in the standard parametrization is exactly equal to $\delta$ 
defined in the present Wolfenstein-like parametrization. 

An interesting point is that the {\it effective} lepton flavor mixing
matrix in matter can similarly be parametrized in terms of four 
matter-corrected parameters 
$\tilde{\Lambda}$, $\tilde{A}$, $\tilde{B}$ and $\tilde{\delta}$:
\footnotesize
\begin{equation}
\tilde{V} \; =\; \left ( \matrix{
\sqrt{1-\tilde{A}^2 \tilde{\Lambda}^4}  & 
\tilde{A} \tilde{\Lambda}^2  & 
\tilde{B}\tilde{\Lambda}^8 e^{-i\tilde{\delta}} \cr\cr
-\tilde{A} \tilde{\Lambda}^2 \sqrt{1-\tilde{\Lambda}^2}  & 
\sqrt{(1-\tilde{\Lambda}^2 ) (1-\tilde{A}^2\tilde{\Lambda}^4 )}  & 
\tilde{\Lambda} \cr\cr
\tilde{\Lambda}^3 \left [\tilde{A} - \tilde{B}\tilde{\Lambda}^5 
\sqrt{(1-\tilde{\Lambda}^2 )
(1-\tilde{A}^2\tilde{\Lambda}^4 )} ~ e^{i\tilde{\delta}} \right ] & 
-\tilde{\Lambda} \sqrt{1-\tilde{A}^2\tilde{\Lambda}^4}  & 
\sqrt{1-\tilde{\Lambda}^2} \cr}
\right ) \; ,
%       (3.63)
\end{equation}
\normalsize
where the Majorana phase matrix $Q$ has been omitted for simplicity.
There also exist simple relations between the {\it effective} 
mixing angles of $\tilde{V}$ (i.e., $\tilde{\theta}_x$, 
$\tilde{\theta}_y$ and $\tilde{\theta}_z$) and the corresponding 
new parameters ($\tilde{\Lambda}$, $\tilde{A}$ and $\tilde{B}$). 
It has been shown in section 3.2.3 that 
$\sin\tilde{\theta}_y \approx \sin\theta_y$ and 
$\sin\tilde{\delta} \approx \sin\delta$ hold to leading order for a 
variety of terrestrial long-baseline neutrino oscillation experiments. 
Therefore, we have
$\tilde{\Lambda} \approx \Lambda$ and $\tilde{\delta} \approx \delta$.
This result implies that $\Lambda$ and $\delta$ are essentially
stable against terrestrial matter effects. Hence the expansion of 
$\tilde{V}$ in powers of $\tilde{\Lambda} \approx \Lambda$ makes sense.
Only $A$ and $B$ in $V$ are sensitive to the matter-induced corrections.
Because of $\tilde{A} \propto \sin\tilde{\theta}_x$ and
$\tilde{B} \propto \sin\tilde{\theta}_z$, two remarkable conclusions 
can be drawn from Ref. \cite{Freund} for our new parameters: 
(a) $\tilde{A}/A$ 
is suppressed up to the order $\Delta m^2_{\rm sun}/\Delta m^2_{\rm atm}$; 
and (b) $\tilde{B}/B$ may have the resonant behavior similar to the
two-neutrino MSW resonance \cite{MSW}.

Finally we give some speculation on the physical meaning of $\Lambda$.
It is well known that the Wolfenstein parameter $\lambda \approx 0.22$ can 
be related to the ratios of quark masses in the Fritzsch ansatz of
quark mass matrices \cite{F79} or its modified versions \cite{40Q}:
\begin{equation}
\lambda \; \approx \; \left |\sqrt{\frac{m_u}{m_c}} - e^{i\phi_\lambda}
\sqrt{\frac{m_d}{m_s}} \right | \; ,
%       (3.64)
\end{equation}
where $\phi_\lambda$ denotes the phase difference between the (1,2)
elements of up- and down-type quark mass matrices. This relation
indicates that the smallness of $\lambda$ is a natural consequence of 
the strong quark mass hierarchy. Could the largeness of $\Lambda$ be 
attributed to a relatively weak hierarchy of three neutrino masses? 
The answer is indeed affirmative for the Fritzsch texture of lepton mass 
matrices in (2.39), which is compatible with current experimental data 
on neutrino oscillations if the masses of three neutrinos 
have a normal but weak hierarchy (typically, $m_1$ : $m_2$ : $m_3$ 
$\approx$ 1 : 3 : 10) \cite{XingF}. In this phenomenological model,
we approximately obtain  
\begin{equation}
\Lambda \; \approx \; \left |\sqrt{\frac{m_2}{m_3}} - e^{i\phi_\Lambda}
\sqrt{\frac{m_\mu}{m_\tau}} \right | \; ,
%       (3.65)
\end{equation}
where $\phi_\Lambda$ denotes the phase difference between the (2,3)
elements of charged lepton and neutrino mass matrices. We find that
$\phi_\Lambda \sim \pm 180^\circ$ is practically favored \cite{XingF},
in order to obtain a sufficiently large $\Lambda$. To illustrate, we 
typically take $m_2/m_3 \sim 0.3$ as well as 
$m_\mu/m_\tau \approx 0.06$ \cite{PDG02}. 
Then we arrive at $\Lambda \sim 0.8$, a result consistent with our
empirical expectation for the order of $\Lambda$ in (3.59)
%%%%%%%%%%%%%%%%%%%%%%%%%%%%
\footnote{Assuming a somewhat stronger mass hierarchy for three neutrinos,
Kaus and Meshkov \cite{Kaus} have proposed a different expansion of
the neutrino mixing matrix in terms of $\Lambda = \sqrt{m_2/m_3} = 
(\Delta m^2_{\rm sun}/\Delta m^2_{\rm atm})^{1/4} \sim 0.37$. This
parameter is associated with $V_{e2}$ instead of $V_{\mu 3}$, therefore
it is sensitive to the matter effect. In contrast, our parametrization
does not rely on the assumption of neutrino mass hierarchy, and its 
expansion parameter is insensitive to the matter-induced corrections.}.
%%%%%%%%%%%%%%%%%%%%%%%%%%%%

\section{CP and T Violation}
\setcounter{equation}{0}
\setcounter{figure}{0}

\subsection{Rephasing Invariants in Matter}

No matter whether massive neutrinos are Dirac or Majorana particles, 
the effects of CP or T violation in 
normal neutrino-neutrino and antineutrino-antineutrino oscillations
are measured by a universal parameter $J$ in vacuum or
$\tilde J$ in matter \cite{Jarlskog}, defined through 
\begin{eqnarray}
{\rm Im} (V_{\alpha i} V_{\beta j} V^*_{\alpha j} V^*_{\beta i}) 
& = & J \sum_{\gamma} \epsilon_{\alpha\beta\gamma} \sum_k \epsilon_{ijk} \; , 
\nonumber \\
{\rm Im} (\tilde{V}_{\alpha i} \tilde{V}_{\beta j} \tilde{V}^*_{\alpha j} 
\tilde{V}^*_{\beta i}) & = & \tilde{J} 
\sum_{\gamma} \epsilon_{\alpha\beta\gamma} \sum_k \epsilon_{ijk} \; ,
%       (4.1)
\end{eqnarray}
where the subscripts ($\alpha, \beta, \gamma$) and ($i, j, k$) run 
respectively over ($e,\mu,\tau$) and (1, 2, 3). Clearly $J$ and $\tilde J$ 
are rephasing-invariant; i.e., they are independent
of any redefinition of the phases for charged lepton and neutrino fields.  
Note that $J$ and $\tilde J$ can be expressed in terms of the moduli of
four independent matrix elements of $V$ and $\tilde V$, respectively, 
as follows \cite{FXReview}
%%%%%%%%%%%%%%%%%%%%%%%%%%
\footnote{Note that there was a typing error in the original formula of
$J^2$ given in Ref. \cite{Sasaki}.}:
%%%%%%%%%%%%%%%%%%%%%%%%%%
\begin{eqnarray}
J^2 & = & |V_{\alpha i}|^2 |V_{\beta j}|^2 |V_{\alpha j}|^2
|V_{\beta i} |^2 - \frac{1}{4} \left ( 1 + |V_{\alpha i}|^2
|V_{\beta j}|^2 + |V_{\alpha j}|^2 |V_{\beta i}|^2 
\right .  \nonumber \\
& & \left . - |V_{\alpha i}|^2 - |V_{\beta j}|^2
- |V_{\alpha j}|^2 - |V_{\beta i}|^2 \right )^2 \; ,
\nonumber \\
\tilde{J}^2 & = & |\tilde{V}_{\alpha i}|^2 |\tilde{V}_{\beta j}|^2 
|\tilde{V}_{\alpha j}|^2 |\tilde{V}_{\beta i} |^2 - 
\frac{1}{4} \left ( 1 + |\tilde{V}_{\alpha i}|^2
|\tilde{V}_{\beta j}|^2 + |\tilde{V}_{\alpha j}|^2 |\tilde{V}_{\beta i}|^2 
\right .  \nonumber \\
& & \left . - |\tilde{V}_{\alpha i}|^2 - |\tilde{V}_{\beta j}|^2
- |\tilde{V}_{\alpha j}|^2 - |\tilde{V}_{\beta i}|^2 \right )^2 \; ,
%	(4.2)
\end{eqnarray}
in which $\alpha \neq \beta$ running over $(e, \mu, \tau)$ and
$i \neq j$ running over $(1, 2, 3)$. The implication of this result is 
obvious: the information about leptonic CP violation can in principle be
extracted from the measured moduli of the flavor mixing matrix
elements. 

Now let us establish the relationship between $\tilde J$ and $J$.
Note that the imaginary parts of the rephasing-invariant quantities 
$Z_{e \mu} Z_{\mu \tau} Z_{\tau e}$ and 
$\tilde{Z}_{e \mu} \tilde{Z}_{\mu \tau} \tilde{Z}_{\tau e}$, 
\begin{eqnarray}
{\rm Im} ( Z_{e \mu} Z_{\mu \tau} Z_{\tau e} )
& = & \sum^3_{i=1} \sum^3_{j=1} \sum^3_{k=1} \left [ m^2_i m^2_j m^2_k 
~ {\rm Im} \left ( V_{e i} V_{\mu j} V_{\tau k} 
V^*_{e k} V^*_{\mu i} V^*_{\tau j} \right ) \right ] \; ,
\nonumber \\
{\rm Im} ( \tilde{Z}_{e \mu} \tilde{Z}_{\mu \tau} 
\tilde{Z}_{\tau e} )
& = & \sum^3_{i=1} \sum^3_{j=1} \sum^3_{k=1} \left [ \tilde{m}^2_i 
\tilde{m}^2_j \tilde{m}^2_k ~ {\rm Im}
\left ( \tilde{V}_{e i} \tilde{V}_{\mu j} \tilde{V}_{\tau k} 
\tilde{V}^*_{e k} \tilde{V}^*_{\mu i} 
\tilde{V}^*_{\tau j} \right ) \right ] \; ,
%	(4.3)
\end{eqnarray}
which do not vanish unless leptonic CP and T are good symmetries, 
amount to each other as a trivial consequence of (3.50).
The right-hand side of (4.3) can be expanded in terms of $J$ and 
$\tilde{J}$. In doing so, one needs to use (4.1) as well as the 
unitarity conditions of $V$ and $\tilde{V}$ frequently. After some 
lengthy but straightforward algebraic calculations, we arrive at an 
elegant relation between the universal CP-violating parameters $J$ in 
vacuum and $\tilde J$ in matter \cite{Xing01}:
\begin{equation}
\tilde{J} \prod_{i<j} \left (\tilde{m}^2_i - \tilde{m}^2_j \right ) 
\; = \; J \prod_{i<j} \left (m^2_i - m^2_j \right ) \; .
%      	(4.4)
\end{equation}
This interesting result was first obtained by Naumov \cite{Naumov}. 
Of course $\tilde{J} = J$ holds if $A = 0$, and $\tilde{J} = 0$ holds if 
$J =0$. It is worth mentioning that (4.4) can also be derived from the 
equality between ${\rm Det} (\tilde{X})$ and ${\rm Det} (X)$ given
in (3.46) \cite{Scott}. Indeed it is easy to show that 
\begin{eqnarray}
{\rm Det}(X) & = & 2 J 
\prod_{\alpha < \beta} \left (m^2_\alpha - m^2_\beta \right )
\prod_{i<j} \left (m^2_i - m^2_j \right ) \; ,
\nonumber \\
{\rm Det}(\tilde{X}) & = & 2 \tilde{J} 
\prod_{\alpha < \beta} \left (m^2_\alpha - m^2_\beta \right )
\prod_{i<j} \left (\tilde{m}^2_i - \tilde{m}^2_j \right ) \; ,
%      	(4.5)
\end{eqnarray}
where the Greek indices run over $(e, \mu, \tau)$; and the
Latin indices run over $(1, 2, 3)$. It should be noted that
the determinant ${\rm Det}(X)$ or ${\rm Det}(\tilde{X})$ contains
the same information about leptonic CP violation as the 
universal parameter $J$ or $\tilde J$, although their expressions 
are somehow different \cite{FX99}. The latter is apparently simpler 
and more instructive for the description of leptonic CP violation in 
neutrino oscillations. Taking the standard parametrization of $V$ in
(3.23) or $\tilde{V}$ in (3.55), we have
\begin{eqnarray}
J & = & \sin\theta_x \cos\theta_x \sin\theta_y \cos\theta_y
\sin\theta_z \cos^2\theta_z \sin\delta \; ,
\nonumber \\
\tilde{J} & = & \sin\tilde{\theta}_x \cos\tilde{\theta}_x 
\sin\tilde{\theta}_y \cos\tilde{\theta}_y \sin\tilde{\theta}_z 
\cos^2\tilde{\theta}_z \sin\tilde{\delta} \; .
%       (4.6)
\end{eqnarray}
Thus the CP-violating phases $\delta$ and $\tilde{\delta}$ can
be related to each other via (4.4) and (4.6).

Once again, the formulas obtained above are valid only for neutrinos 
propagating in vacuum and in matter. They will become valid for 
antineutrinos, if the straightforward replacements 
$V \Longrightarrow V^*$ and $A \Longrightarrow -A$ are made. 

We remark that the effective CP-violating parameter
$\tilde J$ depends not only upon the fundamental CP-violating 
parameter of the lepton flavor mixing matrix ($J$), but also 
upon the mass-squared differences of neutrinos ($\Delta m^2_{21}$
and $\Delta m^2_{31}$) and the matter-induced effect ($A$).
In particular, both the sign and the magnitude of $\tilde J$ are
dependent on the signs of $\Delta m^2_{21}$ and $\Delta m^2_{31}$.
With the help of (3.37) and (3.38), we find that
the effective mass-squared differences $\Delta \tilde{m}^2_{21}$
and $\Delta \tilde{m}^2_{31}$ can keep unchanged in proper 
arrangements of the signs of $\Delta m^2_{21}$, $\Delta m^2_{31}$
and $A$. A careful analysis of (4.4) leads to the following exact 
relations \cite{Xing00a}:
\begin{eqnarray}
\tilde{J} (+\Delta m^2_{21}, +\Delta m^2_{31}, +A )
& = & -\tilde{J} (-\Delta m^2_{21}, -\Delta m^2_{31}, -A ) \; ,
\nonumber \\
\tilde{J} (+\Delta m^2_{21}, -\Delta m^2_{31}, +A )
& = & -\tilde{J} (-\Delta m^2_{21}, +\Delta m^2_{31}, -A ) \; ,
\nonumber \\
\tilde{J} (+\Delta m^2_{21}, +\Delta m^2_{31}, -A )
& = & -\tilde{J} (-\Delta m^2_{21}, -\Delta m^2_{31}, +A ) \; ,
\nonumber \\
\tilde{J} (+\Delta m^2_{21}, -\Delta m^2_{31}, -A )
& = & -\tilde{J} (-\Delta m^2_{21}, +\Delta m^2_{31}, +A ) \; .
%	(4.7)
\end{eqnarray}
The validity of these relations are independent of 
both the neutrino beam energy and the baseline length.
Note that current experimental data on solar neutrino oscillations
favor $\Delta m^2_{21} > 0$, but the sign of $\Delta m^2_{31}$
remains unknown.

It is also worth remarking that the CP- and T-violating effects are
measured by the same parameter ($J$ in vacuum or $\tilde J$ in matter),
as a straightforward consequence of CPT invariance. As for neutrino
oscillations, a signal of CP violation comes from the probability
asymmetry between $\nu_\alpha \rightarrow \nu_\beta$ and
$\overline{\nu}_\alpha \rightarrow \overline{\nu}_\beta$ transitions;
while a signal of T violation is attributed to the probability
asymmetry between $\nu_\alpha \rightarrow \nu_\beta$ and
$\nu_\beta \rightarrow \nu_\alpha$ transitions or between
$\overline{\nu}_\alpha \rightarrow \overline{\nu}_\beta$ and
$\overline{\nu}_\beta \rightarrow \overline{\nu}_\alpha$ transitions
($\alpha \neq \beta$).
The former is roughly associated with the difference between 
$\tilde{J}(J, A)$ and $\tilde{J}(-J, -A)$, in which the $A$-induced
effects essentially add each other; but the latter is 
roughly associated with the difference between $\tilde{J}(J, A)$ and
$\tilde{J}(-J, A)$ or between $\tilde{J}(J, -A)$ and $\tilde{J}(-J, -A)$,
in which the $A$-induced effects essentially cancel each other.
For this reason, the CP-violating
asymmetries in long-baseline neutrino oscillations are rather sensitive 
to the terrestrial matter effect; while the T-violating asymmetries are
almost independent of the terrestrial matter effect 
(see Refs. \cite{Xing00a,Xing00b} for detailed discussions). Measuring 
T violation is practically more difficult than measuring CP violation, 
however \cite{Chen}.

To illustrate the dependence of $\tilde{J}$ on matter effects, we
typically take $\Delta m^2_{21} = 5\times 10^{-5} ~ {\rm eV}^2$ and
$\Delta m^2_{31} = 3\times 10^{-3} ~ {\rm eV}^2$ as well as 
$\theta_x \approx 35^\circ$, $\theta_y \approx 40^\circ$, 
$\theta_z \approx 5^\circ$ and $\delta \approx \pm 90^\circ$ in the 
standard parametrization of $V$. The result is shown in
Fig. 4.1, where $10^{-7} ~ {\rm eV}^2 \leq A \leq 10^{-2} ~ {\rm eV}^2$
has been chosen for the matter parameter $A$. One can see that the 
magnitude of $\tilde{J}$ decreases, when the matter effect becomes 
significant (e.g., $A \geq 10^{-4} ~ {\rm eV}^2$). Nevertheless, this 
feature does not necessarily
imply that the CP-violating asymmetries in realistic long-baseline
neutrino oscillations would be smaller than their values in vacuum.
Very large terrestrial matter effects can significantly modify the 
frequencies of neutrino oscillations and thus enhance (or suppress) 
the genuine signals of CP violation.
%%%%%%%%%%%%%%%%%%%% Fig. 4.1 %%%%%%%%%%%%%%%%
\begin{figure}[t]
\vspace{-2.5cm}
\epsfig{file=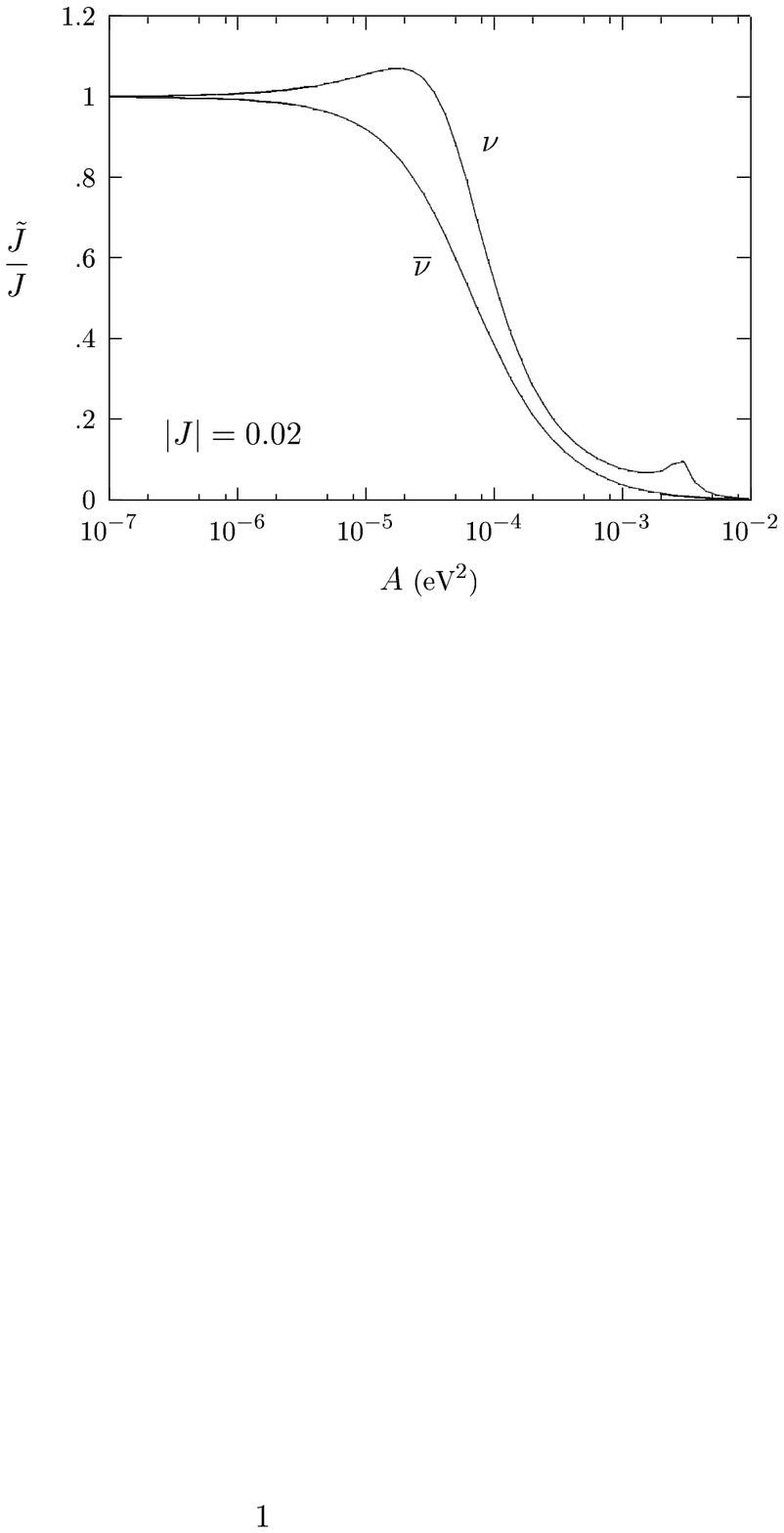,bbllx=3cm,bblly=5.5cm,bburx=21cm,bbury=27.8cm,%
width=15cm,height=19cm,angle=0,clip=}
\vspace{-9.8cm}
\caption{The ratio $\tilde{J}/J$ changing with the matter parameter
$A$ for neutrinos ($\nu$) and antineutrinos ($\overline{\nu}$),
in which $\Delta m^2_{21} = 5\times 10^{-5} ~ {\rm eV}^2$,
$\Delta m^2_{31} = 3\times 10^{-3} ~ {\rm eV}^2$,
$\theta_x \approx 35^\circ$, $\theta_y \approx 40^\circ$, 
$\theta_z \approx 5^\circ$ and $\delta \approx \pm 90^\circ$
have typically been input.}
\end{figure}
%%%%%%%%%%%%%%%%%%%%%%%%%%%%%%%%%%%%%%%%%%%%

\subsection{Leptonic Unitarity Triangles}

A geometric description of the lepton flavor mixing phenomenon 
is physically intuitive and instructive. The unitarity of the 
$3\times 3$ flavor mixing matrix $V$ in vacuum can be expressed
by two sets of orthogonality relations and two sets of 
normalization conditions for its nine matrix elements:
\begin{eqnarray}
\sum_i \left (V^*_{\alpha i} V_{\beta i} \right )
& = & \delta_{\alpha\beta} \; , 
\nonumber \\
\sum_\alpha \left (V^*_{\alpha i}V_{\alpha j} \right ) 
& = & \delta_{ij} \; , 
%   	(4.8)
\end{eqnarray}
where Greek and Latin indices run over $(e, \mu, \tau)$ and $(1, 2, 3)$, 
respectively. In the complex plane the six orthogonality 
relations in (4.8) define six triangles 
$(\triangle_e, \triangle_\mu, \triangle_\tau)$ and 
$(\triangle_1, \triangle_2, \triangle_3)$ shown in Fig. 4.2,
the so-called unitarity triangles. 
In general, these six triangles have eighteen different 
sides and nine different inner (or outer) angles. 
The unitarity requires that all six triangles have the same
area amounting to $J/2$, where $J$ is just the
rephasing-invariant measure of CP violation defined in 
(4.1). If CP were an exact symmetry, $J=0$
would hold and those unitarity triangles would collapse into
lines in the complex plane. Note that the shape and area of each
unitarity triangle are irrelevant to the nature of neutrinos;
i.e., they are the same for Dirac and Majorana neutrinos.
%%%%%%%%%%%%%%%%%%%% Fig. 4.2 %%%%%%%%%%%%%%%%
\begin{figure}[t]
\vspace{0.2cm}
\epsfig{file=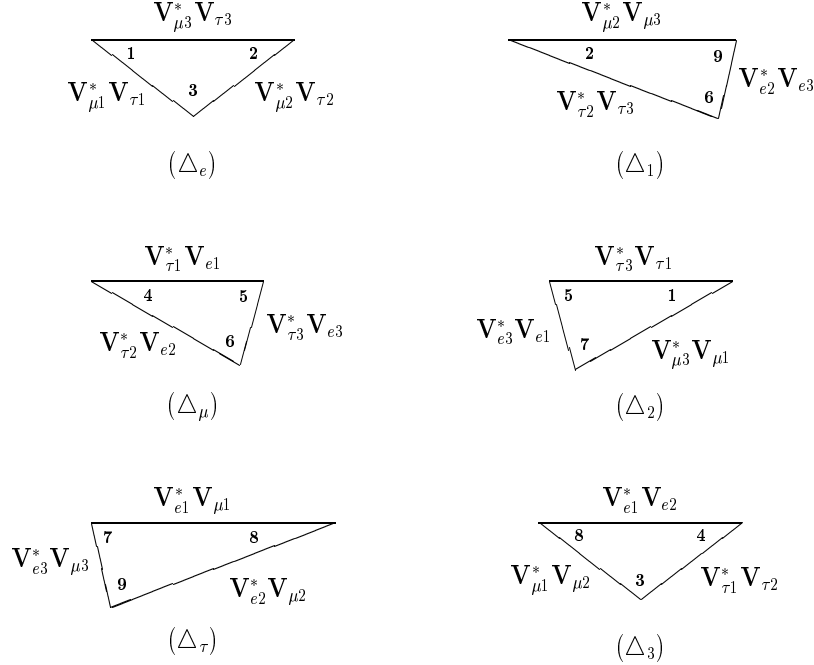,bbllx=0.5cm,bblly=1.5cm,bburx=20cm,bbury=27cm,%
width=15cm,height=22cm,angle=0,clip=}
\vspace{-13cm}
\caption{Unitarity triangles of the lepton flavor mixing matrix 
in the complex plane. Each triangle is named by 
the index that does not manifest in its three sides \cite{FXReview}.}
\end{figure}
%%%%%%%%%%%%%%%%%%%%%%%%%%%%%%%%%%%%%%%%%%%%

In additon to $J$, there exist other two characteristic
quantities of $V$, resulting from its normalization conditions 
in (4.8). They are the off-diagonal asymmetries ${\cal A}_{\rm L}$
and ${\cal A}_{\rm R}$ defined in (3.60). Clearly ${\cal A}_{\rm L} =0$ 
(or ${\cal A}_{\rm R} =0$) would imply that the flavor mixing matrix $V$ 
were symmetric about its $V_{e1}$-$V_{\mu 2}$-$V_{\tau 3}$ 
(or $V_{e3}$-$V_{\mu 2}$-$V_{\tau 1}$) axis. Geometrically this would 
correspond to the congruence between two unitarity triangles; i.e.,
\begin{eqnarray}
{\cal A}_{\rm L} = 0 \; ~~~ \Longrightarrow ~~~ \; \triangle_e & \cong &
\triangle_1 \; , ~~~
\triangle_\mu \; \cong \; \triangle_2 \; , ~~~
\triangle_\tau \; \cong \; \triangle_3 \; ; 
\nonumber \\
{\cal A}_{\rm R} = 0 \; ~~~ \Longrightarrow ~~~ \; \triangle_e & \cong &
\triangle_3 \; , ~~~
\triangle_\mu \; \cong \; \triangle_2 \; , ~~~
\triangle_\tau \; \cong \; \triangle_1 \; .
%    	(4.9)
\end{eqnarray}
Our current knowledge on lepton flavor mixing is rather poor, hence it 
remains difficult to quantify the magnitude of leptonic CP violation.
Nevertheless, we are sure that ${\cal A}_{\rm L} > 0$ definitely holds. 
The possibility for ${\cal A}_{\rm R} = 0$ cannot be ruled out at 
present \cite{Xing02a}. Useful information about the sides and angles 
of the leptonic unitarity triangles will be obtained from the medium- 
and long-baseline neutrino experiments in the near future.

For instance, one might be able to determine three inner angles of 
the unitarity triangle $\triangle_3$, which are defined as
\begin{eqnarray}
\alpha^{~}_l & \equiv & \arg \left ( - \frac{V^*_{e 1}V_{e 2}}
{V^*_{\mu 1}V_{\mu 2}} \right ) \;\; , 
\nonumber \\
\beta_l & \equiv & \arg \left ( - \frac{V^*_{\mu 1}V_{\mu 2}}
{V^*_{\tau 1}V_{\tau 2}} \right ) \;\; ,
\nonumber \\
\gamma^{~}_l & \equiv & \arg \left ( - \frac{V^*_{\tau 1}V_{\tau 2}}
{V^*_{e 1}V_{e 2}} \right ) \;\; .
%   	(4.10)
\end{eqnarray}
Then it should be possible to test the self-consistency of our lepton 
flavor mixing and CP-violating picture 
(e.g., $\alpha^{~}_l + \beta_l + \gamma^{~}_l
\neq \pi$ would imply the presence of new physics which violates
the unitarity of the $3\times 3$ lepton flavor mixing matrix).
The measurement of $\alpha^{~}_l$, $\beta_l$ and $\gamma^{~}_l$ can
in principle be realized in a long-baseline neutrino oscillation
experiment. Taking $\Delta m^2_{21} \approx \Delta m^2_{\rm sun}$ and 
$\Delta m^2_{31} \approx \Delta m^2_{32} \approx \pm 
\Delta m^2_{\rm atm}$, we simplify (3.26) to \cite{FXReview} 
\begin{eqnarray}
P (\nu_e \rightarrow \nu_\mu) & = & 4 |V_{e3}|^2 |V_{\mu 3}|^2
\sin^2 F_{\rm atm} ~ - ~ 4 {\rm Re} \left (V_{e 1} V_{\mu 2}
V^*_{e 2} V^*_{\mu 1} \right ) \sin^2 F_{\rm sun}
\nonumber \\ 
&& ~~~~~~~~~~~~~~~~~~~~~~~~~~~~ - ~ 8 J 
\sin F_{\rm sun} \sin^2 F_{\rm atm} \; ,
\nonumber \\ 
P (\nu_\mu \rightarrow \nu_\tau) & = & 4 |V_{\mu 3}|^2 |V_{\tau 3}|^2
\sin^2 F_{\rm atm} ~ - ~ 4 {\rm Re} \left ( V_{\mu 1} V_{\tau 2}
V^*_{\mu 2} V^*_{\tau 1} \right ) \sin^2 F_{\rm sun}
\nonumber \\
&& ~~~~~~~~~~~~~~~~~~~~~~~~~~~~ - ~ 8 J
\sin F_{\rm sun} \sin^2 F_{\rm atm} \; ,
\nonumber \\ 
P (\nu_\tau \rightarrow \nu_e) & = & 4 |V_{\tau 3}|^2 |V_{e 3}|^2
\sin^2 F_{\rm atm} ~ - ~ 4 {\rm Re} \left ( V_{\tau 1} V_{e 2}
V^*_{\tau 2} V^*_{e 1} \right ) \sin^2 F_{\rm sun}
\nonumber \\
&& ~~~~~~~~~~~~~~~~~~~~~~~~~~~~ - ~ 8 J
\sin F_{\rm sun} \sin^2 F_{\rm atm} \; ,
% 	(4.11)    	
\end{eqnarray}
where $F_{\rm atm} = F_{32} \approx F_{31}$ and $F_{\rm sun} = F_{21}$ 
defined below (3.5) measure the oscillation frequencies of atmospheric 
and solar neutrinos, respectively. Provided the baseline satisfies the 
condition $L \sim E/\Delta m^2_{\rm sun}$ (i.e., $F_{\rm sun} \sim 1$),
the CP-conserving quantities
${\rm Re}(V_{\alpha 1} V_{\beta 2} V^*_{\alpha 2} V^*_{\beta 1})$
and the CP-violating parameter $J$ could both be determined from (4.11) 
for changing values of the neutrino beam energy $E$. Then one may get
\begin{eqnarray}
\tan\alpha^{~}_l & = & - \frac{J}{{\rm Re}(V_{e1} V_{\mu 2}
V^*_{e 2}V^*_{\mu 1})} \;\; ,
\nonumber \\
\tan\beta_l & = & - \frac{J}{{\rm Re}(V_{\mu 1} V_{\tau 2}
V^*_{\mu 2} V^*_{\tau 1})} \;\; ,
\nonumber \\
\tan\gamma^{~}_l & = & - \frac{J}{{\rm Re}(V_{\tau 1} V_{e 2}
V^*_{\tau 2} V^*_{e 1})} \;\; .
%     	(4.12)
\end{eqnarray}
In practice, the earth-induced matter effects are likely to 
fake the genuine CP-violating signals and have to be taken into 
account. If the neutrino beam energy is low enough (e.g., $E \sim 0.1$ 
to 1 GeV) and the baseline length is not too long (e.g., $L \sim 100$ km), 
the terrestrial matter effects might not be very significant for our 
purpose \cite{Sato}.

Another interesting unitarity triangle is $\triangle_\tau$, whose
three sides might more easily be measured. To establish 
$\triangle_\tau$ needs very precise data, which should be able to show
$|V^*_{e1} V_{\mu 1}| + |V^*_{e3} V_{\mu 3}| > |V^*_{e2} V_{\mu 2}|$ or 
$|V^*_{e2} V_{\mu 2}| + |V^*_{e3} V_{\mu 3}| > 
|V^*_{e1} V_{\mu 1}|$ \cite{Smirnov}. Such an accuracy requirement   
is practically a big challenge, because one side of $\triangle_\tau$
is much shorter than its other two sides (i.e.,
$|V^*_{e1}V_{\mu 1}| \sim |V^*_{e2}V_{\mu 2}| \gg |V^*_{e3}V_{\mu 3}|$,
as (3.19) has shown). Note that three sides of the effective
unitarity triangle $\tilde{\triangle}_\tau$ in matter, defined as
\begin{equation}
\tilde{\triangle}_\tau: ~
\tilde{V}^*_{e1} \tilde{V}_{\mu 1} + \tilde{V}^*_{e2} \tilde{V}_{\mu 2} 
+ \tilde{V}^*_{e3} \tilde{V}_{\mu 3} \; = \; 0 \; ,
%	(4.13)
\end{equation}
are possible to be comparable in magnitude. With the help of (3.52),
we arrive at
\begin{eqnarray}
\tilde{V}^*_{e1} \tilde{V}_{\mu 1} & = &
\frac{\Delta \hat{m}^2_{21} \Delta m^2_{31}}
{\Delta \tilde{m}^2_{21} \Delta \tilde{m}^2_{31}} V^*_{e1} V_{\mu 1} ~ + ~
\frac{\Delta \hat{m}^2_{11} \Delta m^2_{32}}
{\Delta \tilde{m}^2_{21} \Delta \tilde{m}^2_{31}} V^*_{e2} V_{\mu 2} \;\; ,
\nonumber \\
\tilde{V}^*_{e2} \tilde{V}_{\mu 2} & = &
\frac{\Delta \hat{m}^2_{32} \Delta m^2_{21}}
{\Delta \tilde{m}^2_{32} \Delta \tilde{m}^2_{21}} V^*_{e2} V_{\mu 2} ~ + ~
\frac{\Delta \hat{m}^2_{22} \Delta m^2_{31}}
{\Delta \tilde{m}^2_{32} \Delta \tilde{m}^2_{21}} V^*_{e3} V_{\mu 3} \;\; ,
\nonumber \\
\tilde{V}^*_{e3} \tilde{V}_{\mu 3} & = &
\frac{\Delta \hat{m}^2_{13} \Delta m^2_{23}}
{\Delta \tilde{m}^2_{31} \Delta \tilde{m}^2_{32}} V^*_{e3} V_{\mu 3} ~ + ~
\frac{\Delta \hat{m}^2_{33} \Delta m^2_{21}}
{\Delta \tilde{m}^2_{31} \Delta \tilde{m}^2_{32}} V^*_{e1} V_{\mu 1} \;\; ,
%	(4.14)
\end{eqnarray}
where $\Delta \hat{m}^2_{ij} \equiv m^2_i - \tilde{m}^2_j$.
The explicit expressions of $\Delta \tilde{m}^2_{ij}$ and 
$\Delta \hat{m}^2_{ij}$, which depend on both $\Delta m^2_{ij}$ and $A$, 
can be obtained from (3.37). The instructive result in (4.14) clearly
shows how three sides of $\triangle_\tau$ get modified in matter.
Of course, one may use (4.14) to evaluate three sides of 
$\tilde{\triangle}_\tau$. It is also possible to get a numerical feeling
of $\triangle_\tau$ and $\tilde{\triangle}_\tau$ directly from Fig. 3.2.
For example, we find that three sides of $\tilde{\triangle}_\tau$ become
comparable in magnitude (of order $|V^*_{e3}V_{\mu 3}|$), when the neutrino 
beam energy $E$ is about 1 GeV or somehow larger. In a similar way, 
we can define and discuss other effective unitarity triangles in matter.

If $|V_{e3}|$, $|V_{\mu 3}|$ and $|V_{\tau 3}|$ are well determined in 
the first-round long-baseline neutrino experiments, 
it should be possible to check one of the three normalization conditions:
$|V_{e3}|^2 + |V_{\mu 3}|^2 + |V_{\tau 3}|^2 =1$.
If $|V_{e1}|$, $|V_{e2}|$ and $|V_{e3}|$ are measured to a good degree
of accuracy, one can test another  
normalization condition $|V_{e1}|^2 + |V_{e2}|^2 + |V_{e3}|^2 =1$.
At this stage the moduli of $V_{\mu 1}$, $V_{\mu 2}$, $V_{\tau 1}$ and
$V_{\tau 2}$ remain unknown. 
To determine the universal CP-violating parameter $J$ and
to make a full test of the unitarity of $V$, much more delicate
long-baseline neutrino experiments are needed.

\subsection{Baryon Asymmetry via Leptogenesis}

In the universe, the density of baryons compared to that of photons 
is extremely small: $\eta \equiv n^{~}_{\rm B}/n^{~}_\gamma =
(2.6 - 6.3) \times 10^{-10}$, extracted from the Big-Bang
nucleosynthesis \cite{PDG02}. This tiny quantity measures
the observed matter-antimatter or baryon-antibaryon asymmetry of 
the universe,
\begin{equation}
Y_{\rm B} \; \equiv \; \frac{n^{~}_{\rm B} - n^{~}_{\rm\bar B}}{\bf s} 
\; \approx \; \frac{\eta}{7.04} \; = \; (3.7 - 8.9) \times 10^{-11} \; ,
%          (4.15)
\end{equation}
where $\bf s$ denotes the entropy density. To produce a net baryon 
asymmetry in the standard Big-Bang model, three Sakharov necessary 
conditions have to be satisfied \cite{Sakharov}: 
(a) baryon number nonconservation, (b) C and CP violation, and 
(c) a departure from thermal equilibrium. Among a number of 
interesting and viable baryogenesis mechanisms proposed in the 
literature \cite{Baryon}, Fukugita and Yanagida's leptogenesis 
mechanism \cite{FY} has recently attracted a lot of attention -- due 
partly to the fact that neutrino physics is entering a flourishing era.

As mentioned in section 2.1.2, a simple extension of the standard 
model to generate neutrino masses is to include one
right-handed neutrino in each of three lepton families, while 
the Lagrangian of electroweak interactions keeps invariant under 
the $\rm SU(2)_{\rm L} \times U(1)_{\rm Y}$ gauge 
transformation. In this case, the Yukawa interactions of leptons
are described by
%%%%%%%%%%%%%%%%%%%%%%%%%%%%
\footnote{One may also consider a similar extension of the minimal
supersymmetric standard model to generate neutrino masses and to 
resolve the hierarchy problem induced by heavy right-handed Majorana
neutrinos \cite{Davidson}. The parameter counting for lepton flavor
mixing and leptogenesis is identical in models with and without 
supersymmetry \cite{Raidal}.}
%%%%%%%%%%%%%%%%%%%%%%%%%%%
\begin{equation}
- {\cal L}_{\rm Y} \; = \; 
\overline{l}_{\rm L} \tilde{\phi} Y_l e^{~}_{\rm R} ~ + ~ 
\overline{l}_{\rm L} \phi Y_\nu \nu^{~}_{\rm R} ~ + ~
\frac{1}{2} \overline{\nu^{\rm c}_{\rm R}} M_{\rm R} \nu_{\rm R} 
~ + ~ {\rm h.c.} \; , 
%          (4.16)
\end{equation}
where $l_{\rm L}$ denotes the left-handed lepton doublet, $e_{\rm R}$
and $\nu_{\rm R}$ stand respectively for the right-handed charged lepton
and Majorana neutrino singlets, and $\phi$ is the Higgs-boson weak 
isodoublet. The lepton number violation induced by the third term of 
${\cal L}_{\rm Y}$ allows decays of the heavy (right-handed) Majorana 
neutrinos $N_i$ (for $i=1,2,3$) to happen; i.e.,  
$N_i \rightarrow l + \phi^\dagger$ vs 
$N_i \rightarrow l^{\rm c} + \phi$, as illustrated in Fig. 4.3.
%%%%%%%%%%%%%%%%%%%%%%%% Fig. 4.3 %%%%%%%%%%%%%%%%%%%%%%%%%%%%%
\begin{figure}[t]
\vspace{-1.5cm}
\epsfig{file=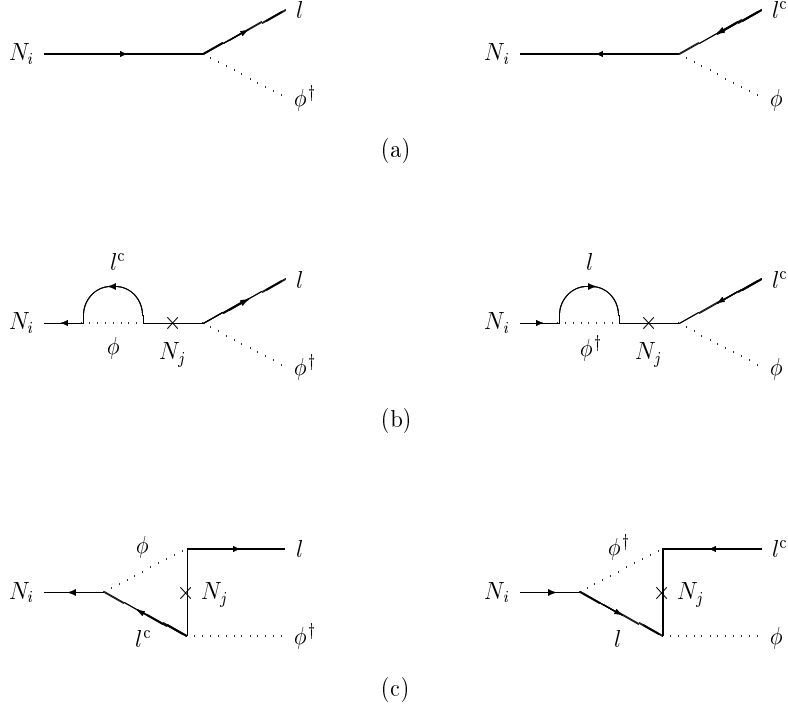,bbllx=0cm,bblly=1.5cm,bburx=20cm,bbury=27cm,%
width=15cm,height=20cm,angle=0,clip=}
\vspace{-8.6cm}
\caption{Feynman diagrams for the decay of a heavy (right-handed)
Majorana neutrino $N_i$: (a) the tree-level contribution; (b) the
one-loop self-energy correction; and (c) the one-loop vertex 
correction. Note that $N_i$'s antiparticle is itself up to a phase
factor, and ``$\times$'' indicates the chirality flip between $N_i$ 
and its antiparticle in the propagator. For simplicity, we have only
taken account of the diagrams which may result in the CP-odd
contributions to the decay rates of $N_i$.}
\end{figure}
%%%%%%%%%%%%%%%%%%%%%%%%%%%%%%%%%%%%%%%%%%%%
Because each decay mode occurs both at the tree level and at the
one-loop level (via the self-energy and vertex corrections), the
interference between the tree-level and one-loop decay amplitudes
can lead to a CP-violating asymmetry $\varepsilon_i$ between the two
CP-conjugated processes \cite{Covi}:
\begin{eqnarray}
\varepsilon_i & \equiv & \frac{\Gamma (N_i \rightarrow l + \phi^\dagger)
- \Gamma (N_i \rightarrow l^{\rm c} + \phi)}
 {\Gamma (N_i \rightarrow l + \phi^\dagger)
+ \Gamma (N_i \rightarrow l^{\rm c} + \phi)} 
\nonumber \\
& = & \frac{1}{8\pi \left (Y^\dagger_\nu Y_\nu \right )_{ii}} 
\sum^3_{j=1} \left \{ {\rm Im} \left [ \left (Y^\dagger_\nu Y_\nu
\right )_{ij} \right ]^2 \left [ f^{~}_{\rm V} \left (\frac{M^2_j}{M^2_i} 
\right ) + f^{~}_{\rm S} \left (\frac{M^2_j}{M^2_i}\right ) \right ] 
\right \} \; ,
%       (4.17)
\end{eqnarray}
where the loop functions $f^{~}_{\rm V}$ (vertex) and $f^{~}_{\rm S}$ 
(self-energy) are given by 
\begin{eqnarray}
f^{~}_{\rm V}(x) & = & \left \{ \matrix{
\displaystyle \sqrt{x} \left [ 1 - (1+x) \ln \frac{1+x}{x} \right ] 
~~~~~ ({\rm SM}) \; , ~~~~ \cr
\displaystyle -\sqrt{x} \ln \frac{1+x}{x} 
~~~~~ ({\rm SUSY}) \; ;  ~~~~~~~~~~~~~~~~ \cr} \right .
\nonumber \\
f^{~}_{\rm S}(x) & = & \left \{ \matrix{
\displaystyle \frac{\sqrt{x}}{1-x} ~~~~~ ({\rm SM}) \; , ~~~~~~~~~~~~~ \cr
\displaystyle \frac{2\sqrt{x}}{1-x} ~~~~~ ({\rm SUSY}) \; . ~~~~~~~~~~ \cr} 
\right .
%       (4.18)
\end{eqnarray}
If $x\gg 1$ holds, we arrive at 
$f^{~}_{\rm S} \approx 2f^{~}_{\rm V} \approx -1/\sqrt{x}$ (SM) or 
$f^{~}_{\rm S} \approx 2f^{~}_{\rm V} \approx -2/\sqrt{x}$ (SUSY). 

The central idea of baryogenesis via leptogenesis is rather 
simple \cite{FY}. In the early universe, the heavy (right-handed) Majorana 
neutrinos $N_i$ are present in the primordial thermal bath. They may 
decay into leptons and scalars as soon as the universe becomes cold 
enough. As shown in Fig. 4.3, the decays of $N_i$ violate both the lepton 
number conservation and the CP symmetry. If such decays occur out of 
thermal equilibrium, a net lepton-antilepton asymmetry can be generated. 
Later on, this net lepton-antilepton asymmetry is converted in part to a 
net baryon-antibaryon asymmetry through the $(B+L)$-violating sphaleron 
processes \cite{Kuzmin}. For simplicity, we consider the case that
three heavy Majorana neutrinos $N_i$ have a hierarchical mass spectrum
($M_1 \ll M_2 \ll M_3$). Then the interactions of $N_1$ are in thermal 
equilibrium when $N_2$ and $N_3$ decay, and the CP-violating asymmetries 
produced in the decays of $N_2$ and $N_3$ (i.e., $\varepsilon_2$ and 
$\varepsilon_3$) can be erased before $N_1$ decays. The CP-violating 
asymmetry $\varepsilon_1$ produced by the out-of-equilibrium decay of 
$N_1$ survives, and it results in a net lepton-antilepton asymmetry 
\begin{equation}
Y_{\rm L} \; \equiv \; \frac{n^{~}_{\rm L} - n^{~}_{\rm\bar L}}{\bf s}
\; = \; \frac{d}{g^{~}_*} \varepsilon_1 \; ,
%       (4.19)
\end{equation}
where $g^{~}_* = 106.75$ (SM) or 228.75 (SUSY) is an effective number 
characterizing the relativistic degrees of freedom which contribute to 
the entropy {\bf s}, and $d$ accounts for the dilution effects induced 
by the lepton-number-violating wash-out processes \cite{Covi}. The 
lepton asymmetry $Y_{\rm L}$ is eventually converted into a net baryon 
asymmetry $Y_{\rm B}$ via nonperturbative sphaleron 
processes \cite{Kuzmin}. The explicit relation between 
$Y_{\rm L}$ (initial) and $Y_{\rm B}$ (equilibrium)
is given as \cite{Turner}
\begin{equation}
Y_{\rm B} \; = \; - c Y_{\rm L} \; = \; 
-c \frac{d}{g^{~}_*} \varepsilon_1 \; ,
%       (4.20)
\end{equation}
where $c=(8{\rm N}_f + 4{\rm N}_\phi)/(22{\rm N}_f + 13{\rm N}_\phi)$
with ${\rm N}_f$ being the number of fermion families and 
${\rm N}_\phi$ being the number of Higgs doublets. Taking
${\rm N}_f = 3$ and ${\rm N}_\phi =1$ for example, we obtain
$c \approx 1/3$. Note that the dilution factor $d$ can be computed by
integrating the full set of Boltzmann equations \cite{Boltzmann}. In
the literature \cite{King}, some useful analytical approximations 
for $d$ have frequently been made.

In the simple mechanism of thermal leptogenesis introduced above, the
baryon asymmetry of the universe is attributed to the 
out-of-equilibrium decay of the {\it lightest} heavy (right-handed) 
Majorana neutrino $N_1$. Another interesting scenario of thermal
leptogenesis is to consider the lepton-number-violating decays of
two heavy Majorana neutrinos, whose masses are approximately 
degenerate \cite{DEG}. Because the self-energy contribution to
$\varepsilon_i$ can significantly be enhanced in the case of
$M_i \approx M_j$, it is possible to generate the observed 
baryon-antibaryon asymmetry $Y_{\rm B}$ via the out-of-equilibrium
decays of two relatively 
light and moderately degenerate $N_i$. Such a scenario could allow the 
smallest mass of $N_i$ to be close to or below the maximum
reheating temperature of the universe after inflation in the generic
supergravity models. Of course, one may abandon the idea of thermal
leptogenesis and interpret the baryon asymmetry through the decays of
heavy (right-handed) Majorana neutrinos produced non-thermally by
the inflaton decay \cite{NON}. The non-thermal leptogenesis 
mechanism seems to be more speculative and less elegant than the
thermal leptogenesis mechanism, however. Thus we continue to focus
on the simple scenario of thermal leptogenesis in the following.

After spontaneous symmetry breaking, ${\cal L}_{\rm Y}$ in (4.16)
becomes ${\cal L}_{\rm mass}$ in (2.6), where 
$M_l = Y_l \langle \phi \rangle$ for the charged lepton mass 
matrix and $M_{\rm D} = Y_\nu \langle \phi \rangle$ for the Dirac
neutrino mass matrix. The scale of $M_l$ and $M_{\rm D}$ is 
characterized by the gauge symmetry breaking scale 
$v \equiv \langle \phi \rangle \approx 174$ GeV; but the scale of 
$M_{\rm R}$ may be much higher than $v$, because right-handed 
neutrinos are $\rm SU(2)_{\rm L}$ singlets and their mass term is 
not subject to the electroweak symmetry breaking. As already shown 
in section 2.1.2, the light (left-handed) neutrino mass matrix 
$M_\nu$ can be given in terms of $M_{\rm D}$ and $M_{\rm R}$ via 
the seesaw relation (2.9). Note that lepton flavor mixing at low 
energy scales stems from a nontrivial mismatch between the 
diagonalizations of $M_\nu$ and $M_l$, while the baryon asymmetry 
at high energy scales depends on complex $Y_\nu$ and $M_{\rm R}$ 
in the thermal leptogenesis mechanism. To see the latter more 
clearly, let us diagonalize the symmetric mass matrix $M_{\rm R}$
by a unitary transformation:
\small
\begin{equation}
U^\dagger_{\rm R} M_{\rm R} U^*_{\rm R} \; = \; \left ( \matrix{
M_1 & 0 & 0 \cr
0 & M_2 & 0 \cr
0 & 0 & M_3 \cr} \right ) \; ,
%	(4.21)
\end{equation}
\normalsize
where $M_i$ are the physical masses of $N_i$. Provided three heavy 
Majorana neutrinos have a strong mass hierarchy 
(i.e., $M_1 \ll M_2 \ll M_3$), the CP-violating asymmetry 
between $N_1 \rightarrow l + \phi^\dagger$ and 
$N_1 \rightarrow l^{\rm c} + \phi$ decays in (4.17) can be 
simplified to \cite{XingLEP}
\begin{equation}
\varepsilon_1 \; \approx \; - \frac{3}{16\pi v^2} \cdot
\frac{M_1}{\left [U^T_{\rm R} M^\dagger_{\rm D} M_{\rm D} 
U^*_{\rm R} \right ]_{11}} 
\sum^3_{j=2} \frac{{\rm Im} 
\left ( \left [U^T_{\rm R} M^\dagger_{\rm D}
M_{\rm D} U^*_{\rm R} \right ]_{1j} \right )^2}{M_j} \; .
%    	(4.22)
\end{equation}
In obtaining (4.22), we have taken into account the point that 
the Dirac neutrino Yukawa coupling matrix $Y_\nu$ takes the form 
$M_{\rm D}U^*_{\rm R}/v$ in the physical basis where $M_{\rm R}$
is diagonal. One can see that $\varepsilon_1$ 
depends on the complex phases of $M_{\rm D}$ and $U_{\rm R}$.
The only possible relationship between $\varepsilon_1$ and the
lepton flavor mixing matrix $V$ at low energy scales is due to 
the seesaw relation (2.9), which links $M_\nu$ to $M_{\rm D}$ 
and $M_{\rm R}$. Therefore we conclude that there is no direct 
connection between CP violation in heavy Majorana neutrino decays 
($\varepsilon_i$) and that in neutrino oscillations ($J$). 
Such a general conclusion was first drawn by Buchm$\rm\ddot{u}$ller 
and Pl$\rm\ddot{u}$macher in Ref. \cite{Buch}. Recently some other 
authors have carried out more delicate analyses and reached 
the same conclusion.

Depending on the specific flavor basis that we choose in model 
building, $\varepsilon_i$ and $V$ can either be completely 
disconnected or somehow connected. To illustrate, we consider two 
extreme cases \cite{ICHEP}:
\begin{itemize}
\item 	In the flavor basis where $M_\nu$ is diagonal (i.e., 
lepton flavor mixing and CP violation at low energy scales arise 
solely from the charged lepton sector \cite{FX96}), 
we find that $\varepsilon_1$ has nothing to do with $V$.
In this special case, less fine-tuning is expected in building a
phenomenological model which can simultaneously interpret the
baryon asymmetry of the universe and lepton flavor mixing at low
energy scales.
\item	In the flavor basis where both $M_l$ and $M_{\rm R}$ are
diagonal (i.e., lepton flavor mixing and CP violation at low
energy scales arise solely from the neutrino sector), we find that
$\varepsilon_1$ can indirectly be linked to $V$ through the seesaw 
relation in (2.9). In this special case, it is highly nontrivial
to build a predictive model or ansatz which can simultaneously 
interpret the observed baryon asymmetry of the universe and current 
neutrino oscillation data.
\end{itemize}
It is worth remarking that the correlation between high-energy
physics and low-energy physics may offer us a valuable opportunity 
to probe the former from the latter, or vice versa \cite{Bari}. 
When a specific model is 
built, however, the textures of $M_{\rm R}$ and $M_{\rm D}$ 
have to be carefully chosen or fine-tuned to guarantee acceptable
agreement between the model predictions and the observational or
experimental data (see Refs. \cite{XingLEP,BW} for example). 
Many different models or ans$\rm\ddot{a}$tze on leptogenesis and 
neutrino oscillations have recently been proposed \cite{LEPreview}, 
but some of them are rather preliminary and speculative.

An important question is whether the leptogenesis mechanism can
experimentally be proved. The answer is unfortunately negative.
But baryogenesis via leptogenesis might become a conceivable and 
acceptable mechanism to interpret the matter-antimatter 
asymmetry of the universe \cite{Murayama02}, if (1) the electroweak 
baryogenesis scenario is ruled out; (2) the Majorana nature of massive
neutrinos is verified (e.g., by measuring the neutrinoless double 
beta decay); and (3) the leptonic CP-violating effect is observed 
(e.g., in the future long-baseline neutrino oscillation experiments).

\section{Concluding Remarks}
\setcounter{equation}{0}
\setcounter{figure}{0}

We have presented an overview of recent progress in the 
phenomenological study of neutrino masses, lepton flavor mixing
and CP violation. Particular attention has been paid to the
model-independent properties of massive neutrinos in vacuum and
in matter. 

With the help of current experimental and observational data,
we have obtained some enlightening information about the neutrino
mass spectrum. To pin down the absolute mass scale of three
active neutrinos, one has to make much more efforts to detect
the tritium beta decay and the neutrinoless double beta decay.
Precise cosmological data are also expected to play an important 
role in determining the absolute values of neutrino masses.
The relative magnitudes of three neutrino masses can be fixed
by means of a variety of long-baseline neutrino oscillation 
experiments in the coming years.

Different from quark flavor mixing, lepton flavor mixing involves
two large mixing angles. Whether the largeness of neutrino mixing
angles is associated with a relativly weak hierarchy of neutrino
masses remains an open question. We hope that both the smallest 
angle of lepton flavor mixing and the Dirac phase of CP violation 
can be measured in the long-baseline neutrino oscillation experiments.
It seems hopeless to separately determine two Majorana
phases of CP violation from the measurements of the neutrinoless
double beta decay and other lepton-number-violating processes.
Nevertheless, it is extremely important to realize such 
measurements in order to identify the Majorana nature of massive
neutrinos and to shed light on the nontrivial features of lepton
number violation.

To conclude, the robust experimental evidence for neutrino masses and 
lepton flavor mixing strongly indicates that the standard electroweak 
model is actually incomplete. This incompleteness motivates us to
open a new window to go beyond the standard model. Although there
have been a number of unresolved questions associated with massive
neutrinos, we are certainly paving the way for satisfactory answers 
to them. A convincing and predictive theory of massive neutrinos 
should be achieved in the foreseeable future. Such a theory, which 
must include new physics of both leptons and quarks at high energy
scales, is expected to offer us a deeper insight into the generation 
of fermion masses, the pattern of flavor mixing and the origin of 
CP violation.

\vspace{0.3cm}

{\bf Acknowledgements:} 
I am indebted to Dr. K.K. Yim for inviting me to write this
review article, and to Dr. A.H. Chan for his enthusiasm for
my progress. I would like to thank H. Fritzsch and W.L. Guo for 
very enjoyable collaboration on the topic of massive neutrinos. 
I am also grateful to K. Hagiwara and the KEK theory group, where 
part of this paper was written. This work was supported in part 
by the National Natural Science Foundation of China.  

\appendix

\section{Mixing between Active and Sterile Neutrinos}

In addition to the robust evidence for atmospheric and solar
neutrino oscillations accumulated from the SK \cite{SK}, 
SNO \cite{SNO}, KamLAND \cite{KM} and K2K \cite{K2K} experiments, 
$\overline{\nu}_\mu \rightarrow \overline{\nu}_e$ and
$\nu_\mu \rightarrow \nu_e$ transitions have been observed
by the LSND Collaboration \cite{LSND}. The LSND data can also be 
interpreted in the assumption of neutrino oscillations,
whose mass-squared difference and mixing factor read as
\begin{equation}
\Delta m^2_{\rm LSND} \; \sim \; 1 ~ {\rm eV}^2 \; ,
~~~~~
\sin^2 2\theta_{\rm LSND} \; \sim \; 10^{-3} - 10^{-2} \; .
%	(A1)
\end{equation}
Because solar, atmospheric and LSND neutrino oscillations
involve three distinct mass-squared differences
($\Delta m^2_{\rm sun} \ll \Delta m^2_{\rm atm} \ll
\Delta m^2_{\rm LSND}$), a simultaneous interpretation of
them requires the introduction of a light {\it sterile} neutrino
%%%%%%%%%%%%%%%%%%%%%%%%
\footnote{A {\it sterile} particle means that it has
vanishing or extremely feeble couplings to the
other particles in the standard model. 
Instead of introducing a light sterile neutrino, a few
more far-fetched ideas (such as the violation of CPT 
symmetry in the neutrino sector \cite{CPT} and the 
lepton-number-violating muon decay \cite{Babu}) have been 
proposed in the literature to simultaneously interpret the 
solar, atmospheric and LSND anomalies within the 
three-neutrino mixing scheme.}.
%%%%%%%%%%%%%%%%%%%%%%%%
There are two typical categories
of four-neutrino mixing schemes and their sample mass spectra
are shown in Fig. A.1: (a) the (2+2) scheme with two pairs of 
nearly degenerate massive neutrinos, whose mass-squared gap
is characterized by $\Delta m^2_{\rm LSND}$; and (b) the 
(3+1) scheme with a triplet of nearly degenerate massive 
neutrinos and an isolated massive neutrino, whose mass-squared
gap is also characterized by $\Delta m^2_{\rm LSND}$. 
Although each four-neutrino mixing scheme consists of three
independent mass-squared differences and six flavor mixing angles,
its parameter space can be tightly constrained by current 
experimental data. The global analyses of all available
neutrino oscillation data \cite{4v} have shown that the (2+2) 
mixing scenario is strongly disfavored, while the (3+1) mixing 
scenario is marginally allowed. However, the recent cosmological 
upper bound on the sum of all light neutrino masses in (2.22)
is too low to be consistent with the LSND result given in (A.1),
at least at the $95\%$ confidence level \cite{MAP}. 
It is argued by Hannestad \cite{Hannestad} and Giunti \cite{MAP} 
that the (3+1) mixing scheme may still survive in a very tiny 
parameter space, if the relevant cosmological data are considered 
at a higher confidence level. The upcoming 
MiniBooNE experiment \cite{MB} is crucial to firmly confirm or
disprove the LSND measurement. 
%%%%%%%%%%%%%%%%%%%% Fig. A.1 %%%%%%%%%%%%%%%%
\begin{figure}[t]
\vspace{-7cm}
\epsfig{file=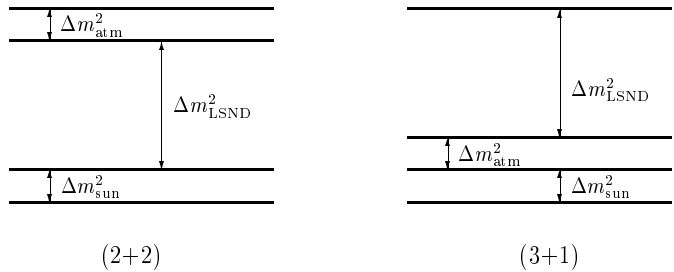,bbllx=0.5cm,bblly=0cm,bburx=20cm,bbury=27cm,%
width=15cm,height=22cm,angle=0,clip=}
\vspace{-11.3cm}
\caption{Sample mass spectra of two four-neutrino mixing schemes, where
each line denotes $m^2_i$ (for $i=0,1,2,3$).}
\end{figure}
%%%%%%%%%%%%%%%%%%%%%%%%%%%%%%%%%%%%%%%%%%%%

Before a definitely negative conclusion can be drawn from 
MiniBooNE, we do not think that the LSND data should be completely
discarded. In particular, it is worthwhile to study the 
four-neutrino mixing scenarios in a way without special theoretical
biases and (or) empirical assumptions. Starting from such a point
of view, we present some model-independent results on the 
description of four-neutrino mixing and CP violation in this Appendix.

\subsection{Standard Parametrization}

Let us begin with a generic $\rm SU(2)_L \times U(1)_Y$ model of 
electroweak interactions, in which there exist $n$ charged 
leptons belonging to isodoublets, $n$ active neutrinos belonging 
to isodoublets, and $n'$ sterile neutrinos belonging to isosinglets. 
The charged-current weak interactions of leptons are then associated 
with a rectangular flavor mixing matrix of $n$ rows and $(n+n')$ 
columns \cite{Valle}. Without loss of generality, one may choose to 
identify the flavor eigenstates of charged leptons with their mass 
eigenstates. In this specific basis, the $n\times (n+n')$ lepton 
mixing matrix links the neutrino flavor eigenstates directly to the 
neutrino mass eigenstates. Although sterile neutrinos do not 
participate in normal weak interactions, they may oscillate among 
themselves and with active neutrinos. Once the latter is concerned 
we are led to a more general $(n+n')\times (n+n')$ lepton
flavor mixing matrix \cite{FXReview}, defined as $V$ in the chosen 
flavor basis. For the mixing of one sterile neutrino ($\nu_s$) 
and three active neutrinos ($\nu_e, \nu_\mu, \nu_\tau$), the explicit 
form of $V$ can be written as 
\begin{equation}
\left ( \matrix{
\nu_s \cr
\nu_e \cr
\nu_\mu \cr
\nu_\tau \cr} \right ) \; = \; \left ( \matrix{
V_{s0} & V_{s1} & V_{s2} & V_{s3} \cr
V_{e0} & V_{e1} & V_{e2} & V_{e3} \cr
V_{\mu 0} & V_{\mu 1} & V_{\mu 2} & V_{\mu 3} \cr
V_{\tau 0} & V_{\tau 1} & V_{\tau 2} & V_{\tau 3} \cr} \right ) 
\left ( \matrix{
\nu_0 \cr 
\nu_1 \cr
\nu_2 \cr
\nu_3 \cr} \right ) \; ,
%       (A2)
\end{equation}
where $\nu_i$ (for $i=0,1,2,3$) denote the mass eigenstates of four 
neutrinos. If neutrinos are Dirac particles, $V$ can be parametrized 
in terms of six mixing angles and three phase angles. If neutrinos 
are Majorana particles, however, three additional phase angles are 
required to get a full parametrization of $V$. 
We totally need six complex rotation matrices, denoted as $R_{01}$, 
$R_{02}$, $R_{03}$, $R_{12}$, $R_{13}$ and $R_{23}$, which
correspond to simple rotations with angles $\theta_{ij}$
in the (0,1), (0,2), (0,3), (1,2), (1,3) and (2,3) planes. 
For simplicity, we assume that each $R_{ij}$ involves only a single 
phase angle $\delta_{ij}$ associated with its $\sin\theta_{ij}$ term. 
Explicitly, we have
\begin{eqnarray}
R_{01}(\theta_{01}, \delta_{01}) & = &
\left ( \matrix{
c_{01}  & \hat{s}^*_{01}        & 0     & 0 \cr
- \hat{s}_{01}        & c_{01}        & 0     & 0 \cr
0       & 0     & 1     & 0 \cr
0       & 0     & 0     & 1 \cr} \right ) \; ,
\nonumber \\
R_{02}(\theta_{02}, \delta_{02}) & = &
\left ( \matrix{
c_{02}  & 0     & \hat{s}^*_{02}        & 0 \cr
0       & 1     & 0     & 0 \cr
- \hat{s}_{02}      & 0     & c_{02}  & 0 \cr
0       & 0     & 0     & 1 \cr} \right ) \; ,
\nonumber \\
R_{03}(\theta_{03}, \delta_{03}) & = &
\left ( \matrix{
c_{03}  & 0     & 0     & \hat{s}^*_{03} \cr
0       & 1     & 0     & 0 \cr
0       & 0     & 1     & 0 \cr
- \hat{s}_{03} & 0     & 0     & c_{03} \cr } \right ) \; , 
\nonumber \\
R_{12}(\theta_{12}, \delta_{12}) & = &
\left ( \matrix{
1       & 0     & 0     & 0 \cr
0       & c_{12}        & \hat{s}^*_{12}        & 0 \cr
0       & - \hat{s}_{12}       & c_{12}        & 0 \cr
0       & 0     & 0     & 1 \cr} \right ) \; , 
\nonumber \\
R_{13}(\theta_{13}, \delta_{13}) & = &
\left ( \matrix{
1       & 0     & 0     & 0 \cr
0       & c_{13}        & 0     & \hat{s}^*_{13} \cr
0       & 0     & 1     & 0 \cr
0       & - \hat{s}_{13}       & 0     & c_{13} \cr} \right ) \; , 
\nonumber \\
R_{23}(\theta_{23}, \delta_{23}) & = &
\left ( \matrix{
1       & 0     & 0     & 0 \cr
0       & 1     & 0     & 0 \cr
0       & 0     & c_{23}        & \hat{s}^*_{23} \cr
0       & 0     & - \hat{s}_{23}       & c_{23} \cr } \right ) \; ,
%       (A3) 
\end{eqnarray}
where $\hat{s}_{ij} \equiv \sin\theta_{ij} ~ e^{i\delta_{ij}}$ and
$c_{ij} \equiv \cos\theta_{ij}$. There exist numerous different 
ways to arrange the products of these rotation matrices \cite{FX98}, 
such that the resultant flavor mixing matrix $V$
covers the whole $4\times 4$ space. After a proper assignment of the
relevant rotation angles, one can find that only sixteen 
parametrizations of $V$ are structurally distinct: four of them 
have the $\cos\theta_{ij}$ terms in the $(i,i)$ positions
of $V$; and twelve of them have the $\sin\theta_{ij}$ terms
in the twelve different $(i,j)$ positions of $V$ (for $i\neq j$). 

To avoid unnecessary complication, we shall not write out a lengthy
list of the sixteen parametrizations of $V$. Instead
we only take an instructive example for illustration.
Similar to the representation in (3.23), the $4\times 4$ neutrino 
mixing matrix $V$ can be parametrized as 
$V = R_{23} \otimes R_{13} \otimes R_{03} \otimes R_{12} 
\otimes R_{02} \otimes R_{01}$ \cite{Dai}; i.e., 
\begin{equation}
V \; = \; \left ( \matrix{
c_{01}c_{02}c_{03} 
& c_{02}c_{03}\hat{s}_{01}^*
& c_{03}\hat{s}_{02}^* 
& \hat{s}_{03}^* 
\cr\cr
-c_{01}c_{02}\hat{s}_{03}\hat{s}_{13}^*
& -c_{02}\hat{s}_{01}^*\hat{s}_{03}\hat{s}_{13}^*
& -\hat{s}_{02}^*\hat{s}_{03}\hat{s}_{13}^*
& c_{03}\hat{s}_{13}^* 
\cr
-c_{01}c_{13}\hat{s}_{02}\hat{s}_{12}^*
& -c_{13}\hat{s}_{01}^*\hat{s}_{02}\hat{s}_{12}^*
& +c_{02}c_{13}\hat{s}_{12}^*
& 
\cr
-c_{12}c_{13}\hat{s}_{01}
& +c_{01}c_{12}c_{13}
&
&
\cr\cr
-c_{01}c_{02}c_{13}\hat{s}_{03}\hat{s}_{23}^*
& -c_{02}c_{13}\hat{s}_{01}^*\hat{s}_{03}\hat{s}_{23}^*
& -c_{13}\hat{s}_{02}^*\hat{s}_{03}\hat{s}_{23}^*
& c_{03}c_{13}\hat{s}_{23}^*
\cr
+c_{01}\hat{s}_{02}\hat{s}_{12}^*\hat{s}_{13}\hat{s}_{23}^*
& +\hat{s}_{01}^*\hat{s}_{02}\hat{s}_{12}^*\hat{s}_{13}\hat{s}_{23}^*
& -c_{02}\hat{s}_{12}^*\hat{s}_{13}\hat{s}_{23}^*
&
\cr
-c_{01}c_{12}c_{23}\hat{s}_{02}
& -c_{12}c_{23}\hat{s}_{01}^*\hat{s}_{02}
& +c_{02}c_{12}c_{23}
&
\cr
+c_{12}\hat{s}_{01}\hat{s}_{13}\hat{s}_{23}^*
& -c_{01}c_{12}\hat{s}_{13}\hat{s}_{23}^*
&
&
\cr 
+c_{23}\hat{s}_{01}\hat{s}_{12} 
& -c_{01}c_{23}\hat{s}_{12}
&
&
\cr\cr
-c_{01}c_{02}c_{13}c_{23}\hat{s}_{03}
& -c_{02}c_{13}c_{23}\hat{s}_{01}^*\hat{s}_{03}
& -c_{13}c_{23}\hat{s}_{02}^*\hat{s}_{03}
& c_{03}c_{13}c_{23}
\cr
+c_{01}c_{23}\hat{s}_{02}\hat{s}_{12}^*\hat{s}_{13}
& +c_{23}\hat{s}_{01}^*\hat{s}_{02}\hat{s}_{12}^*\hat{s}_{13}
& -c_{02}c_{23}\hat{s}_{12}^*\hat{s}_{13}
&
\cr
+c_{01}c_{12}\hat{s}_{02}\hat{s}_{23}
& +c_{12}\hat{s}_{01}^*\hat{s}_{02}\hat{s}_{23}
& -c_{02}c_{12}\hat{s}_{23}
&
\cr
+c_{12}c_{23}\hat{s}_{01}\hat{s}_{13}
& -c_{01}c_{12}c_{23}\hat{s}_{13}
&
&
\cr
-\hat{s}_{01}\hat{s}_{12}\hat{s}_{23}
& +c_{01}\hat{s}_{12}\hat{s}_{23}
&
&
\cr } \right ) \; .
%	(A4)
\end{equation}
Without loss of generality, the six mixing angles $\theta_{ij}$ can 
all be arranged to lie in the first quadrant. The six CP-violating 
phases $\delta_{ij}$ may take arbitrary values between $0$ and $2\pi$.
Note that $V$ can also be decomposed into a Dirac-like flavor 
mixing matrix with six rotation angles and three CP-violating 
phases, multiplied by a diagonal phase matrix with three Majorana 
phases. In normal neutrino-neutrino and antineutrino-antineutrino 
oscillations, CP-violating phases of the Majorana nature are unable 
to be measured.

\subsection{Invariants of CP Violation}

No matter whether neutrinos are Dirac or Majorana particles, one 
may define the Jarlskog rephasing invariants of four-neutrino 
mixing \cite{Jarlskog}, which govern the CP- and T-violating effects 
in normal neutrino-neutrino and antineutrino-antineutrino oscillations.
To be explicit, we have
\begin{equation}
J^{ij}_{\alpha\beta} \; \equiv \; {\rm Im} \left ( V_{\alpha i} V_{\beta j}
V^*_{\alpha j} V^*_{\beta i} \right ) \; ,
%	(A5)
\end{equation}
where the Greek subscripts run over $(s, e, \mu, \tau)$ and the Latin 
superscripts run over $(0, 1, 2, 3)$. Of course,
$J^{ii}_{\alpha\beta} = J^{ij}_{\alpha\alpha} =0$ and
$J^{ij}_{\alpha\beta} = -J^{ji}_{\alpha\beta} 
= -J^{ij}_{\beta\alpha} = J^{ji}_{\beta\alpha}$ hold by definition. 
The unitarity of $V$ leads to the following correlation equations of 
$J^{ij}_{\alpha \beta}$:
\begin{equation}
\sum_i J^{ij}_{\alpha\beta} \; = \; 
\sum_j J^{ij}_{\alpha\beta} \; = \; 
\sum_\alpha J^{ij}_{\alpha\beta} \; = \;
\sum_\beta J^{ij}_{\alpha\beta} \; =\; 0 \; .
%	(A6)
\end{equation}
Hence there are totally nine independent $J^{ij}_{\alpha\beta}$, whose
magnitudes depend only upon three of the six CP-violating phases or
their combinations in a specific parametrization of $V$. 

We use the standard parametrization of $V$ in (A.4) to calculate
$J^{ij}_{\alpha\beta}$. After some lengthy but straightforward
calculations, we obtain the exact analytical  expressions of nine 
independent $J^{ij}_{\alpha \beta}$ \cite{GX02a}:
\footnotesize
\begin{eqnarray}
J^{02}_{\tau s} & = & c^2_{01} c_{02} c^2_{03} c_{12} c_{13} c_{23}
s_{02} s_{03} s_{23} \sin \phi_x + c_{01} c_{02} c^2_{03} c_{12} c_{13}
c^2_{23} s_{01} s^2_{02} s_{03} s_{13} \sin \phi_y 
\nonumber \\
&& + \left (c^2_{23} s^2_{13} - s^2_{23} \right ) c_{01} c^2_{02} c^2_{03}
c_{12} s_{01} s_{02} s_{12} \sin \phi_z - c_{01} c^2_{02} c^2_{03}
c^2_{12} c_{23} s_{01} s_{02} s_{13} s_{23} \sin (\phi_x - \phi_y)
\nonumber \\
&& - c_{01} c_{02} c^2_{03} c_{13} c_{23} s_{01} s^2_{02} s_{03}
s_{12} s_{23} \sin (\phi_x + \phi_z) + c^2_{01} c_{02} c^2_{03} c_{13}
c^2_{23} s_{02} s_{03} s_{12} s_{13} \sin (\phi_y - \phi_z)
\nonumber \\
&& - c_{01} c^2_{02} c^2_{03} c_{23} s_{01} s_{02} s^2_{12}
s_{13} s_{23} \sin (\phi_x - \phi_y + 2\phi_z) \; ,
\nonumber \\
J^{03}_{\tau s} & = & -c^2_{01} c_{02} c^2_{03} c_{12} c_{13}
c_{23} s_{02} s_{03} s_{23} \sin \phi_x -
c_{01} c_{02} c^2_{03} c_{12} c_{13} c^2_{23} s_{01} s_{03} s_{13}
\sin \phi_y 
\nonumber \\
&& + c_{01} c_{02} c^2_{03} c_{13} c_{23} s_{01} s_{03} s_{12}
s_{23} \sin (\phi_x + \phi_z) - c^2_{01} c_{02} c^2_{03} c_{13} c^2_{23}
s_{02} s_{03} s_{12} s_{13} \sin (\phi_y - \phi_z) \; ,
\nonumber \\
J^{23}_{\tau s} & = & c_{02} c^2_{03} c_{12} c_{13} c_{23} s_{02}
s_{03} s_{23} \sin \phi_x + c_{02} c^2_{03} c_{13} c^2_{23}
s_{02} s_{03} s_{12} s_{13} \sin (\phi_y - \phi_z) \; ;
%	(A7)
\end{eqnarray}
\normalsize
and 
\footnotesize
\begin{eqnarray}
J^{02}_{se} & = & c_{01} c_{02} c^2_{03} c_{12} c_{13} s_{01} 
s^2_{02} s_{03} s_{13} \sin \phi_y - c_{01} c^2_{02} c^2_{03}
c_{12} c^2_{13} s_{01} s_{02} s_{12} \sin \phi_z
\nonumber \\
&& + c^2_{01} c_{02} c^2_{03} c_{13} s_{02} s_{03} s_{12} s_{13}
\sin (\phi_y - \phi_z) \; ,
\nonumber \\
J^{13}_{se} & = & c_{01} c_{02} c^2_{03} c_{12} c_{13} s_{01} s_{03}
s_{13} \sin \phi_y - c_{02} c^2_{03} c_{13} s^2_{01} s_{02} s_{03}
s_{12} s_{13} \sin (\phi_y - \phi_z) \; ,
\nonumber \\
J^{23}_{se} & = & c_{02} c^2_{03} c_{13} s_{02} s_{03} s_{12} s_{13} 
\sin (\phi_y - \phi_z) \; ;
%	(A8)
\end{eqnarray}
\normalsize
as well as
\footnotesize
\begin{eqnarray}
J^{12}_{e\mu} & = & - \left ( c^2_{01} c^2_{12} s^2_{13} -
c^2_{01} c^2_{13} s^2_{12} - s^2_{01} s^2_{03} s^2_{13} + 
s^2_{01} s^2_{12} \right ) c_{02} c_{12} c_{13} c_{23} s_{02}
s_{03} s_{23} \sin \phi_x
\nonumber \\
&& + \left ( c^2_{02} c^2_{23} s^2_{12} - c^2_{12} c^2_{23} s^2_{02}
+ s^2_{02} s^2_{03} s^2_{23} - s^2_{12} s^2_{23} \right )
c_{01} c_{02} c_{12} c_{13} s_{01} s_{03} s_{13} \sin \phi_y
\nonumber \\
&& + \left ( c^2_{13} c^2_{23} - c^2_{13} s^2_{03} s^2_{23}
- c^2_{23} s^2_{03} s^2_{13} + s^2_{03} s^2_{13} s^2_{23}  \right )
c_{01} c^2_{02} c_{12} s_{01} s_{02} s_{12} \sin \phi_z 
\nonumber \\
&& + \left ( c^2_{02} s^2_{13} - c^2_{13} s^2_{02} \right )
c_{01} c^2_{12} c_{23} s_{01} s_{02} s^2_{03} s_{13} s_{23}
\sin (\phi_x - \phi_y)
\nonumber \\
&& + \left ( c^2_{12} - c^2_{02} s^2_{13} - c^2_{13} s^2_{02} + 
s^2_{02} s^2_{03} s^2_{13} \right ) c_{01} c_{02} c_{13} c_{23} s_{01} s_{03}
s_{12} s_{23} \sin (\phi_x + \phi_z)
\nonumber \\
&& - \left ( c^2_{01} c^2_{12} c^2_{23} - c^2_{01} c^2_{12} s^2_{23}
- c^2_{12} c^2_{23} s^2_{01} + s^2_{01} s^2_{03} s^2_{23}
- s^2_{01} s^2_{12} s^2_{23} \right ) c_{02} c_{13} s_{02} s_{03}
s_{12} s_{13} \sin (\phi_y - \phi_z)
\nonumber \\
&& + \left ( c^2_{01} c^2_{02} c^2_{13} - c^2_{01} c^2_{13} s^2_{02} s^2_{03}
- c^2_{02} c^2_{13} s^2_{01} s^2_{03} + c^2_{13} s^2_{01} s^2_{02} 
s^2_{03} \right ) c_{12} c_{23} s_{12} s_{13} s_{23} 
\sin (\phi_x - \phi_y + \phi_z) 
\nonumber \\
&& - \left ( c^2_{02} c^2_{13} s^2_{12} - 
c^2_{02} s^2_{03} s^2_{12} s^2_{13} - 
c^2_{13} s^2_{02} s^2_{03} s^2_{12} \right ) 
c_{01} c_{23} s_{01} s_{02} s_{13} s_{23} 
\sin (\phi_x - \phi_y + 2\phi_z)
\nonumber \\
&& - c_{01} c^2_{02} c^2_{13} c_{23} s_{01} s_{02} s^2_{03} s_{13} s_{23}
\sin (\phi_x + \phi_y)
+ c_{01} c_{02} c^2_{12} c_{13} c_{23} s_{01} s^2_{02} s_{03} 
s_{12} s_{23} \sin (\phi_x - \phi_z)
\nonumber \\
&& + \left ( c^2_{23} - s^2_{23} \right ) c_{01} c_{02} c_{12} c_{13}
s_{01} s^2_{02} s_{03} s^2_{12} s_{13} \sin (\phi_y - 2\phi_z)
\nonumber \\
&& - \left (c^2_{01} - s^2_{01} \right ) c_{02} c_{12} c_{13} c_{23}
s_{02} s_{03} s^2_{12} s^2_{13} s_{23} 
\sin (\phi_x - 2\phi_y + 2\phi_z)
\nonumber \\
&& - c_{01} c_{02} c^2_{12} c_{13} c_{23} s_{01} s_{03} s_{12} s^2_{13}
s_{23} \sin (\phi_x - 2\phi_y + \phi_z)
\nonumber \\
&& + c_{01} c_{02} c_{13} c_{23} s_{01} s^2_{02} s_{03} s^2_{12}
s^2_{13} s_{23} \sin (\phi_x - 2\phi_y + 3\phi_z) \; ,
\nonumber \\
J^{13}_{e\mu} & = & c_{02} c^2_{03} c_{12} c_{13} c_{23} s^2_{01}
s_{02} s_{03} s^2_{13} s_{23} \sin \phi_x +
c_{01} c_{02} c^2_{03} c_{12} c_{13} s_{01} s_{03} s_{13} s^2_{23}
\sin \phi_y
\nonumber \\
&& - c_{01} c^2_{03} c^2_{12} c^2_{13} c_{23} s_{01} s_{02} s_{13}
s_{23} \sin (\phi_x - \phi_y) + c_{01} c_{02} c^2_{03} c_{13} c_{23} s_{01}
s_{03} s_{12} s^2_{13} s_{23} \sin (\phi_x + \phi_z) 
\nonumber \\
&& - c_{02} c^2_{03} c_{13} s^2_{01} s_{02} s_{03} s_{12} s_{13}
s^2_{23} \sin (\phi_y - \phi_z) 
+ c_{01} c^2_{03} c^2_{13} c_{23} s_{01} s_{02} s^2_{12} 
s_{13} s_{23} \sin (\phi_x - \phi_y + 2\phi_z) 
\nonumber \\
&& - \left (c^2_{01} - s^2_{01} s^2_{02} \right )
c^2_{03} c_{12} c^2_{13} c_{23} s_{12} s_{13} s_{23} 
\sin (\phi_x - \phi_y + \phi_z) \; ,
\nonumber \\
J^{23}_{e\mu} & = & - c_{02} c^2_{03} c_{12} c_{13} c_{23} s_{02}
s_{03} s^2_{13} s_{23} \sin \phi_x + c_{02} c^2_{03} c_{13} s_{02}
s_{03} s_{12} s_{13} s^2_{23} \sin (\phi_y - \phi_z)
\nonumber \\
&& + c^2_{02} c^2_{03} c_{12} c^2_{13} c_{23} s_{12} s_{13} s_{23}
\sin (\phi_x - \phi_y + \phi_z) \; ,
%	(A9)
\end{eqnarray}
\normalsize
where 
\begin{eqnarray}
\phi_x & \equiv & \delta_{03} - \delta_{02} - \delta_{23} \; ,
\nonumber \\
\phi_y & \equiv & \delta_{03} - \delta_{01} - \delta_{13} \; ,
\nonumber \\
\phi_z & \equiv & \delta_{02} - \delta_{01} - \delta_{12} \; .
%	(A10)
\end{eqnarray}
With the help of (A.6), one may easily derive the 
expressions of all the other rephasing invariants of CP and T
violation from (A.7), (A.8) and (A.9). The results obtained 
above are expected to be very useful for a systematic
study of CP- and T-violating effects in the four-neutrino mixing 
models. The same results are also applicable for the discussion of
CP and T violation in the four-quark mixing models \cite{Chau,FP}.

Note that all CP- and T-violating observables in neutrino 
oscillations must be related linearly to $J^{ij}_{\alpha\beta}$. 
To see this point more clearly, we consider that a neutrino $\nu_\alpha$ 
converts to another neutrino $\nu^{~}_\beta$ in vacuum.
The probability of this conversion is given by 
\small
\begin{equation}
P(\nu_\alpha \rightarrow \nu^{~}_\beta) \; = \;
\delta_{\alpha\beta} - 4 \sum_{i<j} \left [ {\rm Re} \left (
V_{\alpha i} V_{\beta j} V^*_{\alpha j} V^*_{\beta i} \right ) 
\sin^2 F_{ji} \right ] - 2 \sum_{i<j} \left ( J^{ij}_{\alpha\beta}
\sin 2 F_{ji} \right ) \; ,
%	(A11)
\end{equation}
\normalsize
where $F_{ji} \equiv 1.27 \Delta m^2_{ji} L/E$ with
$\Delta m^2_{ji} \equiv m^2_j - m^2_i$, $L$ stands for the baseline 
length (in unit of km), and $E$ is the neutrino beam energy (in unit 
of GeV). CPT invariance assures that the transition probabilities 
$P(\nu^{~}_\beta \rightarrow \nu_\alpha)$ and 
$P(\overline{\nu}_\alpha \rightarrow \overline{\nu}^{~}_\beta)$ are
identical, and they can directly be read off from (A.11) through the 
replacement $J^{ij}_{\alpha\beta} \Longrightarrow -J^{ij}_{\alpha\beta}$
(i.e., $V \Longrightarrow V^*$). Thus the CP-violating asymmetry
between $P(\nu_\alpha \rightarrow \nu^{~}_\beta)$ and
$P(\overline{\nu}_\alpha \rightarrow \overline{\nu}^{~}_\beta)$
is equal to the T-violating asymmetry between
$P(\nu_\alpha \rightarrow \nu^{~}_\beta)$ and
$P(\nu^{~}_\beta \rightarrow \nu_\alpha)$. The latter can be explicitly
and compactly expressed as follows \cite{GX02a}:
\begin{eqnarray}
\Delta P_{\alpha\beta} & \equiv &
P(\nu^{~}_\beta \rightarrow \nu^{~}_\alpha) ~ - ~ 
P(\nu^{~}_\alpha \rightarrow \nu^{~}_\beta) \;
\nonumber \\
& = & 16\left (J^{12}_{\alpha\beta} \sin F_{21} \sin F_{31} \sin F_{32}
+ J^{01}_{\alpha\beta} \sin F_{10} \sin F_{30} \sin F_{31}
\right . 
\nonumber \\
&& ~ + \left . J^{02}_{\alpha\beta} \sin F_{20} \sin F_{30} \sin F_{32} 
\right ) \; .
%	(A12)
\end{eqnarray}
Equivalently, one may obtain
\begin{eqnarray}
\Delta P_{\alpha\beta} & = & 
16\left (J^{23}_{\alpha\beta} \sin F_{21} \sin F_{31} \sin F_{32}
- J^{02}_{\alpha\beta} \sin F_{10} \sin F_{20} \sin F_{21}
\right . 
\nonumber \\
&& ~ - \left . J^{03}_{\alpha\beta} \sin F_{10} \sin F_{30} \sin F_{31} 
\right ) \; ,
%	(A13)
\end{eqnarray}
or
\begin{eqnarray}
\Delta P_{\alpha\beta} & = & 
16\left (J^{31}_{\alpha\beta} \sin F_{21} \sin F_{31} \sin F_{32}
+ J^{01}_{\alpha\beta} \sin F_{10} \sin F_{20} \sin F_{21}
\right . 
\nonumber \\
&&  ~ - \left . J^{03}_{\alpha\beta} \sin F_{20} \sin F_{30} \sin F_{32} 
\right ) \; .
%	(A14) 
\end{eqnarray}
In getting Eqs. (A.12)--(A.14), the equality
\begin{equation}
\sin 2 F_{ij} + \sin 2 F_{jk} + \sin 2 F_{ki} 
\; = \; -4 \sin F_{ij} \sin F_{jk} \sin F_{ki}
%       (A15)
\end{equation}
and (A.6) have been used.
Only three of the twelve asymmetries $\Delta P_{\alpha\beta}$
are independent, and they probe three of the six CP-violating phases
(or their combinations) of $V$. Since only the transitions between 
active neutrinos can in practice be measured, the useful 
probability asymmetries for the study of leptonic CP and T violation
in long-baseline neutrino oscillation experiments are
$\Delta P_{e \mu}$, $\Delta P_{\mu \tau}$ and $\Delta P_{\tau e}$. 

\subsection{Leptonic Unitarity Quadrangles}

The unitarity of $V$ implies that there exist twelve orthogonality
relations and eight normalization conditions among its sixteen matrix
elements. The former corresponds to twelve quadrangles in the 
complex plane, the so-called unitarity quadrangles. To be
explicit, let us write out the twelve orthogonality relations and
name their corresponding quadrangles \cite{GX02b}:
\begin{eqnarray}
{\rm Q}_{se}: & ~ &
V_{s0}V_{e0}^{\ast}+V_{s1}V_{e1}^{\ast}+V_{s2}V_{e2}^{\ast}
+V_{s3}V_{e3}^{\ast}=0 \; ,
\nonumber \\
{\rm Q}_{s\mu}: & ~ &
V_{s0}V_{\mu0}^{\ast}+V_{s1}V_{\mu1}^{\ast}+V_{s2}V_{\mu2}^{\ast}
+V_{s3}V_{\mu3}^{\ast}=0 \; ,
\nonumber \\
{\rm Q}_{s\tau}: & ~ &
V_{s0}V_{\tau0}^{\ast}+V_{s1}V_{\tau1}^{\ast}+V_{s2}V_{\tau2}^{\ast}
+V_{s3}V_{\tau3}^{\ast}=0 \; ,
\nonumber \\ 
{\rm Q}_{e\mu}: & ~ &
V_{e0}V_{\mu0}^{\ast}+V_{e1}V_{\mu1}^{\ast}+V_{e2}V_{\mu2}^{\ast}
+V_{e3}V_{\mu3}^{\ast}=0 \; ,
\nonumber \\ 
{\rm Q}_{e\tau}: & ~ &
V_{e0}V_{\tau0}^{\ast}+V_{e1}V_{\tau1}^{\ast}+V_{e2}V_{\tau2}^{\ast}
+V_{e3}V_{\tau3}^{\ast}=0 \; ,
\nonumber \\
{\rm Q}_{\mu\tau}: & ~ &
V_{\mu0}V_{\tau0}^{\ast}+V_{\mu1}V_{\tau1}^{\ast}+V_{\mu2}V_{\tau2}^{\ast}
+V_{\mu3}V_{\tau3}^{\ast}=0 \; ;
%        (A16)
\end{eqnarray}
and
\begin{eqnarray}
{\rm Q}_{01}: & ~ &
V_{s0}V_{s1}^{\ast}+V_{e0}V_{e1}^{\ast}+V_{\mu0}V_{\mu1}^{\ast}
+V_{\tau0}V_{\tau1}^{\ast}=0 \; ,
\nonumber \\
{\rm Q}_{02}: & ~ &
V_{s0}V_{s2}^{\ast}+V_{e0}V_{e2}^{\ast}+V_{\mu0}V_{\mu2}^{\ast}
+V_{\tau0}V_{\tau2}^{\ast}=0 \; ,
\nonumber \\ 
{\rm Q}_{03}: & ~ &
V_{s0}V_{s3}^{\ast}+V_{e0}V_{e3}^{\ast}+V_{\mu0}V_{\mu3}^{\ast}
+V_{\tau0}V_{\tau3}^{\ast}=0 \; ,
\nonumber \\ 
{\rm Q}_{12}: & ~ &
V_{s1}V_{s2}^{\ast}+V_{e1}V_{e2}^{\ast}+V_{\mu1}V_{\mu2}^{\ast}
+V_{\tau1}V_{\tau2}^{\ast}=0 \; ,
\nonumber \\
{\rm Q}_{13}: & ~ &
V_{s1}V_{s3}^{\ast}+V_{e1}V_{e3}^{\ast}+V_{\mu1}V_{\mu3}^{\ast}
+V_{\tau1}V_{\tau3}^{\ast}=0 \; ,
\nonumber \\
{\rm Q}_{23}: & ~ &
V_{s2}V_{s3}^{\ast}+V_{e2}V_{e3}^{\ast}+V_{\mu2}V_{\mu3}^{\ast}
+V_{\tau2}V_{\tau3}^{\ast}=0 \; .
\end{eqnarray}
%      (A17)
If six mixing angles and six CP-violating phases of $V$ are all
known, one can plot twelve unitarity quadrangles without ambiguities.
Note, however, that each quadrangle has three distinct topologies
in the complex plane. For illustration, we take quadrangle ${\rm Q}_{se}$
for example and show its three topologies in Fig. A.2, where 
the sizes and phases of $V_{si}V^*_{ei}$ (for $i=0,1,2,3$) have 
been fixed. One can see that different topologies of quadrangle
${\rm Q}_{se}$ arise from different orderings of its four sides,
and their areas are apparently different from one another. As a
whole, there are totally thirty-six different topologies among twelve 
unitarity quadrangles.
%%%%%%%%%%%%%%%%%%%% Fig. A.2 %%%%%%%%%%%%%%%%
\begin{figure}
\vspace{-3.5cm}
\epsfig{file=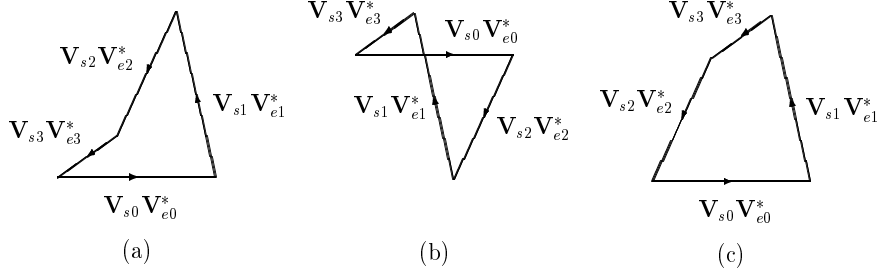,bbllx=-1cm,bblly=4cm,bburx=19cm,bbury=32cm,%
width=15cm,height=22cm,angle=0,clip=}
\vspace{-14.4cm}
\caption{Three distinct topologies of unitarity quadrangle 
${\rm Q}_{se}$ .}
\end{figure}
%%%%%%%%%%%%%%%%%%%%%%%%%%%%%%%%%%%%%%%%%%%

Now we calculate the areas of all unitarity triangles and
relate them to the rephasing invariants of CP violation 
$J^{ij}_{\alpha\beta}$. The results for  
thirty-six different topologies of twelve unitarity quadrangles 
are summarized as follows
%%%%%%%%%%%%%%%%%%%%%%%%%%%
\footnote{It should be noted that the areas of unitarity quadrangles 
under discussion are ``algebraic areas'', namely, they can be either 
positive or negative. Of course, it is always possible to take 
$S^a_{se} = (|J^{10}_{se}|+|J^{21}_{se}|+|J^{32}_{se}|+|J^{03}_{se}|)/4$
or $S^a_{se} = |J^{10}_{se}+J^{21}_{se}+J^{32}_{se}+J^{03}_{se}|/4$,
such that $S^a_{se}$ is definitely positive.
We find, however, that the language of ``algebraic areas'' is simpler and
more convenient in the description of unitarty quadrangles.
In particular, the algebraic area of unitarity quadrangle ${\rm Q}_{se}$
in the case of topology (b) means an algebraic sum of the areas of
its two disassociated triangles, which have opposite signs. Hence 
both $S^b_{se}=0$ and $S^b_{se}<0$ are in general allowed. 
As for topologies (a) and (c) of ${\rm Q}_{se}$, $S^a_{se}>0$ and 
$S^c_{se}>0$ are simply a matter of sign or phase convention.}:
%%%%%%%%%%%%%%%%%%%%%%%%%%%%
\begin{eqnarray}
S_{\alpha\beta}^{a} & = & \frac{1}{4} \left (J_{\alpha\beta}^{10}
+J_{\alpha\beta}^{21}+J_{\alpha\beta}^{32}+J_{\alpha\beta}^{03} \right ) \; ,
\nonumber \\
S_{ij}^{a} & = & \frac{1}{4} \left (J_{es}^{ij}+J_{\mu
e}^{ij}+J_{\tau\mu}^{ij}+J_{s\tau}^{ij} \right ) \; ;
\nonumber \\
S_{\alpha\beta}^{b} & = & \frac{1}{4} \left (J_{\alpha\beta}^{20}
+J_{\alpha\beta}^{12}+J_{\alpha\beta}^{31}+J_{\alpha\beta}^{03} \right ) \; ,
\nonumber \\
S_{ij}^{b} & = & \frac{1}{4} \left (J_{\mu
s}^{ij}+J_{e\mu }^{ij}+J_{\tau e}^{ij}+J_{s\tau}^{ij} \right ) \; ;
\nonumber \\
S_{\alpha\beta}^{c} & = & \frac{1}{4} \left (J_{\alpha\beta}^{10}+
J_{\alpha\beta}^{31}+J_{\alpha\beta}^{23}+J_{\alpha\beta}^{02} \right ) \; ,
\nonumber \\
S_{ij}^{c} & = & \frac{1}{4} \left (J_{e
s}^{ij}+J_{\tau e}^{ij}+J_{\mu\tau}^{ij}+J_{s\mu}^{ij} \right ) \; ,
%      (A18)
\end{eqnarray}
where the subscripts $\alpha\beta = se$, $s\mu$, $s\tau$,
$e\mu$, $e\tau$ or $\mu\tau$, and $ij = 01$, $02$, $03$, $12$, $13$ or 
$23$. Note that the correlation of $J^{ij}_{\alpha\beta}$ allows us to
simplify (A.18). Then each $S^q_{\alpha\beta}$ or 
$S^q_{ij}$ (for $q=a, b, c$) can be expressed as a sum of two independent 
Jarlskog invariants. Such simplified expressions of $S^q_{\alpha\beta}$
and $S^q_{ij}$ depend on the choice of independent $J^{ij}_{\alpha\beta}$,
therefore they may have many different forms. If nine independent
Jarlskog invariants are fixed, however, some expressions of 
$S^q_{\alpha\beta}$ and $S^q_{ij}$ must consist of three 
$J^{ij}_{\alpha\beta}$. This point will become clear later on.

As $J^{ij}_{\alpha\beta} = -J^{ji}_{\alpha\beta} =
-J^{ij}_{\beta\alpha} = J^{ji}_{\beta\alpha}$ holds by definition, 
one may easily obtain $S^q_{\alpha\beta} = -S^q_{\beta\alpha}$ and
$S^q_{ij} = -S^q_{ji}$ (for $q=a$, $b$, $c$). From 
the sum rule in (A.6), one can also find
\begin{equation}
\sum_\alpha S_{\alpha\beta}^q \; = \; 
\sum_\beta S_{\alpha\beta}^q \; = \;
\sum_i S_{ij}^q \; = \; \sum_j S_{ij}^q \; = \; 0 \; ,
%	(A19)
\end{equation}
where $\alpha$ or $\beta$ runs over $(s, e, \mu, \tau)$, and $i$ or $j$
runs over $(0, 1, 2, 3)$. In addition to (A.19), the following
relations can be derived from (A.18):
\begin{eqnarray}
-S_{se}^{a}-S_{e\mu}^{a}-S_{\mu\tau}^{a}+S_{s\tau}^{a} & = &
-S_{01}^{a}-S_{12}^{a}-S_{23}^{a}+S_{03}^{a} \; ,
\nonumber \\
-S_{s\mu}^{a}+S_{e\mu}^{a}-S_{e\tau}^{a}+S_{s\tau}^{a} & = &
-S_{01}^{b}-S_{12}^{b}-S_{23}^{b}+S_{03}^{b} \; ,
\nonumber \\
-S_{se}^{a}-S_{e\tau}^{a}+S_{\mu\tau}^{a}+S_{s\mu}^{a} & = & 
-S_{01}^{c}-S_{12}^{c}-S_{23}^{c}+S_{03}^{c} \; ;
\nonumber \\
-S_{se}^{b}-S_{e\mu}^{b}-S_{\mu\tau}^{b}+S_{s\tau}^{b} & = & 
-S_{02}^{a}+S_{12}^{a}-S_{13}^{a}+S_{03}^{a} \; ,
\nonumber \\
-S_{s\mu}^{b}+S_{e\mu}^{b}-S_{e\tau}^{b}+S_{s\tau}^{b} & = &
-S_{02}^{b}+S_{12}^{b}-S_{13}^{b}+S_{03}^{b} \; ,
\nonumber \\
-S_{se}^{b}-S_{e\tau}^{b}+S_{\mu\tau}^{b}+S_{s\mu}^{b} & = &
-S_{02}^{c}+S_{12}^{c}-S_{13}^{c}+S_{03}^{c} \; ;
\nonumber \\
-S_{se}^{c}-S_{e\mu}^{c}-S_{\mu\tau}^{c}+S_{s\tau}^{c} & = &
-S_{01}^{a}-S_{13}^{a}+S_{23}^{a}+S_{02}^{a} \; ,
\nonumber \\
-S_{s\mu}^{c}+S_{e\mu}^{c}-S_{e\tau}^{c}+S_{s\tau}^{c} & = &
-S_{01}^{b}-S_{13}^{b}+S_{23}^{b}+S_{02}^{b} \; ,
\nonumber \\
-S_{se}^{c}-S_{e\tau}^{c}+S_{\mu\tau}^{c}+S_{s\mu}^{c} & = &
-S_{01}^{c}-S_{13}^{c}+S_{23}^{c}+S_{02}^{c} \; .
%	(A20)
\end{eqnarray}
The correlative relations in (A.19) and (A.20) indicate that there 
are only nine independent $S^q_{\alpha\beta}$ and (or) $S^q_{ij}$, 
corresponding to nine independent $J^{ij}_{\alpha\beta}$.  

Without loss of generality, let us choose the following nine
independent $S^q_{\alpha\beta}$ \cite{GX02b}:
\begin{eqnarray}
S_{se}^{a} & = & \frac{1}{2} 
\left (J_{se}^{02} - J_{se}^{13} - 2J_{se}^{23} \right ) \; ,
\nonumber \\
S_{s\tau}^{a} & = & \frac{1}{2} 
\left (J_{s\tau}^{02} + J_{s\tau}^{03}- J_{s\tau}^{23} \right ) \; ,
\nonumber \\
S_{e\mu}^{a} & = & \frac{1}{2} 
\left (-J_{e\mu}^{12} - J_{e\mu}^{13}- J_{e\mu}^{23} \right ) \; ,
\nonumber \\
S_{se}^{b} & = & \frac{1}{2} 
\left (-J_{se}^{02} - J_{se}^{13} \right ) \; ,
\nonumber \\
S_{s\tau}^{b} & = & \frac{1}{2} 
\left (-J_{s\tau}^{02} + J_{s\tau}^{03}+ J_{s\tau}^{23} \right ) \; ,
\nonumber \\
S_{e\mu}^{b} & = & \frac{1}{2} 
\left (J_{e\mu}^{12} - J_{e\mu}^{13}- J_{e\mu}^{23} \right ) \; ,
\nonumber \\
S_{se}^{c} & = & \frac{1}{2} 
\left (J_{se}^{02} - J_{se}^{13} \right ) \; ,
\nonumber \\
S_{s\tau}^{c} & = & \frac{1}{2} 
\left (J_{s\tau}^{02} + J_{s\tau}^{03}+ J_{s\tau}^{23} \right ) \; ,
\nonumber \\
S_{e\mu}^{c} & = & \frac{1}{2} 
\left (-J_{e\mu}^{12} - J_{e\mu}^{13} + J_{e\mu}^{23} \right ) \; .
%	(A21)
\end{eqnarray}
In terms of six flavor mixing angles and three independent phase 
combinations of $V$, we have expressed the nine independent 
$J^{ij}_{\alpha\beta}$ appearing on the right-hand side 
of (A.21) in section A.2. Then one may directly obtain the explicit 
expressions of the nine-independent $S^q_{\alpha\beta}$ in terms of the 
same mixing angles and CP-violating phases. 

\subsection{Matter Effects on T Violation}

The effective Hamiltonians responsible for the propagation of active
and sterile neutrinos in vacuum and in matter can respectively be 
written as \cite{MSW}
\begin{eqnarray}
{\cal H}_{\rm eff} & = & \frac{1}{2E} \left (M_\nu M^\dagger_\nu \right )
\; =\; \frac{1}{2E} \left (V D^2_\nu V^\dagger \right ) \; ,
\nonumber \\
\tilde{\cal H}_{\rm eff} & = & \frac{1}{2E}
\left (\tilde{M}_\nu \tilde{M}^\dagger_\nu \right ) \; =\; \frac{1}{2E}
\left (\tilde{V} \tilde{D}^2_\nu \tilde{V}^\dagger \right ) \; ,
%	(A22)
\end{eqnarray}
where $D_\nu \equiv {\rm Diag}\{m_0, m_1, m_2, m_3 \}$ with $m_i$ being
the fundamental neutrino masses in vacuum,
$\tilde{D}_\nu \equiv {\rm Diag}
\{\tilde{m}_0, \tilde{m}_1, \tilde{m}_2, \tilde{m}_3 \}$ with
$\tilde{m}_i$ being the effective neutrino masses in matter,
and $E \gg m_i$ denotes the neutrino beam energy. 
The deviation of $\tilde{\cal H}_{\rm eff}$ from ${\cal H}_{\rm eff}$
is given by
\begin{equation}
\Delta {\cal H}_{\rm eff} \; \equiv \; \tilde{\cal H}_{\rm eff}
- {\cal H}_{\rm eff}  =
\left ( \matrix{
a' & 0 & 0 & 0 \cr
0 & a & 0 & 0 \cr
0 & 0 & 0 & 0 \cr
0 & 0 & 0 & 0 \cr} \right ) \; ,
%	(A23)
\end{equation}
where $a = \sqrt{2} ~ G_{\rm F} N_e$ and
$a' = \sqrt{2} ~ G_{\rm F} N_n/2$ with 
$N_e$ and $N_n$ being the background densities of electrons and 
neutrons \cite{Grimus}, respectively. In the literature one often
assumes a constant earth density profile (i.e., $N_e$ = constant and
$N_n$ = constant), which is a good approximation for all of the 
presently-proposed long-baseline neutrino experiments.

Now let us introduce the commutators of $4\times 4$ lepton mass matrices to
describe the mixing of one sterile and three active neutrinos. 
Without loss of any generality, we continue to work in the afore-chosen 
flavor basis, where $M_l$ takes the diagonal form
$D_l = {\rm Diag} \{ m_s, m_e, m_\mu, m_\tau \}$ 
with $m_s = 0$. Note that we have assumed the $(1, 1)$ element of 
$D_l$ to be zero, because there is no counterpart of the sterile 
neutrino $\nu_s$ in the charged lepton sector. We shall see later on that
our physical results are completely independent of $m_s$, no matter what 
value it may take. The commutator of lepton mass matrices in vacuum and 
that in matter can then be defined as
\begin{eqnarray}
C & \equiv & i \left [ M_\nu M^\dagger_\nu ~ , M_l M^\dagger_l \right ]
= i \left [V D^2_\nu V^\dagger , D^2_l \right ] \; , 
\nonumber \\
\tilde{C} & \equiv & i \left [ \tilde{M}_\nu \tilde{M}^\dagger_\nu ~ , 
M_l M^\dagger_l \right ] =
i \left [ \tilde{V} \tilde{D}^2_\nu \tilde{V}^\dagger , 
D^2_l \right ] \; .
%	(A24)
\end{eqnarray}
Obviously $C$ and $\tilde C$ are traceless Hermitian matrices. In terms
of neutrino masses and flavor mixing matrix elements, we obtain the
explicit expressions of $C$ and $\tilde C$ as follows:
\begin{eqnarray}
C & = & i \left ( \matrix{
0 & \Delta_{es} Z_{se} & \Delta_{\mu s} Z_{s \mu} & 
\Delta_{\tau s} Z_{s \tau} \cr
\Delta_{se} Z_{es} & 0 & \Delta_{\mu e} Z_{e \mu} & 
\Delta_{\tau e} Z_{e \tau} \cr
\Delta_{s \mu} Z_{\mu s} & \Delta_{e \mu} Z_{\mu e} & 0 
& \Delta_{\tau \mu} Z_{\mu \tau} \cr
\Delta_{s \tau} Z_{\tau s} & \Delta_{e \tau} Z_{\tau e} 
& \Delta_{\mu \tau} Z_{\tau \mu} & 0 \cr} \right ) \; ,
\nonumber \\ \nonumber \\
\tilde{C} & = & i \left ( \matrix{
0 & \Delta_{es} \tilde{Z}_{se} & \Delta_{\mu s} \tilde{Z}_{s \mu} 
& \Delta_{\tau s} \tilde{Z}_{s \tau} \cr
\Delta_{se} \tilde{Z}_{es} & 0 & \Delta_{\mu e} \tilde{Z}_{e \mu} 
& \Delta_{\tau e} \tilde{Z}_{e \tau} \cr
\Delta_{s \mu} \tilde{Z}_{\mu s} & \Delta_{e \mu} \tilde{Z}_{\mu e} & 0 
& \Delta_{\tau \mu} \tilde{Z}_{\mu \tau} \cr
\Delta_{s \tau} \tilde{Z}_{\tau s} & \Delta_{e \tau} \tilde{Z}_{\tau e} 
& \Delta_{\mu \tau} \tilde{Z}_{\tau \mu} & 0 \cr} \right ) \; ,
%	(A25)
\end{eqnarray}
where $\Delta_{\alpha \beta} \equiv m^2_\alpha - m^2_\beta$ for
$\alpha \neq \beta$ running over $(s, e, \mu, \tau)$, and
\begin{eqnarray}
Z_{\alpha \beta} & \equiv & \sum^3_{i=0} \left ( m^2_i V_{\alpha i} 
V^*_{\beta i} \right ) \; ,
\nonumber \\
\tilde{Z}_{\alpha \beta} & \equiv & \sum^3_{i=0} \left ( \tilde{m}^2_i
\tilde{V}_{\alpha i} \tilde{V}^*_{\beta i} \right ) \; .
%	(A26)
\end{eqnarray}
One can see that $\Delta_{\beta \alpha} = - \Delta_{\alpha \beta}$,
$Z_{\beta \alpha} = Z^*_{\alpha \beta}$ and
$\tilde{Z}_{\beta \alpha} = \tilde{Z}^*_{\alpha \beta}$ hold.
To find out how $\tilde{Z}_{\alpha \beta}$ is related to 
$Z_{\alpha \beta}$,
we need to establish the relation between $\tilde C$ and $C$. 
Taking account of (A.22) and (A.23), we immediately obtain
\begin{equation}
\tilde{C} \; = \; 2iE
\left [ \tilde{\cal H}_{\rm eff} ~ , D^2_l \right ] 
= C + 2iE \left [ \Delta {\cal H}_{\rm eff} ~ , D^2_l \right ]  
= C \; .
%	(A27)
\end{equation}
This interesting result indicates that the commutator
of lepton mass matrices in vacuum is invariant under terrestrial 
matter effects.
As a straightforward consequence of $\tilde{C} = C$, we arrive at
$\tilde{Z}_{\alpha \beta} = Z_{\alpha \beta}$ from (A.25); i.e.,
\begin{equation}
\sum^3_{i=0} \left ( \tilde{m}^2_i \tilde{V}_{\alpha i} 
\tilde{V}^*_{\beta i} \right ) \; = \; 
\sum^3_{i=0} \left ( m^2_i V_{\alpha i} V^*_{\beta i} \right ) \; ;
%	(A28)
\end{equation}
or equivalently 
\begin{eqnarray}
~~~ \sum^3_{i=1} \left (\tilde{\Delta}_{i0}
\tilde{V}_{\alpha i} \tilde{V}^*_{\beta i} \right ) \; = \;
\sum^3_{i=1} \left ( \Delta_{i0} V_{\alpha i} V^*_{\beta i} \right ) \; ,
%	(A29)
\end{eqnarray}
where $\Delta_{i0} \equiv m^2_i - m^2_0$ and 
$\tilde{\Delta}_{i0} \equiv \tilde{m}^2_i - \tilde{m}^2_0$ for 
$i=1,2,3$. It becomes obvious that the validity of (A.28) or (A.29) 
has nothing to do with the assumption of $m_s = 0$ in the charged 
lepton sector. Note that the results obtained above are only valid for
neutrinos propagating in vacuum and in matter. As for antineutrinos, 
the corresponding sum rule can straightforwardly be written out from 
(A.28) or (A.29) through the replacements $V\Longrightarrow V^*$, 
$a \Longrightarrow -a$ and $a' \Longrightarrow -a'$.

To describe CP or T violation in neutrino oscillations, 
we consider the rephasing-invariant relationship
$\tilde{Z}_{\alpha \beta} \tilde{Z}_{\beta \gamma} \tilde{Z}_{\gamma \alpha}
= Z_{\alpha \beta} Z_{\beta \gamma} Z_{\gamma \alpha}$ 
for $\alpha \neq \beta \neq \gamma$ running over $(s, e, \mu, \tau)$.
The imaginary parts of $Z_{\alpha \beta} Z_{\beta \gamma} Z_{\gamma \alpha}$
and 
$\tilde{Z}_{\alpha \beta} \tilde{Z}_{\beta \gamma} \tilde{Z}_{\gamma \alpha}$
read explicitly as
\begin{eqnarray}
{\rm Im} ( Z_{\alpha\beta} Z_{\beta\gamma} Z_{\gamma\alpha} )
& = & \sum^3_{i=1} \sum^3_{j=1} \sum^3_{k=1} \left [ \Delta_{i0} \Delta_{j0}
\Delta_{k0} ~ {\rm Im} \left ( V_{\alpha i} V_{\beta j} V_{\gamma k} 
V^*_{\alpha k} V^*_{\beta i} V^*_{\gamma j} \right ) \right ] \; ,
\nonumber \\
{\rm Im} ( \tilde{Z}_{\alpha\beta} \tilde{Z}_{\beta\gamma} 
\tilde{Z}_{\gamma\alpha} )
& = & \sum^3_{i=1} \sum^3_{j=1} \sum^3_{k=1} \left [ \tilde{\Delta}_{i0} 
\tilde{\Delta}_{j0} \tilde{\Delta}_{k0} ~ {\rm Im}
\left ( \tilde{V}_{\alpha i} \tilde{V}_{\beta j} \tilde{V}_{\gamma k} 
\tilde{V}^*_{\alpha k} \tilde{V}^*_{\beta i} 
\tilde{V}^*_{\gamma j} \right ) \right ] \; ,
%	(A30)
\end{eqnarray}
and their equality allows us to derive an interesting relation between 
the Jarlskog invariant $J^{ij}_{\alpha\beta}$ in vacuum and its effective 
counterpart in matter,
\begin{equation}
\tilde{J}^{ij}_{\alpha\beta} \; \equiv \; 
{\rm Im} \left ( \tilde{V}_{\alpha i} \tilde{V}_{\beta j} 
\tilde{V}^*_{\alpha j} \tilde{V}^*_{\beta i} \right ) \; ,
%       (A31)
\end{equation}
where the Greek subscripts run over $(s, e, \mu, \tau)$ and the Latin
superscripts run over $(0, 1, 2, 3)$. The result is \cite{Xing01}
\begin{eqnarray}
& & \tilde{\Delta}_{10} \tilde{\Delta}_{20} \tilde{\Delta}_{30} \sum^3_{i=1}
\left ( \tilde{J}^{0i}_{\alpha\beta} |\tilde{V}_{\gamma i}|^2 + 
\tilde{J}^{0i}_{\beta\gamma} |\tilde{V}_{\alpha i}|^2 + 
\tilde{J}^{0i}_{\gamma\alpha} |\tilde{V}_{\beta i}|^2 
\right ) 
\nonumber \\
& & + \sum^3_{i=1} \sum^3_{j=1} \left [ \tilde{\Delta}_{i0} 
\tilde{\Delta}^2_{j0} \left (
\tilde{J}^{ij}_{\alpha\beta} |\tilde{V}_{\gamma j}|^2 + 
\tilde{J}^{ij}_{\beta\gamma} |\tilde{V}_{\alpha j}|^2 + 
\tilde{J}^{ij}_{\gamma\alpha} |\tilde{V}_{\beta j}|^2 
\right ) \right ] 
\nonumber \\
& = & \Delta_{10} \Delta_{20} \Delta_{30} \sum^3_{i=1}
\left ( J^{0i}_{\alpha\beta} |V_{\gamma i}|^2 + 
J^{0i}_{\beta\gamma} |V_{\alpha i}|^2 + J^{0i}_{\gamma\alpha} |V_{\beta i}|^2 
\right ) 
\nonumber \\
& & + \sum^3_{i=1} \sum^3_{j=1} \left [ \Delta_{i0} \Delta^2_{j0} \left (
J^{ij}_{\alpha\beta} |V_{\gamma j}|^2 + 
J^{ij}_{\beta\gamma} |V_{\alpha j}|^2 + J^{ij}_{\gamma\alpha} |V_{\beta j}|^2 
\right ) \right ] \; .
%	(A32)
\end{eqnarray}
If one ``switches off'' the mass of the sterile neutrino and its mixing 
with active neutrinos (i.e., $a' = 0$,
$\Delta_{i0} = m^2_i$, $\tilde{\Delta}_{i0} = \tilde{m}^2_i$,
$J^{0i}_{\alpha\beta} =0$, and $\tilde{J}^{0i}_{\alpha\beta} =0$),
then (A.32) is simplified to the elegant Naumov form \cite{Naumov},
as shown in (4.4).

The matter-corrected CP-violating parameters $\tilde{J}^{ij}_{\alpha\beta}$
can, at least in principle, be determined from the measurement of 
CP- and T-violating effects in a variety of long-baseline neutrino 
oscillation experiments. The conversion probability of a neutrino 
$\nu_\alpha$ to another neutrino $\nu^{~}_\beta$ in matter is given as
\begin{equation}
\tilde{P}(\nu_\alpha \rightarrow \nu^{~}_\beta) \; = \;
\delta_{\alpha\beta} - 4 \sum_{i<j} \left [ {\rm Re} \left (
\tilde{V}_{\alpha i} \tilde{V}_{\beta j} \tilde{V}^*_{\alpha j} 
\tilde{V}^*_{\beta i} \right ) 
\sin^2 \tilde{F}_{ji} \right ] - 2 \sum_{i<j} 
\left [ \tilde{J}^{ij}_{\alpha\beta} \sin (2 \tilde{F}_{ji}) \right ] \; ,
%	(A33)
\end{equation}
where $\tilde{F}_{ji} \equiv 1.27 \tilde{\Delta}_{ji} L/E$ with
$\tilde{\Delta}_{ji} \equiv \tilde{m}^2_j - \tilde{m}^2_i$.
The transition probability $\tilde{P}(\nu^{~}_\beta \rightarrow \nu_\alpha)$ 
can directly be read off from (A.33), if the replacements 
$\tilde{J}^{ij}_{\alpha\beta} \Longrightarrow -\tilde{J}^{ij}_{\alpha\beta}$ 
are made
%%%%%%%%%%%%%%%%%%%%%%%
\footnote{Note that the differences of effective neutrino masses 
$\tilde{\Delta}_{i0}$ (for $i=1,2,3$), which must be CP-conserving,
keep unchanged for the replacements 
$J^{ij}_{\alpha\beta} \Longrightarrow -J^{ij}_{\alpha\beta}$. Therefore the
sign flip of $J^{ij}_{\alpha\beta}$ results in that of
$\tilde{J}^{ij}_{\alpha\beta}$, 
as indicated by the sum rules given in (A.32).}.
%%%%%%%%%%%%%%%%%%%%%%% 
To obtain the probability
$\tilde{P}(\overline{\nu}_\alpha \rightarrow \overline{\nu}^{~}_\beta)$,
however, both the replacements 
$J^{ij}_{\alpha\beta} \Longrightarrow -J^{ij}_{\alpha\beta}$
and $(a, a') \Longrightarrow (-a, -a')$ need be made for (A.33). 
In this case, 
$\tilde{P}(\overline{\nu}_\alpha \rightarrow \overline{\nu}^{~}_\beta)$
is not equal to $\tilde{P}(\nu^{~}_\beta \rightarrow \nu_\alpha)$. The
difference between 
$\tilde{P}(\overline{\nu}_\alpha \rightarrow \overline{\nu}_\beta)$ and
$\tilde{P}(\nu_\beta \rightarrow \nu_\alpha)$ is a false signal of 
CPT violation, induced actually by the matter effect \cite{XingCPT}. 
Thus the CP-violating asymmetry between
$\tilde{P}(\nu_\alpha \rightarrow \nu^{~}_\beta)$ and
$\tilde{P}(\overline{\nu}_\alpha \rightarrow \overline{\nu}^{~}_\beta)$
is in general different from the T-violating asymmetry between
$\tilde{P}(\nu_\alpha \rightarrow \nu^{~}_\beta)$ and
$\tilde{P}(\nu^{~}_\beta \rightarrow \nu_\alpha)$. The latter
can be explicitly expressed as follows:
\begin{eqnarray}
\Delta \tilde{P}_{\alpha\beta} & \equiv & 
\tilde{P}(\nu^{~}_\beta \rightarrow \nu^{~}_\alpha) ~ - ~ 
\tilde{P}(\nu^{~}_\alpha \rightarrow \nu^{~}_\beta) \;
\nonumber \\
& = & 16\left (\tilde{J}^{23}_{\alpha\beta} \sin \tilde{F}_{21} 
\sin \tilde{F}_{31} \sin \tilde{F}_{32} - \tilde{J}^{02}_{\alpha\beta} 
\sin \tilde{F}_{10} \sin \tilde{F}_{20} \sin \tilde{F}_{21}
\right . 
\nonumber \\
&& ~ - \left . \tilde{J}^{03}_{\alpha\beta} \sin \tilde{F}_{10} 
\sin \tilde{F}_{30} \sin \tilde{F}_{31} \right ) \; .
%	(A34)
\end{eqnarray}
If the hierarchical patterns of neutrino masses and flavor mixing
angles are assumed, the expression of $\Delta \tilde{P}_{\alpha\beta}$ 
may somehow be simplified \cite{Tanimoto}. Note that only three of 
the twelve nonvanishing asymmetries $\Delta \tilde{P}_{\alpha\beta}$
are independent, as a consequence of the unitarity of $\tilde V$ or 
the correlation of $\tilde{J}^{ij}_{\alpha\beta}$. Since
only the transition probabilities of active neutrinos can be 
realistically measured, we are more interested in the T-violating
asymmetries $\Delta \tilde{P}_{e \mu}$, $\Delta \tilde{P}_{\mu \tau}$ 
and $\Delta \tilde{P}_{\tau e}$. 
The overall matter contamination residing in $\Delta \tilde{P}_{\alpha\beta}$ 
is usually expected to be insignificant. The reason is simply that the 
terrestrial matter effects in $\tilde{P}(\nu_\alpha \rightarrow \nu^{~}_\beta)$
and $\tilde{P}(\nu^{~}_\beta \rightarrow \nu_\alpha)$, 
which both depend on the parameters $(a, a')$, may partly (even essentially) 
cancel each other in the T-violating asymmetry 
$\Delta \tilde{P}_{\alpha\beta}$. In contrast, 
$\tilde{P}(\nu_\alpha \rightarrow \nu^{~}_\beta)$ and
$\tilde{P}(\overline{\nu}_\alpha \rightarrow \overline{\nu}^{~}_\beta)$
are associated respectively with $(+a, +a')$ and $(-a, -a')$, thus
there should not have large cancellation of matter effects in the 
corresponding CP-violating asymmetries.

\end{document}